\documentclass[preprint,2p,onecolumn,numbers,sort&compress]{elsarticle}

\makeatletter
\protected\def\xvcenter{%
  \hbox\bgroup$\everyvbox{\everyvbox{}\aftergroup\m@th\aftergroup$\aftergroup\egroup}%
  \vcenter
}

\DeclareRobustCommand{\midscript}[1]{
  \mathchoice{\mid@script\scriptstyle{#1}}
    {\mid@script\scriptstyle{#1}}
    {\mid@script\scriptscriptstyle{#1}}
    {\mid@script\scriptscriptstyle{#1}}
}
\newcommand{\mid@script}[2]{
  \vcenter{\hbox{$\m@th#1#2$}}
}

\DeclareRobustCommand{\textmidscript}[1]{%
  \xvcenter{\hbox{\scriptsize#1}}%
}
\makeatother

\def\appendixname{}
\makeatletter
\renewcommand\appendix{\par
  \setcounter{section}{0}%
  \setcounter{subsection}{0}%
  \setcounter{equation}{0}
  \gdef\thefigure{\@Alph\c@section.\arabic{figure}}%
  \gdef\thetable{\@Alph\c@section.\arabic{table}}%
  \gdef\thesection{\appendixname\@Alph\c@section}%
  \@addtoreset{equation}{section}%
  \gdef\theequation{\@Alph\c@section.\arabic{equation}}%
  \addtocontents{toc}{\string\let\string\numberline\string\tmptocnumberline}{}{}
}
\makeatother

\usepackage{pdflscape}
\usepackage{afterpage}
\usepackage{fancyhdr}
\fancypagestyle{lscape}{%
\fancyhf{} 
\fancyfoot[LE]{}
\fancyfoot[LO] {}

}

\usepackage{rotating}
\usepackage{pdflscape}
\usepackage{eso-pic}
\usepackage{zref-user}
\usepackage{scalerel}

\makeatletter
\newcounter{cntsideways}
\AddToShipoutPictureBG{%
 \ifnum\zref@extractdefault{rotate\number\value{page}}{page}{0}=0
  \PLS@RemoveRotate
 \else
  \PLS@AddRotate{90}%
 \fi}

\newcommand\rotatesidewayslabel{\stepcounter{cntsideways}%
 \zlabel{tmp\thecntsideways}\zlabel{rotate\zref@extractdefault{tmp\thecntsideways}{page}{0}}}
\makeatother

\usepackage{amsmath}
\usepackage{amssymb}
\usepackage{booktabs}
\usepackage[inline]{enumitem}
\usepackage{geometry}
\usepackage{hyperref}

\usepackage{cleveref}
\crefname{lstlisting}{listing}{listings}
\Crefname{lstlisting}{Listing}{Listings}
\crefname{appsec}{appendix}{appendices}
\Crefname{appsec}{Appendix}{Appendices}

\usepackage[nowatermark]{fixmetodonotes}
\usepackage{isotope}
\usepackage{verbatim}

\usepackage{siunitx}
\DeclareSIUnit\clight{\text{\ensuremath{c}}}
\DeclareSIUnit\atom{\text{atom}}
\DeclareSIUnit\ton{\text{t}}
\DeclareSIUnit\parsec{\text{pc}}
\usepackage{xcolor}
\definecolor{lightgray}{rgb}{.9,.9,.9}
\definecolor{darkgray}{rgb}{.4,.4,.4}
\definecolor{purple}{rgb}{0.65, 0.12, 0.82}
\definecolor{darkred}{rgb}{0.64, 0.0, 0.0}

\usepackage{listings}
\lstdefinelanguage{MARLEYConfig}{
  keywords={seed, structure, reactions, source, direction, log,
    executable_settings,},
  keywordstyle=\color{black}\bfseries,
  ndkeywords={type, neutrino, Emin, Emax, temperature, eta, x, y, z,
    file, level, overwrite, events, output, format, mode},
  ndkeywordstyle=\color{black}\bfseries,
  identifierstyle=\color{black},
  sensitive=false,
  comment=[l]{//},
  morecomment=[s]{/*}{*/},
  commentstyle=\color{blue}\ttfamily,
  stringstyle=\color{darkred}\ttfamily,
  escapechar=|,
  morestring=[b]"
}

\DeclareFontFamily{OT1}{pzc}{}
\DeclareFontShape{OT1}{pzc}{m}{it}{<-> s * [1.10] pzcmi7t}{}
\DeclareMathAlphabet{\mathpzc}{OT1}{pzc}{m}{it}

\usepackage[mathlines]{lineno}

\newcounter{bla}

\journal{Computer Physics Communications}

\begin{document}

\newcommand{\marley}{\texttt{MARLEY}}
\newcommand{\rootcern}{ROOT}
\newcommand{\ascii}{ASCII}
\newcommand{\hepevt}{HEPEVT}
\newcommand{\cpp}{C\textmidscript{++}} 
\newcommand{\version}{1.2.0}
\newcommand{\minGCCversion}{4.9.4}
\newcommand{\minClangVersion}{3.5.2}
\newcommand{\myRepo}{\url{http://github.com/MARLEY-MC/marley}}
\newcommand{\myDOI}{\href{http://doi.org/10.5281/zenodo.3905443}
{10.5281/zenodo.3905443}}
\newcommand{\totKinE}{\varepsilon}
\newcommand{\fragmentKinELab}{ \varepsilon_\text{lab} }
\newcommand{\totKinEmax}{\varepsilon_\text{max}}
\newcommand{\totWidth}{\Gamma}
\newcommand{\SMatrixElement}{\left< S_{\ell j} \right>}
\newcommand{\xmin}{a}
\newcommand{\xmax}{b}
\newcommand{\fmax}{f_\text{max}}
\newcommand{\myN}{N}
\newcommand{\fragment}{ \ensuremath{a} }
\newcommand{\matchThresh}{ V_\text{thresh} }
\newcommand{\matchScale}{ \mathcal{S} }
\newcommand{\pdgCodeVar}{ \mathrm{P_{ID}} }
\newcommand{\levelIndex}{ \Lambda }
\newcommand{\ExCut}{ E_x^{\,\text{cut}} }
\newcommand{\decayProductIndex}{ \ensuremath{u} }
\newcommand{\exitChannelIndex}{ \ensuremath{e} }
\newcommand{\termIndex}{ \ensuremath{b} }
\newcommand{\FluxAvgTotXSec}{ \left<\sigma\right> }
\newcommand{\FluxAvgDiffXSec}{ \left<\frac{d\sigma}{dx}\right> }
\newcommand{\fragmentMomCM}{ \mathpzc{k} }

\newcommand{\FNRpLep}{\mathpzc{K}}
\newcommand{\FNReLep}{\mathpzc{E}}
\newcommand{\FNRpLepEff}{\FNRpLep_\text{\,\,eff}}
\newcommand{\FNReLepEff}{\FNReLep_\text{\,eff}}

\begin{frontmatter}



\title{Simulating low-energy neutrino interactions with \marley}


\author[a,b]{S. Gardiner}

\cortext[author] {Corresponding author.\\\textit{E-mail address:}
  gardiner@fnal.gov}
\address[a]{Fermi National Accelerator Laboratory, P.O. Box 500, Batavia,
  IL 60510, USA}
\address[b]{University of California, Davis, One Shields Avenue, Davis,
  CA 95616, USA}
\begin{abstract}

Monte Carlo event generators are a critical tool for the interpretation of data
obtained by neutrino experiments. Several modern event generators are available
which are well-suited to the GeV energy scale used in studies of accelerator
neutrinos. However, theoretical modeling differences make their immediate
application to lower energies difficult. In this paper, I present a new event
generator, \marley, which is designed to better address the simulation needs of
the low-energy (tens of MeV and below) neutrino community. The code is written
in \cpp 14 with an optional interface to the popular \rootcern\ data analysis
framework. The current release of \marley\ (version \version) emphasizes
simulations of the reaction $\isotope[40]{Ar}(\nu_e, e^{-})\isotope[40]{K}^{*}$
but is extensible to other channels with suitable user input. This paper
provides detailed documentation of \marley's implementation and usage,
including guidance on how generated events may be analyzed and how \marley\ may
be interfaced with external codes such as Geant4. Further information about
\marley\ is available on the official website at \url{http://www.marleygen.org}.

\end{abstract}


\end{frontmatter}




\noindent
{\bf PROGRAM SUMMARY} \\
\begin{small}
\noindent
{\em Program Title:} \marley\ \version                     \\
{\em Developer's respository link:} \myRepo\ \\
{\em Licensing provisions:} GNU General Public License 3.0 \\
{\em Programming language:} \cpp 14                         \\
{\em External routines/libraries used:}
GNU Scientific Library \cite{Galassi2009,GSLWebsite} (required),
\rootcern\ \cite{Brun1997,ROOTWebsite} (optional) \\
{\em Nature of problem:} Simulation of neutrino-nucleus scattering events at
energies of tens-of-\si{\MeV} and below \\
{\em Solution method:} Initial two-to-two scattering kinematics are sampled
using the allowed approximation differential cross section and tables of
precomputed nuclear matrix elements. Subsequent de-excitations of the remnant
nucleus are simulated using a Monte Carlo implementation of the Hauser-Feshbach
statistical model and tabulated $\gamma$-ray decay schemes for discrete nuclear
levels. \\
{\em Restrictions:} Input data are provided with the code that are suitable for
producing simulations of the charged-current reaction $\isotope[40]{Ar}(\nu_e,
e^{-})\isotope[40]{K}^{*}$, coherent elastic neutrino-nucleus scattering on
spin-zero target nuclei, and neutrino-electron elastic scattering on any atomic
target. Preparation of new reaction input files (whose format is documented in
\cref{sec:reaction_file_format}) would enable other reaction channels and
nuclear targets to be handled by the existing code framework. Although there is
no maximum neutrino energy enforced by the code itself, realistic
neutrino-nucleus scattering events may be generated up to roughly
\SI{50}{\MeV}. Above this energy, the effects of forbidden nuclear transitions,
which are neglected in the current treatment of the cross sections (see
\cref{sec:xsec_theory}), become increasingly important.
\end{small}

\section{Introduction}
\label{sec:intro}

Monte Carlo event generators are a widely-used tool in nuclear and particle
physics. These computer programs implement probabilistic models of physics
processes and produce corresponding sets of \textit{events}: lists of particles
(represented by their charges, 4-momenta, etc.) involved in simulated
interactions.

While helpful as an aid to theoretical calculations, event generators are also
often used by experiments for designing detectors, estimating efficiencies and
backgrounds, assessing systematic uncertainties, and interpreting the results
of measurements. Examples of event generators include PYTHIA
\cite{Sjostrand2015} and Herwig \cite{Herwig7} for high-energy particle
collisions, HYDJET++ \cite{Lokhtin2009} for relativistic heavy ion collisions
specifically, FREYA \cite{Verbeke2015} for fission, and DECAY4
\cite{Ponkratenko2000} for radioactive decays of unstable isotopes.

For studies of neutrinos, much community effort has been directed toward the
development of event generators suitable for use by accelerator-based
experiments at facilities like J-PARC \cite{Nagamiya2012} and Fermilab
\cite{Fava2018}. These experiments employ beams of primarily muon-flavor
neutrinos which are produced over a broad energy range between the low hundreds
of \si{\MeV} to the tens of \si{\GeV}. Neutrino scattering on atomic nuclei is
the primary means of detection, and several modern event generators provide
widely-used models of the relevant physics, including GENIE \cite{GENIE}, GiBUU
\cite{Buss2012}, NEUT \cite{Hayato2009}, and NuWro \cite{Golan2012}.

Although important differences exist between each of these generators, all
share a similar conceptual treatment of neutrino-nucleus interactions. Each
scattering event on a complex nucleus is taken to involve a neutrino striking a
single bound nucleon.\footnote{Two particle-two hole interactions, which
involve a pair of nucleons, are included in modern generators \cite{Katori2015}
but neglected here for simplicity. I likewise neglect coherent processes, which
involve the nucleus as a whole.} A removal energy and initial 3-momentum are
assigned to the struck nucleon using a model of the nuclear ground state.
Traditionally this is done using a variant of the Fermi gas model (e.g., that
of ref.~\cite{BodekRitchie}), but implementations of more sophisticated
treatments, such as the Correlated Basis Function approach
\cite{Benhar1989,Benhar1994}, are beginning to become available
\cite{King2020}.

With the initial state fully defined, the generator then simulates
production of the particles that emerge from the neutrino-nucleon
interaction vertex. A variety of models must be used at this stage due to
competition between multiple nucleon-level processes (e.g, quasi-elastic and
deep inelastic scattering) which may occur at accelerator neutrino energies.

Because they are subject to the strong force, outgoing hadrons from the primary
neutrino interaction will often rescatter within the nuclear medium. These
\textit{final-state interactions} (FSIs) can have a pronounced effect on the
kinematics and multiplicities of the hadrons that ultimately exit the nucleus.
Intranuclear hadron transport and FSIs are handled in GENIE, NEUT, and NuWro
using variations of the intranuclear cascade (INC) model
\cite{Dytman2009,Ma2017,Golan2012b}. This model assumes that propagation of a
hadron $h$ through the nucleus may be described in terms of a mean free path
\begin{equation}
\lambda(h,E,r) = \frac{ 1 }{ \rho(r) \, \sigma_{hN}(E) } \,,
\end{equation}
where $E$ is the hadron energy, $r$ is its radial position within the nucleus,
and $\rho$ is the number density of nucleons. The total cross section for $h$
scattering on a free nucleon, $\sigma_{hN}$, may be used directly but is
typically modified with approximate corrections for nuclear effects.

The GiBUU code simulates intranuclear hadron transport using a semi-classical
model which considers the time evolution of the phase space density for each
hadronic species. The equations describing the behavior of distinct kinds of
hadrons are coupled through the common nuclear mean field and through a
collision term which represents the influence of FSIs. The numerical
implementation adopts a test-particle ansatz to solve the relativistic
Boltzmann-Uehling-Uhlenbeck (BUU) equation. Extensive documentation of GiBUU's
treatment of various nuclear reactions, including neutrino-nucleus scattering,
is available in refs.~\cite{Buss2012,GiBUUpubs}. The INC models used by other
generators for FSIs may largely be regarded as approximate simplifications of
the BUU approach \cite[sec.~2]{Mosel2019}.

When all produced particles either escape the nucleus or are re-absorbed by it,
the hadron transport stage of the simulation is complete. This typically marks
the end of the physics workflow needed to generate a single neutrino-nucleus
scattering event. The finished events may be used both for standalone
calculations of kinematic distributions and as input to later stages of an
experiment's detector simulation chain.\footnote{Ref.~\cite{NOvAchain}
describes a GENIE-based example used by the NOvA experiment.}

In addition to accelerator-based neutrino oscillation experiments, there is
also worldwide interest in pursuing detailed measurements of lower-energy
(tens-of-\si{\MeV} and below) neutrinos produced by supernovae
\cite{Raffelt2011,Scholberg2012,Horiuchi2018}, by the Sun
\cite{DUNEsolar,Bakhti2020}, and by terrestrial facilities via pion and muon
decays at rest \cite{Harnik2020,Suekane2016,Rott2020,COH2018,Grant2015}.
Several distinct reaction modes will enable neutrino detection in these
measurements. Elastic scattering on electrons and protons \cite{Beacom2002} is
flavor-blind but produces signal events down to arbitrarily low neutrino
energies. Coherent elastic neutrino-nucleus scattering (CEvNS), a
neutral-current (NC) process in which a neutrino scatters off of a complex
nucleus and leaves it in its ground state \cite{Freedman1974}, shares these
properties and was recently observed for the first time by the COHERENT
experiment \cite{Akimov2017,Akimov2021}. Apart from reactions involving complex
nuclei, the only inelastic channel available at tens-of-\si{\MeV} energies is
charged-current (CC) absorption of $\bar{\nu}_e$ on free protons via the
inverse beta decay (IBD) reaction\footnote{The analogous reaction $\nu_e + n
\rightarrow p + e^{-}$ is also possible. However, it is not practical for use
in a neutrino detector due to the instability of free neutrons.}
\begin{equation}
\bar{\nu}_e + p \rightarrow n + e^{+} \,.
\end{equation}
The IBD cross section is precisely known \cite{Ankowski2016b,Strumia2003} and
dominates the expected signal for supernova neutrinos in water Cherenkov and
liquid scintillator detectors \cite{Scholberg2012}.

The low-energy neutrino interaction modes listed so far do not present any
special difficulties from an event generator perspective. Relatively simple
expressions are adequate to compute differential cross sections for all but the
most precise calculations of these processes, and, with the recent addition of
a CEvNS model \cite{GENIEv32preview}, the GENIE event generator currently
provides an implementation of all of them that may be suitable for use by
low-energy neutrino experiments.

In contrast to other low-energy processes, inelastic neutrino scattering on all
but the lightest complex nuclei (e.g., deuterium) is theoretically cumbersome,
requiring an elaborate treatment of nuclear physics in order to fully describe
the interactions. Despite the significant challenges involved, however,
obtaining realistic simulations of low-energy inelastic neutrino-nucleus
scattering is highly desirable for a variety of scientific applications.
Principal among these is detection of low-energy astrophysical neutrinos by the
upcoming Deep Underground Neutrino Experiment (DUNE) \cite{dunetdrvol2}. Thanks
to the experiment's planned use of four ten-kiloton liquid argon time
projection chambers (LArTPCs), DUNE has the potential to perform detailed
measurements of these neutrinos via the charged-current reaction
\begin{equation}
\label{eq:CC40Ar}
\nu_e + \isotope[40]{Ar} \rightarrow e^{-} + \isotope[40]{K}^* \,.
\end{equation}
This process is anticipated to provide most of the signal in DUNE for the
neutrino burst associated with a galactic core-collapse supernova, granting the
experiment the potential for unique sensitivity among large detectors to the
$\nu_e$ component of the supernova neutrino flux \cite{Abi2020,Ankowski2016}.
DUNE also shows substantial promise as a detector for studying solar neutrinos
\cite{DUNEsolar}.

Primary sensitivity to low-energy $\nu_e$ is shared with DUNE by the 79-ton
Helium and Lead Observatory (HALO) \cite{Duba2008,Zuber2015,HALOwebsite}, but
in this case the detection technique is indirect. Inelastic
neutrino reactions on lead nuclei, such as
\begin{equation}
\label{eq:CC208Pb}
\nu_e + \isotope[208]{Pb} \rightarrow e^{-} + \isotope[208]{Bi}^* \,,
\end{equation}
will sometimes lead to the production of neutrino-induced neutrons (NINs) via
de-excitations of the residual nucleus. HALO's lead neutrino target is
instrumented with \isotope[3]{He}-based neutron counters, which will be used to
search for NINs in the event of a nearby core-collapse supernova. Due to its
widespread use as a radiation shielding material, lead is also a potential
source of background NINs in precision CEvNS measurements. To better constrain
theoretical modeling of this background, the COHERENT experiment is pursuing
direct measurements of NIN production on lead, iron, and copper \cite{COH2018}.

Beyond the examples mentioned here, inelastic scattering on a variety of other
nuclear targets is of interest either as a means of low-energy neutrino
detection\footnote{The pioneering Homestake solar neutrino experiment famously
employed the $\isotope[37]{Cl}(\nu_e, e^{-})\isotope[37]{Ar}$ reaction for this
purpose \cite{Davis1968}.} or as a source of background. Additional nuclei for
which models of these processes have been studied in detail include carbon
\cite{Suzuki2019,Samana2011,Suzuki2006,Hayes2000,Volpe2000,Kolbe1999}, oxygen
\cite{Reen2020,Nakazato2018,Suzuki2018,Langanke1996,Kolbe1993}, molybdenum
\cite{Ydrefors2013,Ydrefors2012a,Ydrefors2012b,Balasi2011,Ejiri2002,Ejiri2000},
and xenon \cite{Haselschwardt2020,Pirinen2019,Divari2013}, among others
\cite{Shulgina1993,Raghavan1997,Almosly2014a,Almosly2014b,Ejiri2014,Divari2018,Divari2020,Vyborov2019}.

Despite the successes of standard neutrino event generators in describing
accelerator neutrino data, their prevailing treatment of inelastic
neutrino-nucleus scattering is likely to be inadequate when applied to
interactions at energies of tens of \si{\MeV} and below. This is due in part to
approximations made in their modeling of nuclear structure. For few-\si{\MeV}
neutrinos, inelastic neutrino-nucleus cross sections are governed by the
energetically-accessible transitions to low-lying discrete energy levels of the
daughter nucleus. At somewhat higher energies, excitations of collective
vibrational modes of the nucleus, known as \textit{giant resonances}
\cite{Harakeh2001,Goeke1982}, also begin to play a major role. Both of these
details are entirely missing from the conventional Fermi gas model of the
nuclear ground state. While these deficiencies of the Fermi gas model become
less problematic as the neutrino energy increases to several hundred \si{\MeV}
and beyond, it has been pointed out that low-energy nuclear excitations are
still expected to exert a noticeable influence on $\sim$\SI{1}{\GeV}
neutrino-nucleus differential cross sections at very forward scattering angles
\cite{Pandey2016,Pandey2015}.

A second modeling difference that limits the suitability of typical neutrino
event generators for the low-energy regime is the description of hadronic
final-state interactions. Rather than taking an INC- or BUU-like dynamical
approach to FSIs, in which intranuclear scattering is explicitly modeled,
typical low-energy calculations
\cite{Langanke1996,Balasi2015,Kolbe1992,Kolbe2000,Kolbe2001,Kolbe2003b,Cheoun2012b,Bandyopadhyay2017,Vale2016}
opt instead for a statistical treatment, in which only bulk properties of the
nuclear system (such as its excitation energy and spin-parity) are employed to
predict its behavior. The latter strategy is usually justified by assuming that
the struck nucleon from the primary neutrino interaction will scatter
repeatedly within the nuclear medium without being directly knocked out. These
multiple intranuclear collisions lead to the transferred energy being widely
shared among the constituent nucleons and thus to thermal equilibration:
de-excitations of the resultant \textit{compound nucleus} may be treated
independently of the manner in which it was formed.

While theoretical predictions of neutrino-nucleus cross sections at high (low)
energies tend to rely exclusively on a dynamical (statistical) model of FSIs, a
more complete treatment may be achieved by combining the two approaches. The
current release of GiBUU enables such calculations by providing an optional
interface to the Statistical Multifragmentation Model (SMM) \cite{Bondorf1995}
code. Disintegration of the residual nucleus is simulated by SMM in a
post-processing step which takes otherwise complete GiBUU events as input.
Although the two codes have been used together to examine other processes
\cite{Gaitanos2008,Gaitanos2009,Gaitanos2009a,Larionov2020}, their joint
application to neutrino interactions remains unstudied.

At present, native support for statistical nuclear de-excitation models in the
other three neutrino generators mentioned above is limited to some simple
approximations used in GENIE's treatment of nucleon emission
\cite{Dytman2009a}. However, official GENIE interfaces to INCL++
\cite{Leray2013,INCLwebsite} and to the Bertini Cascade implementation
\cite{Wright2015} in Geant4 \cite{Geant4,Geant4Website} are in the late stages
of development \cite{GENIEinclG4,GENIEinclDoc}. These will provide enhancements
to GENIE FSI modeling which are similar in scope to those obtained with SMM for
GiBUU. Unofficial interfaces are also being explored by outside groups, with at
least one attempt \cite{Cheng2020} having been made to apply the TALYS
\cite{Talys1,Talys2012} nuclear de-excitation model to events generated using
both GENIE and NuWro.

Although interfacing with these external tools provides an FSI treatment more
compatible with standard low-energy approaches, a key omission remains
problematic. With the exception of TALYS, the usual configurations of all of
the remaining codes\footnote{Initializing the main Geant4 Bertini Cascade class
(\texttt{G4CascadeInterface}) using non-default settings
\cite[sec.~4.1]{Wright2015} may enable simulation of discrete $\gamma$-ray
emission via the \texttt{G4ExcitationHandler} class \cite{Quesada2011}.} lack a
means of simulating discrete $\gamma$-ray transitions between low-lying nuclear
energy levels.\footnote{I note, however, that both GENIE and NEUT directly
implement simple models of de-excitation $\gamma$-ray production that are
specific to \isotope[16]{O}.} Gamma-rays created in this way represent a major
component of the final state for inelastic scattering of solar neutrinos on
complex nuclei. For inelastic NC reactions, de-excitation $\gamma$-rays may
often be the only final-state particles which are experimentally observable.

In this work, I present a new neutrino event generator,
\marley,\footnote{\marley\ is an acronym for \textit{Model of Argon Reaction
Low Energy Yields}. Although originally conceived as a tool to simulate the
specific process $\isotope[40]{Ar}(\nu_e,e^{-})\isotope[40]{K}^*$, \marley\ is
moving toward becoming a more general-purpose low-energy neutrino interaction
generator.} which implements a model of inelastic neutrino-nucleus scattering
designed specifically for the low-energy regime. The early version of \marley\
described herein is primarily focused on simulations of charged-current
absorption of $\nu_e$ on \isotope[40]{Ar} \cite{marleyPRC}. However,
preparation of additional input data would allow other reaction channels and
nuclear targets to be handled without difficulty within the existing code
framework. \Cref{sec:theory_summary} provides an overview of the theoretical
treatment of neutrino scattering used in \marley. \Cref{sec:random_sampling}
presents the \marley\ approach to random sampling, which makes ample use of
modern improvements to the \cpp\ language and, as discussed in
\cref{sec:inverse_transform_sampling}, implements a new inverse transform
sampling algorithm \cite{Olver2013} in a physics event generator for the first
time. \Cref{sec:implementation} discusses the \marley\ event generation
workflow and implementation details. Sections
\ref{sec:install}--\ref{sec:output} outline how to install, configure, run, and
interpret the output of the code. \Cref{sec:external_interfaces} explains how
\marley\ can be interfaced with external software toolkits, using the popular
Geant4 \cite{Geant4,Geant4Website} particle transport package as an example.
Finally, \cref{sec:prospects} considers prospects for future improvements to
\marley.

\section{MARLEY treatment of neutrino scattering}
\label{sec:theory_summary}

This section provides a brief overview of the physics models currently
implemented in \marley, with more details available in ref.~\cite{marleyPRC}.
Natural units ($\hbar = c = 1$) are used throughout.

As has been done in many previous calculations of low-energy neutrino cross
sections \cite{Langanke1996,Balasi2015,Kolbe1992,Kolbe2000,Kolbe2001,
Kolbe2003b,Cheoun2012b,Bandyopadhyay2017,Vale2016}, \marley\ treats
neutrino-nucleus scattering events as proceeding via a two-step process. In the
first step, a two-to-two scattering reaction involving the neutrino and the
target nucleus is simulated, and the final nucleus is left in a state with a
well-defined excitation energy, spin, and parity. In the second step, which is
handled independently from the first, the final nucleus de-excites. At low
excitation energies, where discrete level data are available, the
de-excitations are modeled using tabulated $\gamma$-ray branching ratios. At
higher excitation energies, where the final nucleus becomes unbound, the
formation of a compound nucleus is assumed, and the decay widths for all open
channels are calculated using the Hauser-Feshbach statistical model
\cite{Hauser1952}.

\subsection{ Nuclear $\text{2} \rightarrow \text{2}$ scattering model }
\label{sec:xsec_theory}
To simulate two-to-two neutrino-nucleus scattering processes at low energies,
\marley\ \version\ evaluates the nuclear matrix elements in the long-wavelength
limit (in which the four-momentum transfer $q \to 0$) and the slow-nucleon
limit (in which $|\mathbf{p}_N|/m_N \to 0$, where $\mathbf{p}_N$ is the initial
3-momentum of the struck nucleon and $m_N$ is its mass). The combination of
these two limits is sometimes referred to as the \textit{allowed
approximation}. Under this approach, the differential cross section in the
center-of-momentum (CM) frame for a transition to a particular nuclear final
state is given by the expression
\begin{equation}
\label{eq:diff_xsec}
\frac{ d\sigma }{d\cos \theta_{\ell}} =
\; \frac{ G_F^2 }{ 2 \, \pi } \,
\mathcal{F}_\text{\tiny CC} \, \bigg[ \frac{ E_i \, E_f }{ s } \bigg]
\, E_\ell \left|\mathbf{p}_\ell\right|
\bigg[ \Big(1 + \beta_\ell \cos \theta_{\ell} \Big)\, B(\mathrm{F})
+\, \Big(1 - \frac{1}{3} \, \beta_\ell \cos \theta_{\ell} \Big)
\, B(\mathrm{GT}) \bigg] \,.
\end{equation}
Here $G_F$ is the Fermi constant, Mandelstam $s$ is the square of the total
energy in the CM frame, and $E_i$ ($E_f$) is the total energy of the initial
(final) nucleus. The final-state lepton has total energy $E_\ell$, 3-momentum
$\mathbf{p}_\ell$, speed $\beta_\ell = |\mathbf{p}_\ell| / E_\ell$, and
scattering angle $\theta_\ell$, which is defined with respect to the incident
neutrino direction. The first factor in square brackets, $E_i \hspace{0.12em}
E_f / s$, arises due to nuclear recoil and is often neglected. The symbol
$\mathcal{F}_\text{\tiny CC}$ is used to introduce extra factors needed when
computing the cross section for charged-current scattering. Discussion of
$\mathcal{F}_\text{\tiny CC}$ is deferred to \cref{sec:coulomb_corrections}.

The spin-reduced Fermi and Gamow-Teller nuclear matrix elements may be written
in the form
\begin{align}
\label{eq:BF}
B(\mathrm{F}) &\equiv \frac{ g_V^2 }{ 2J_i + 1 } \Big|\big< J_f \, \big\lVert
\, \mathcal{O}_\mathrm{F} \, \big\rVert \, J_i \big> \Big|^2
\\[0.5\baselineskip]
\label{eq:BGT}
B(\mathrm{GT}) &\equiv \frac{ g_A^2 }{ 2J_i + 1 } \Big|\big< J_f \, \big\lVert
\, \mathcal{O}_{\mathrm{GT}} \,
\big\rVert \, J_i \big> \Big|^2
\end{align}
where $g_V$ ($g_A$) is the vector (axial-vector) weak coupling constant of the
nucleon, and $J_i$ ($J_f$) is the initial (final) nuclear spin.
The Fermi matrix element $B(\mathrm{F})$ is subject to the spin-parity
selection rule
\begin{equation}
\label{eq:Fermi_sel_rule}
B(\mathrm{F}) = 0 \text{ unless } J_f = J_i \text{ and } \Pi_f = \Pi_i
\end{equation}
where $\Pi_i$ ($\Pi_f$) is the initial (final) nuclear parity. The Gamow-Teller
matrix element $B(\mathrm{GT})$ likewise obeys the selection rule
\begin{equation}
\label{eq:GT_sel_rule}
B(\mathrm{GT}) = 0 \text{ unless } |J_i - 1| \leq J_f \leq J_i + 1
\text{ and } \Pi_f = \Pi_i\,.
\end{equation}

The Fermi ($\mathcal{O}_\mathrm{F}$) and Gamow-Teller
($\mathcal{O}_{\mathrm{GT}}$) operators from \cref{eq:BF,eq:BGT} are defined
for charged-current and neutral-current scattering processes via the relations
\begin{align}
\label{eq:F_and_GT_operators}
\mathcal{O}_\mathrm{F} &\equiv \begin{cases}
\sum_{n=1}^A t_{\pm}(n) & \mathrm{CC} \\
& \\
Q_W / \, 2 & \mathrm{NC} \\
\end{cases}
&
\mathcal{O}_{\mathrm{GT}} &\equiv \begin{cases}
\sum_{n=1}^A \boldsymbol{\sigma}(n) \, t_{\pm}(n) & \mathrm{CC} \\
& \\
\sum_{n=1}^A \boldsymbol{\sigma}(n) \, t_{3}(n) & \mathrm{NC} \\
\end{cases}
\end{align}
where $\boldsymbol{\sigma}$ is the Pauli vector, $A$ is the nucleon number, and
the isospin lowering (raising) operator $t_{-}$ ($t_{+}$) should be
chosen\footnote{I take the neutron to be the isospin-up state of the nucleon,
i.e., $t_{-} |n\big> = |p\big>$.} for an incident neutrino (antineutrino). The
symbol $t_3$ denotes the third component of isospin, and an operator suffixed
by $(n)$ is understood to act only on the $n$th nucleon. The weak nuclear
charge $Q_W$ for a nucleus with neutron number $N$ and proton number $Z$ is
given in terms of the weak mixing angle $\theta_W$ by
\begin{equation}
Q_W = N - [1 - 4\,\sin^2\theta_W] \, Z \,.
\end{equation}

\subsubsection{Coherent elastic neutrino-nucleus scattering}
\label{sec:CEvNS}

For NC reactions on a spin-zero ($J_i = 0$) target nucleus, the differential
cross section from \cref{eq:diff_xsec} reduces to a particularly simple form
when describing scattering that leaves the nucleus in its ground state. The
Gamow-Teller selection rule (\cref{eq:GT_sel_rule}) ensures that
$B(\mathrm{GT})$ vanishes identically, while the Fermi matrix element for the
ground-state-to-ground-state transition becomes
\begin{equation}
B(\mathrm{F}) = \frac{ g_V^2 \, Q_W^2 }{ 4 }\,.
\end{equation}
For this special case, \cref{eq:diff_xsec} may be rewritten in terms of
the lab-frame kinetic energy $T_f$ of the recoiling ground-state
nucleus as
\begin{equation}
\frac{ d\sigma_\text{CEvNS} }{ dT_f } =
\frac{ G_F^2 \, Q_W^2 \, g_V^2 \, M }{ 4 \pi }
R(s)
\bigg[ 1 - \frac{ T_f }{ T_f^\text{max} } \bigg]
\end{equation}
where $M$ is the mass of the nuclear target. The maximum value
of $T_f$ allowed by the reaction kinematics is given
in terms of the initial lab-frame neutrino energy $E_\nu$
by
\begin{equation}
T_f^\text{max} = \frac{ 2E_\nu^2 }
{ 2E_\nu + M }\,.
\end{equation}

This process is known as \textit{coherent elastic neutrino-nucleus scattering}
(CEvNS) \cite{Akimov2017,Freedman1974}. The nuclear recoil correction factor
\begin{equation}
R(s) \equiv \frac{ (s + M^2)^2 }{ 4s^2 }
\end{equation}
is usually neglected\footnote{Approximating $R(s) \approx 1$ is accurate to
zeroth order in $E_{\nu} / M$.} in the CEvNS literature (see, e.g.,
\cite{Lindner2017}). Due to its adoption of the allowed approximation, the
\marley\ treatment of CEvNS currently does not include a $q^2$-dependent
nuclear form factor that accounts for imperfect coherence in the
cross section \cite{Kerman2016}.

\subsubsection{Coulomb corrections}
\label{sec:coulomb_corrections}

In \cref{eq:diff_xsec}, the symbol $\mathcal{F}_\text{\tiny CC}$ is
used to include extra factors needed solely for CC scattering. It is defined
by
\begin{equation}
\mathcal{F}_\text{\tiny CC} \equiv \begin{cases}
|V_\mathrm{ud}|^2 \, F_C & \mathrm{CC} \\
1 & \mathrm{NC}
\end{cases}
\end{equation}
where $V_\mathrm{ud}$ is the Cabibbo–Kobayashi–Maskawa matrix element
connecting the up and down quarks. The Coulomb correction factor $F_C$ accounts
for the electromagnetic interaction between the outgoing lepton and nucleus in
an approximate way. Three prescriptions for computing $F_C$ are available in
\marley: the Fermi function, the effective momentum approximation (EMA), and
the modified effective momentum approximation (MEMA).

\paragraph{Fermi function}
For very low energies of the outgoing lepton (such as those observed in beta
decay), using the \textit{Fermi function}
\cite{Fermi1934Original,Fermi1934Translation} as the Coulomb correction factor
is a standard approach. A minor complication emerges, however, because the
derivation of the Fermi function\footnote{See, e.g., ref.~\cite{Hoyle1938}.}
assumes that the final nucleus is at rest, while the differential cross section
in \cref{eq:diff_xsec} is evaluated in the CM frame and accounts for nuclear
recoil. To work around this discrepancy, \marley\ uses the relative speed
$\beta_{\text{rel}}$ of the two final-state particles \cite{Cannoni2017}
\begin{align}
\label{eq:Lorentz_invariant_relative_speed}
\beta_{\text{rel}} &= \frac{ \sqrt{ (k^\prime \cdot p^\prime)^2
- m_{\ell}^2 \, m_{f}^2 } }{ k^\prime \cdot p^\prime }
&
\gamma_{\text{rel}} &\equiv \left(1 -
\beta^2_{\text{rel}}\right)^{-1/2},
\end{align}
to evaluate the Fermi function in the rest frame of the final nucleus using the
Lorentz-invariant expression
\begin{align}
\label{eq:Lorentz_invariant_Fermi_function}
F_\text{Fermi} = \frac{2(1+S)}{\big[\Gamma(1+2S)\big]^2}
\, (2 \, \gamma_{\text{rel}} \, \beta_{\text{rel}} \, m_\ell \, R)^{2S-2} \,
e^{-\pi\,\eta} \, \left|\Gamma\left( S - i\eta \right) \right|^2.
\end{align}
In \cref{eq:Lorentz_invariant_relative_speed,eq:Lorentz_invariant_Fermi_function},
$p^\prime$, $m_f$, and $Z_f$ denote the 4-momentum, mass, and proton number of
the final nucleus. Likewise, $k^\prime$, $m_\ell$, and $z_\ell$ represent the
4-momentum, mass, and electric charge (in units of the elementary charge) of
the final-state lepton. The quantity $S$ is defined in terms of the fine
structure constant $\alpha$ by
\begin{equation}
S \equiv \sqrt{1 - \alpha^2 Z_f^2\,}
\end{equation}
while
\begin{equation}
\label{eq:nuclear_radius}
R \approx \frac{ 1.2 \, A^{1/3} \, \text{fm} }{ \hbar\,c }
\end{equation}
is the nuclear radius (in natural units), and the Sommerfeld parameter
$\eta$ is given by
\begin{equation}
\label{eq:Sommerfeld_param}
\eta = \frac{ \alpha \, Z_f \, z_\ell }{ \beta_{\text{rel}} }.
\end{equation}

\paragraph{Effective momentum approximation}

For higher energies of the outgoing lepton, the Fermi function is known to
overestimate the magnitude of the Coulomb corrections, and an alternative
approach called the \textit{effective momentum approximation} (EMA) becomes
more appropriate \cite{Engel1998}. Let the symbol $\FNRpLep$
($\FNReLep$) denote the momentum (total energy) of the outgoing lepton
in the rest frame of the final nucleus:
\begin{align}
\FNReLep &\equiv \gamma_\text{rel} \, m_\ell
& \FNRpLep &\equiv \beta_\text{rel} \, \FNReLep\,.
\end{align}
Then the \textit{effective} values of these variables
\begin{align}
\label{eq:eff_vars}
\FNRpLepEff &\equiv \sqrt{ \FNReLepEff^2 - m_\ell^2 } &
\FNReLepEff &\equiv \FNReLep - V_C(0),
\end{align}
are those that exist in the presence of the nuclear Coulomb potential, which
is taken to be that at the center of a uniformly-charged sphere:
\begin{equation}
V_C(0) \approx \frac{3 \, Z_f \, z_\ell \, \alpha}{2\,R} \,.
\end{equation}
In the \marley\ implementation of the EMA,
the Coulomb correction factor is given by the ratio\footnote{The original EMA
treatment also involves the use of an effective value of the 4-momentum
transfer while computing the scattering amplitude. However, since the nuclear
matrix elements in \cref{eq:diff_xsec} are already evaluated in the limit of
zero momentum transfer, \marley\ neglects this additional correction.}
\begin{equation}
\label{eq:EMA}
F_\text{EMA} \equiv \frac{ \FNRpLepEff }{ \FNRpLep } \,.
\end{equation}

\paragraph{Modified effective momentum approximation}
In ref.~\cite{Engel1998}, an adjustment to the standard EMA prescription
is proposed which improves the accuracy of the approximation in cases where
the final lepton mass cannot be neglected. Under this \textit{modified
effective momentum approximation} (MEMA), the Coulomb correction factor
defined in \cref{eq:EMA} is replaced by
\begin{equation}
\label{eq:FMEMA}
F_\text{MEMA} = \frac{ \FNRpLepEff \, \FNReLepEff }
{ \FNRpLep \, \FNReLep } \,.
\end{equation}

\paragraph{Default behavior}

The final-state Coulomb interaction increases the charged-current cross section
for neutrinos and decreases it for antineutrinos. Because the (M)EMA is known
to overestimate the size of this effect at low energies while the Fermi
function does the same at high energies, previous calculations
\cite{Volpe2002,Ydrefors2012a} have adopted a simple prescription for combining
the two approaches: in any particular case, adopt the method that yields the
smallest Coulomb correction. For \marley, this amounts to defining the Coulomb
correction factor $F_C$ by
\begin{equation}
\label{eq:FC}
F_C \equiv \begin{cases}
  F_\text{Fermi} & |F_\text{Fermi} - 1| < |F_\text{MEMA} - 1| \\
  F_\text{MEMA} & \text{otherwise}
\end{cases}
\end{equation}
Although \cref{eq:FC} represents the default \marley\ approach to Coulomb
corrections, this behavior may be altered by the user in the job configuration
file (see \cref{sec:coulomb_mode}).

\subsubsection{Total cross section}

With the definitions given above, integration of \cref{eq:diff_xsec}
over $\cos\theta_\ell$ becomes trivial, leading to the total cross section
\begin{equation}
\label{eq:tot_xsec}
\sigma = \frac{ G_F^2 }{ \pi } \,
\mathcal{F}_\text{\tiny CC} \, \bigg[ \frac{ E_i \, E_f }{ s } \bigg]
\, E_\ell \left|\mathbf{p}_\ell\right|
\bigg[ B(\mathrm{F}) + B(\mathrm{GT}) \bigg].
\end{equation}

\subsection{Nuclear de-excitation model}
\label{sec:HFSM}

De-excitations from high-lying nuclear levels are simulated in \marley\ using
the Hauser-Feshbach statistical model (HFSM) \cite{Hauser1952}. This model
assumes that the decaying nuclear state may be adequately described as a
thermally-equilibrated compound nucleus with a definite excitation energy
($E_x$), spin ($J$), and parity ($\Pi$). In the \marley\ HFSM implementation,
emissions of $\gamma$-rays and light nuclear fragments ($1 \leq A \leq 4$) are
treated as a sequence of binary decays while fission, production of heavy
nucleon clusters ($A \geq 5$), and simultaneous multiparticle evaporation are
neglected.

\subsubsection{Nuclear fragment emission}
\label{sec:nuc_fragments}

According to the HFSM, the distribution of final excitation
energies $E^\prime_x$ that may result from the emission of a fragment $\fragment$
(with parity $\pi_\fragment$ and separation energy $S_\fragment$) from
the compound nucleus is described by the differential decay width
\begin{equation}
\label{eq:fragment_diff_decay_width}
\frac{ d\Gamma_{\fragment} }{ dE_x^\prime }
= \frac{1}{2 \, \pi \, \rho_i(E_x, J, \Pi) }
\sum_{\ell = 0}^{ \ell_\text{max} }
\;
\sum_{j = |\ell - s|}^{\ell + s}
\;
\sum_{J^\prime = |J - j|}^{J + j}
T_{\ell j}(\totKinE) \, \rho_f(E_x^\prime, J^\prime, \Pi^\prime)
\end{equation}
where $J^\prime$ is the final nuclear spin; $s$, $\ell$, and $j$
are the spin, orbital, and total angular momentum quantum numbers
of the emitted fragment;
\begin{equation}
\label{eq:HF_parity_conservation}
\Pi^\prime = (-1)^\ell \, \pi_\fragment \, \Pi
\end{equation}
is the value of the final-state nuclear parity needed to enforce
parity conservation; and $\totKinE$
is the total kinetic energy of the decay products in the rest frame of the
initial nucleus. The maximum accessible final-state excitation energy
\begin{equation}
E_x^{\prime\,\text{max}} = E_x - S_\fragment
\end{equation}
is related to the total kinetic energy $\totKinE$ via
\begin{equation}
\totKinE = E_x^{\prime\,\text{max}} - E_x^\prime.
\end{equation}
The functions $\rho_i$ and $\rho_f$ represent the density of nuclear levels in
the vicinity of the initial and final states, while the transmission
coefficient $T_{\ell j}$ quantifies how readily the fragment may be emitted
from the nucleus. Because the value of $T_{\ell j}$ at fixed $\totKinE$ falls
off rapidly with increasing $\ell$, \marley\ truncates the infinite sum over
orbital angular momenta using an upper limit $\ell_\text{max}$. By default,
$\ell_\text{max} = 5$ is used. This value may be adjusted by the user as
described in \cref{sec:l_max}.

At low excitation energies, where individual nuclear levels can be resolved,
\marley\ treats the level density $\rho_f$ as
a sum of delta functions, with one term per level. For a specific nuclear level,
the partial decay width for emission of fragment $\fragment$ may be written
in the form
\begin{equation}
\label{eq:fragment_discrete_decay_width}
\Gamma_\fragment = \frac{1}{2 \, \pi \, \rho_i(E_x, J, \Pi) }
\sum_{j = |J - J^\prime|}^{J + J^\prime} \;
\sum_{\ell = |j - s|}^{j + s} \delta_\pi^\ell \,
T_{\ell j}(\totKinE)
\end{equation}
where the symbol $\delta_\pi^\ell$, which enforces parity conservation,
is equal to one if \cref{eq:HF_parity_conservation} is satisfied and
zero if it is not.

At higher excitation energies, \marley\ computes $\rho_f$ according to the
Back-shifted Fermi gas model (BFM) from version 3 of the Reference Input
Parameter Library (RIPL-3) \cite{RIPL3}. The ``BFM effective'' values of the
level density parameters for this model are adopted from a global fit of
nuclear level data for 289 nuclides reported in ref.~\cite{Koning2008}. A full
description of the BFM as implemented in \marley\ is given in appendix B of
ref.~\cite{marleyPRC}. The initial level density $\rho_i$ is always evaluated
according to the BFM regardless of excitation energy.

The partial decay width for a transition to the continuum of nuclear levels
via emission of a fragment $\fragment$ may be computed by integrating
\cref{eq:fragment_diff_decay_width} over the final excitation energy interval
$[E_x^{\prime\,\text{min,c}}, E_x^{\prime\,\text{max}}]$. The lower bound
of the continuum $E_x^{\prime\,\text{min,c}}$ is taken by \marley\ to be
the excitation energy of the highest tabulated discrete nuclear level
for the nuclide of interest. In cases where no such level data are available,
the continuum is taken to start at the nuclear ground state
($E_x^{\prime\,\text{min,c}} = 0$).

\subsubsection{Fragment transmission coefficients}
\label{sec:fragment_tr_coeff}

The fragment transmission coefficients $T_{\ell j}$ that appear in
\cref{eq:fragment_diff_decay_width,eq:fragment_discrete_decay_width}
are computed by numerically solving the radial Schr\"{o}dinger equation
\begin{equation}
\label{eq:rad_schrod}
\left[\frac{d^2}{dr^2} + \fragmentMomCM^2 - \frac{\ell(\ell + 1)}{r^2}
- \frac{\fragmentMomCM^2}{\totKinE}\,\mathcal{U}(r, \fragmentKinELab, \ell, j)
\right]u_{\ell j}(r) = 0
\end{equation}
where $u_{\ell j}$ is the fragment's radial wavefunction,
\begin{align}
\label{eq:fragment_kin_E_lab}
\fragmentKinELab &= \frac{ \totKinE^2 + 2\, (m_\fragment + M^\prime)
\, \totKinE } { 2 M^\prime }
\intertext{is its kinetic energy in the rest frame of the final
nucleus,\footnotemark~and}
\label{eq:fragment_momentum_CM}
\fragmentMomCM &= \sqrt{ \frac{ (2 \, m_\fragment + \fragmentKinELab)
\, {M^\prime}^2 \, \fragmentKinELab } { (m_\fragment + M^{\prime})^2
+ 2 \, M^\prime \, \fragmentKinELab } }
\end{align}
is the magnitude of its 3-momentum in the rest frame of the initial nucleus.
In \cref{eq:fragment_kin_E_lab,eq:fragment_momentum_CM},
$m_\fragment$ ($M^\prime$) denotes the mass of the emitted
fragment (final nucleus).
\footnotetext{The label \textit{lab} for this quantity reflects it status as
the laboratory-frame kinetic energy in the time-reversed process wherein the
fragment is absorbed to form the compound nucleus. See
appendix A of ref.~\cite{marleyPRC}.}

The optical potential $\mathcal{U}$ used by \marley\ for nucleon emission
is the global parameterization of Koning and Delaroche \cite{Koning2003}. For
complex nuclear fragments, a folding approach similar to that of Madland
\cite{Madland1988} is used to
construct the optical potential by weighting the individual neutron and proton
potentials. More details about the \marley\ nuclear optical potential are
available in appendix C of ref.~\cite{marleyPRC}.

Far from the nucleus, the optical potential approaches the Coulomb
potential, and the fragment radial wavefunction approaches the limiting form
\begin{equation}
\label{eq:asymptotic_radial_wavefunction}
\lim_{r \to \infty} u_{\ell j}(r) =
\frac{i}{2}\left[H_\ell^-(\eta, \fragmentMomCM r)
- \SMatrixElement H_\ell^+(\eta, \fragmentMomCM r)\right]
\end{equation}
where $H_\ell^\pm$ are the Coulomb wavefunctions
\cite[ch.~33]{NIST:DLMF}. The Sommerfeld parameter
\begin{equation}
\eta \equiv \frac{z \, Z^\prime \,\alpha}{\beta_\text{rel}}
\end{equation}
is evaluated in terms of the proton number $z$ ($Z^\prime$) of the
emitted fragment (final nucleus) and the relative speed
\begin{equation}
\beta_\text{rel} = \frac{
\sqrt{ \fragmentKinELab^2 + 2 \, m_\fragment \, \fragmentKinELab  } }
{ m_\fragment + \fragmentKinELab }\,.
\end{equation}
The energy-averaged S-matrix element $\SMatrixElement$ that appears in
\cref{eq:asymptotic_radial_wavefunction} is related to the transmission
coefficient $T_{\ell j}$ via
\begin{equation}
T_{\ell j} \equiv 1 - \left|\SMatrixElement\right|^2.
\end{equation}
To approximate $\SMatrixElement$, \marley\ first obtains a numerical solution
$u_{\ell j}(r)$ of \cref{eq:rad_schrod} using Numerov's method
\cite{Numerov1924,Numerov1927,Thijssen2007}.
This method computes $u_{\ell j}$ iteratively on a regular grid with fixed
radial step size $\Delta$. If one defines the function $a_{\ell j}(r)$ to be
equal to the non-derivative terms enclosed in square brackets in
\cref{eq:rad_schrod}, i.e.,
\begin{equation}
a_{\ell j}(r) \equiv \fragmentMomCM^2
- \frac{ \ell (\ell + 1) }{ r^2 }
- \frac{ \fragmentMomCM^2 }{ \totKinE }
\, \mathcal{U}(r, \fragmentKinELab, \ell, j),
\end{equation}
then the Numerov solution (accurate to order $\Delta^4$)
$u_n \approx u_{\ell j}(r_n)$ at the $n$th grid point
\begin{align}
r_n &= n\,\Delta & n \in & \{ 0,1,2,\dots \}
\end{align}
is given by the recurrence relation
\begin{align}
u_n &= \frac{ (2 - 10 \, h \, a_{n-1}) \, u_{n-1}
- (1 + h \, a_{n-2}) \, u_{n-2} } { (1 + h \, a_n) } & n \geq 2
\end{align}
and boundary conditions
\begin{align}
u_0 &= 0 \\
u_1 &= \Delta^{ \ell + 1 }.
\end{align}
Here I have defined the abbreviations
$h \equiv \Delta^2 / 12$ and
$a_n \equiv a_{\ell j}(r_n)$ for $n \geq 1$,
with $a_0 \equiv 0$.

The \marley\ calculation of $\SMatrixElement$ proceeds through
iterations of the Numerov method until the first grid point $r_{\!A}$
is encountered such that
\begin{equation}
\big|\, \mathcal{U}(r_{\!A}) - V_C(r_{\!A}) \,\big| \leq \matchThresh \,,
\end{equation}
that is, the difference between the nuclear optical
potential $\mathcal{U}$ and the Coulomb potential $V_C$
falls below a small threshold $\matchThresh$. Iterations continue
further until a second grid point $r_{\!B}$ is encountered such that
\begin{equation}
r_{\!B} \geq \matchScale \, r_{\!A} \,.
\end{equation}
In \marley\ \version, the parameter values
$\Delta = \SI{0.1}{\femto\meter} \, / \, \hbar \, c$,
$\matchThresh = \SI{1}{\keV}$
and $\matchScale = 1.2$ are used.

Comparing the full solution obtained in this way to the asymptotic form from
\cref{eq:asymptotic_radial_wavefunction} at $r_{\!A}$ and $r_{\!B}$ leads to
the expression
\begin{equation}
\SMatrixElement \approx \frac{ u_{\ell j}(r_{\!A})
\, H^-(\eta, \fragmentMomCM r_{\!B})
- u_{\ell j}(r_{\!B}) \, H^-(\eta, \fragmentMomCM r_{\!A})}
{ u_{\ell j}(r_{\!A}) \, H^+(\eta, \fragmentMomCM r_{\!B}) - u_{\ell j}(r_{\!B})
\, H^+(\eta, \fragmentMomCM r_{\!A})}\,.
\end{equation}
Numerical values of the Coulomb wavefunctions are obtained by interfacing with
the GNU Scientific Library \cite{Galassi2009,GSLWebsite}.

\subsubsection{Gamma-ray emission}
\label{sec:gamma_emission}

In the Hauser-Feshbach formalism, $\gamma$-ray emission is described by the
differential decay width
\begin{equation}
\label{eq:gamma_diff_decay_width}
\frac{ d\Gamma_\gamma }{ dE_x^\prime }
= \frac{1}{2 \, \pi \, \rho_i(E_x, J, \Pi) }
\sum_{\lambda = 1}^{\lambda_\text{max}} \;
\sum_{J^\prime = |J - \lambda|}^{J + \lambda} \;
\sum_{ \Pi^\prime \in \{-1, 1\} }
T_{X\lambda}(E_\gamma)
\, \rho_f(E_x^\prime, J^\prime, \Pi^\prime)
\end{equation}
where $\lambda \geq 1$ is the multipolarity,
$E_\gamma \approx E_x - E_x^\prime$ is the
energy of the emitted $\gamma$-ray,\footnote{Although
\marley\ uses the approximate
expression for $E_\gamma$ given here to compute $\gamma$-ray decay widths,
corrections for nuclear recoil are handled exactly when
the actual emission is simulated.}
and
\begin{equation}
X = \begin{cases}
\mathrm{E} & \Pi = (-1)^\lambda \, \Pi^\prime \\
\mathrm{M} & \Pi = (-1)^{\lambda + 1} \, \Pi^\prime
\end{cases}
\end{equation}
labels the type of transition as either electric ($\mathrm{E}$)
or magnetic ($\mathrm{M}$). The infinite sum over multipolarities
in \cref{eq:gamma_diff_decay_width} is truncated at $\lambda_\text{max}$.
The default cutoff value $\lambda_\text{max} = 5$ may be configured by the
user as described in \cref{sec:l_max}.

The calculation of the level densities $\rho_i$ and $\rho_f$ is identical to
the approach used for nuclear fragment emission (see \cref{sec:nuc_fragments}).
In particular, the level density $\rho_f$ used for $\gamma$-ray transitions to
discrete levels is once again treated as a sum of delta functions. The partial
decay width for $\gamma$-ray emission to a particular nuclear level then
becomes
\begin{equation}
\label{eq:gamma_partial_width}
\Gamma_\gamma =
\frac{ 1 }{ 2 \, \pi \, \rho_i(E_x, J, \Pi) }
\sum_{\lambda = \max(1, |J - J^\prime|)}^{J + J^\prime}
T_{X\lambda}(E_\gamma)\,.
\end{equation}
If $J+J^\prime < 1$, the width $\Gamma_\gamma$ vanishes.

Similarly to the fragment emission case, calculation of the partial decay width
for $\gamma$-ray transitions to the continuum of nuclear levels is performed by
integrating \cref{eq:gamma_diff_decay_width} over the interval
$[E_x^{\prime\,\text{min,c}}, E_x^{\prime\,\text{max}}]$. The continuum lower
bound $E_x^{\prime\,\text{min,c}}$ is handled as in \cref{sec:nuc_fragments},
while the upper bound becomes
\begin{equation}
E_x^{\prime\,\text{max}} = E_x \,.
\end{equation}

\subsubsection{Gamma-ray transmission coefficients}
\label{sec:gamma_tr_coeff}

The transmission coefficients $T_{X \lambda}$ that appear in
\cref{eq:gamma_diff_decay_width,eq:gamma_partial_width}
are typically expressed in terms of a \textit{strength function}
$f_{X \lambda}(E_\gamma)$ such that
\begin{equation}
T_{X \lambda}(E_\gamma) = 2 \, \pi \, E_\gamma^{2\lambda + 1}
\, f_{X \lambda}(E_\gamma) \,.
\end{equation}
In \marley\ \version, the expression used to evaluate the strength function
\begin{equation}
\label{eq:SLM_strength_function}
f_{X \lambda}(E_\gamma) = \frac{ \sigma_{X \lambda} }
{ (2\lambda + 1) \, \pi^2 } \bigg[ \frac{ \Gamma^2_{X \lambda}
\, E_\gamma^{3 - 2\lambda} }{ (E_\gamma^2 - E_{X \lambda}^2)^2 + E_\gamma^2 \,
\Gamma^2_{X \lambda} } \bigg]
\end{equation}
is taken from the RIPL-3 Standard Lorentzian model \cite{RIPL3}. According to
this model, $\gamma$-ray emissions of type $X \lambda$ are assumed to take
place via de-excitation of the corresponding giant multipole resonance, which
has centroid excitation energy $E_{X \lambda}$, width $\Gamma_{X \lambda}$, and
peak cross section $\sigma_{X \lambda}$. A full listing of the values of these
parameters is given in Table II of ref.~\cite{marleyPRC}.

\subsection{Neutrino-electron elastic scattering}
\label{sec:nu_electron_xsec}

In addition to neutrino-nucleus scattering, \marley\ is also capable of
simulating neutrino-electron elastic scattering. For a target atom with proton
number $Z$, the differential cross section in the CM frame for this process is
computed according to
\begin{equation}
\label{eq:diff_xsec_nu_e}
\frac{ d\sigma }{ d \cos\theta_\nu }
= \frac{ 2Z G_F^2 E_\nu^2 }{ \pi }
\bigg[ g_1^2 + \frac{ g_1 \, g_2 \, m_e^2 }{ s } \, (\cos\theta_\nu - 1)
+ g_2^2 \bigg(1 + \frac{1}{2}\Big[ 1 - \frac{ m_e^2 }{ s } \Big]
\Big[ \cos\theta_\nu - 1 \Big] \bigg)^{\!2} \, \bigg]
\end{equation}
where $m_e$ is the electron mass and $E_\nu$ ($\theta_\nu$) is the energy
(scattering angle) of the neutrino. The coupling constants $g_1$ and $g_2$
depend on the neutrino species and are given in \cref{tab:nu_e_constants}.
Electron binding energies are neglected.

\begin{table}
\centering
\begin{tabular}{ccc}
\toprule
Neutrino & $g_1$ & $g_2$ \\
\midrule
$\nu_e$ & $1/2 + \sin^2\theta_W$ & $\sin^2\theta_W$ \\
$\bar{\nu}_e$ & $\sin^2\theta_W$ & $1/2 + \sin^2\theta_W$ \\
$\nu_\mu, \nu_\tau$ & $-1/2 + \sin^2\theta_W$ & $\sin^2\theta_W$ \\
$\bar{\nu}_\mu, \bar{\nu}_\tau$ & $\sin^2\theta_W$ & $-1/2 + \sin^2\theta_W$ \\
\bottomrule
\end{tabular}
\caption{Coupling constants used in calculations of neutrino-electron
elastic scattering. Values are given in terms of the weak mixing angle
$\theta_W$.}
\label{tab:nu_e_constants}
\end{table}

\section{Random sampling implementation}
\label{sec:random_sampling}

Like any other Monte Carlo event generator, \marley\ must make extensive use of
pseudorandom numbers and produce samples from a variety of discrete and
continuous probability distributions. With the advent of \cpp 11, a suite of
high-quality random number generation tools were adopted as part of the \cpp\
standard library \cite{Brown2013}, and \marley\ relies heavily on these new
features. All random numbers used by \marley\ are obtained using the \cpp\
standard library object \texttt{std::mt19937\allowbreak\_64}, which provides a
64-bit implementation of the Mersenne Twister algorithm developed by M.
Matsumoto and T. Nishimura \cite{Matsumoto1998,Nishimura2000}.

\subsection{Discrete distributions}

All sampling from discrete distributions is handled using instances of
the \cpp\ standard library object \texttt{std::\allowbreak discrete\allowbreak
\_distribution}. In cases where the sampling weights for each possible outcome
already exist in memory, the usual method for initializing this object
is to supply iterators that point to the beginning and end
of a collection of sampling weights. For example, line 5 of the code
snippet\footnote{The examples given in this section make use of a \cpp 17
feature (class template argument deduction) in order to avoid code clutter that
is unimportant for a conceptual understanding. In the \marley\ \version\ source
code, the \texttt{std::\allowbreak discrete\_\allowbreak distribution} and
\texttt{marley::Iterator\allowbreak To\allowbreak Member} class templates are
used in a manner that is compatible with \cpp 14.}
\begin{lstlisting}[name=dummy, linewidth=\textwidth, language=C++,
 keywordstyle=\color{black}\bfseries]
#include <random>
#include <vector>

std::vector<double> weights = { 1., 2. };
std::discrete_distribution dist1( weights.begin(), weights.end() );
\end{lstlisting}
initializes a \texttt{std::discrete\_distribution} object \texttt{dist1}
which will sample \texttt{int} values of 0 (1) with probability
$1/3$ ($2/3$).

When the sampling weights are stored as object data members, however,
initializing the distribution becomes more complicated. In particular, a
na\"{i}ve attempt using the approach from the example above
\begin{lstlisting}[name=dummy, linewidth=\textwidth, language=C++,
keywordstyle=\color{black}\bfseries]
struct A {
  A(double w) : weight(w) {}
  double weight;
};

std::vector<A> As = { A(1.), A(2.) };
\end{lstlisting}
\vspace{-0.877\baselineskip}
\begin{lstlisting}[name=dummy, linewidth=\textwidth, language=C++,
basicstyle=\small\ttfamily\color{darkred},
identifierstyle=\color{darkred}]
std::discrete_distribution dist2( As.begin(), As.end() );
\end{lstlisting}
triggers a compilation error on line 12. Because the iterators returned by
\texttt{As.begin()} and \texttt{As.end()} refer to objects of type \texttt{A}
instead of the weights (of type \texttt{double}), they cannot be used
to initialize \texttt{dist2}.

\marley\ works around this difficulty by implementing the
\texttt{iterator\_to\_member} interface proposed by T. Becker \cite{Becker2001}.
This interface converts an iterator that points to an object (\texttt{A}) into
an iterator that points to one of that object's data
members (\texttt{weight}). In this example, including the appropriate \marley\ header
file (\texttt{include/marley/IteratorToMember.hh}) and replacing line 12 with
\begin{lstlisting}[firstnumber=12, linewidth=\textwidth, language=C++,
keywordstyle=\color{black}\bfseries]
marley::IteratorToMember w_begin( As.begin(), &A::weight );
marley::IteratorToMember w_end( As.end(), &A::weight );
std::discrete_distribution dist2( w_begin, w_end );
\end{lstlisting}
compiles successfully and yields the same sampling behavior for \texttt{dist2}
as for \texttt{dist}. For cases in which the input iterators refer to object
pointers instead of the objects themsleves, \marley\ provides a similar
interface via the \texttt{marley::Iterator\allowbreak To\allowbreak
Pointer\allowbreak Member} class template, which is compatible with both bare (e.g.,
\texttt{A*}) and smart pointers (e.g., \texttt{std::unique\_ptr<A>}).

This approach is used in several places in the \marley\ source
code to initialize discrete distributions using sampling weights stored as
object data members, e.g., relative intensities owned by
\texttt{marley::\allowbreak Gamma} objects representing distinct
$\gamma$-ray de-excitations from a particular nuclear energy level.

\subsection{Continuous 1D distributions: accept/reject approach}
\label{sec:rejection_sampling}

To sample from continuous one-dimensional distributions, \marley\ implements
two general schemes, both of which take the bounds of a sampling interval
$[\xmin, \xmax]$ and an arbitrary probability density function $f(x)$ (for
which no particular normalization is assumed) as input. The first scheme uses a
simple rejection method which relies on an accurate knowledge of the global
maximum $\fmax$ of $f(x)$ within the sampling interval. If $\fmax$ is known in
advance, it may be supplied along with the other input parameters. Otherwise,
it is estimated within a specified tolerance by minimizing $-f(x)$ using
Brent's method \cite{Brent1973}. Once the value of $\fmax$ has been obtained,
pairs of uniformly-distributed variables $x \in [\xmin, \xmax]$ and $y \in [0,
\fmax]$ are repeatedly sampled. This continues until $y \leq f(x)$, at which
point the sampled $x$ value is accepted.

\subsection{Continuous 1D distributions: inverse transform approach}
\label{sec:inverse_transform_sampling}

In cases where the global maximum $\fmax$ is unknown and
its estimation via Brent's method is either unreliable (because a local
maximum may be found instead of the global one)
or inefficient (because $f$ is computationally expensive to evaluate),
\marley\ employs a second sampling scheme based on the ``fast
inverse transform sampling'' algorithm originally proposed in ref.
\cite{Olver2013}. Although \marley\ uses numerical techniques similar to those
in the original M\textsc{atlab} code \cite{Olver2013Code}, which was written
as an extension of the Chebfun package \cite{Driscoll2014,
Battles2004}, the \cpp\ implementation described herein is original. To the
author's knowledge, \marley\ represents the first application of this algorithm
to physics event generation.

\subsubsection{Algorithm}
To obtain a random sample $x$ from $f(x)$ using inverse transform sampling,
\marley\ first constructs a polynomial approximant $\tilde{f}$ of $f$ using
a grid of $\myN + 1$ ordered pairs $(x_j, f_j)$ for
$j \in \{0,1,\dots,N\}$, where the $x_j$ are the
Chebyshev points of the second kind
\begin{equation}
x_j = \frac{ \xmin + \xmax + (\xmax - \xmin)\cos( \pi j / \myN ) }{ 2 }
\end{equation}
and the $f_j$ are the function values
\begin{equation}
f_j \equiv f(x_j).
\end{equation}
At points other than the $x_j$ (where $f_j$ is used directly), the polynomial
approximant $\tilde{f}$ is given by the barycentric formula
\cite{Berrut2004}
\begin{equation}
\label{eq:barycentric_formula}
\tilde{f}(x) \equiv \left[\sum\limits_{j = 0}^{\myN}
\dfrac{(-1)^{ \,j } \, w_j^N }{x - x_j} f_j\right] \Bigg/
\left[ \sum\limits_{j = 0}^{\myN} \dfrac{ (-1)^{ \,j } \, w_j^N }{x - x_j} \right]
\end{equation}
with weights
\begin{equation}
w_j^N \equiv \begin{cases} 1/2 & j = 0\text{ or }j = N \\[0.5em]
1 & \text{otherwise.}
\end{cases}
\end{equation}
The $f_j$ are related to the coefficients $\alpha_k$ that appear in an $\myN$th
order expansion of $f$ in Chebyshev polynomials $T_k$, i.e.,
\begin{align}
\label{eq:Chebyshev_expansion_1}
f(x) &\approx \sum_{k = 0}^N w_k^N \, \alpha_k \, T_k(u(x))
& u(x) &\equiv 2\left(\frac{x - \xmin}
{\xmax - \xmin}\right) - 1,
\end{align}
by the type-I discrete cosine transform (DCT-I)
\begin{equation}
\label{eq:DCTI-1}
\alpha_k = \frac{ f_0 + (-1)^k f_\myN }{ \myN } + \frac{2}{\myN}\sum_{j = 1}^{\myN - 1}
f_j \cos\!\left( \frac{ \pi j k }{ \myN } \right).
\end{equation}
To approximate the cumulative density function (CDF)
\begin{equation}
F(z) \equiv \int_{\xmin}^{z} f(x) \, dx,
\;\; z \in [a, b]
\end{equation}
\marley\ uses the formulas
\begin{equation}
\int_{-1}^y T_k(u) \, du = \begin{cases}
T_1(y) + 1 & k = 0 \\[0.5em]
\frac{1}{4}\left[ T_2(y) - 1 \right] & k = 1 \\[0.5em]
\frac{1}{2} \! \left[ \frac{ T_{k + 1}(y) }{ k + 1 }
- \frac{ T_{k - 1}(y) }{ k - 1 } \right]
+ \frac{ (-1)^{ k + 1 } }{ k^2 - 1 } & k \geq 2
\end{cases}
\end{equation}
to integrate \cref{eq:Chebyshev_expansion_1} term-by-term, yielding an
$(N+1)$th order Chebyshev expansion
\begin{equation}
F(z) \approx \sum_{k = 0}^{\myN+1} w_k^{N+1} \, \beta_k \, T_k(u(z))
\end{equation}
where the coefficients $\beta_k$ are given by
\begin{equation}
\beta_k = \left(\frac{\xmax - \xmin}{2}\right)B_k
\end{equation}
with
\begin{equation}
B_k \equiv \begin{cases}
\alpha_0 - \frac{1}{2}\alpha_1 + 2\!\sum\limits_{j=2}^{\myN}\dfrac{\alpha_j (-1)^{\,j}}
{1 - j^2} & k = 0 \\[1em]
\frac{1}{2k}\left(\alpha_{k-1} - \alpha_{k+1}\right) & k > 0\text{ and } k < \myN
\\[0.5em]
\frac{1}{2k}\alpha_{k-1} & k = N\text{ or } k = \myN + 1. \\
\end{cases}
\end{equation}
One may obtain a polynomial approximant $\tilde{F}$ of the cumulative density
function $F$ by applying a second DCT-I (which is its own inverse) to the
expansion coefficients $\beta_k$:
\begin{equation}
\label{eq:DCTI-2}
F_\ell = \frac{ \beta_0 + (-1)^\ell \beta_{\myN + 1} }{ 2 } +
\sum_{k = 1}^{\myN} \beta_k \cos\!\left( \frac{ \pi k \ell }{ \myN + 1 } \right).
\end{equation}
Using the $N+1$ Chebyshev points
\begin{equation}
x_\ell = \frac{ \xmin + \xmax + (\xmax - \xmin)\cos( \pi \ell / [\myN + 1] ) }
{ 2 },
\end{equation}
one may approximate $F(x)$ for any $x \in [\xmin, \xmax]$ using
\cref{eq:barycentric_formula} with the substitutions $f \rightarrow F$,
$j \rightarrow \ell$, and $N \rightarrow N + 1$.

With the approximate CDF $\tilde{F}(x)$ constructed in
this manner, \marley\ obtains a random sample $x$ from $f(x)$ by generating a
uniform random number $\xi \in [0,1]$. Bisection is then used to find the sampled value
of $x \in [a,b]$ which satisfies the relation
\begin{equation}
\xi = \tilde{F}(x) \big/ \tilde{F}(b) \,.
\end{equation}

\subsubsection{Validation}

\begin{figure}
\centering
\includegraphics[height=0.28\textheight]{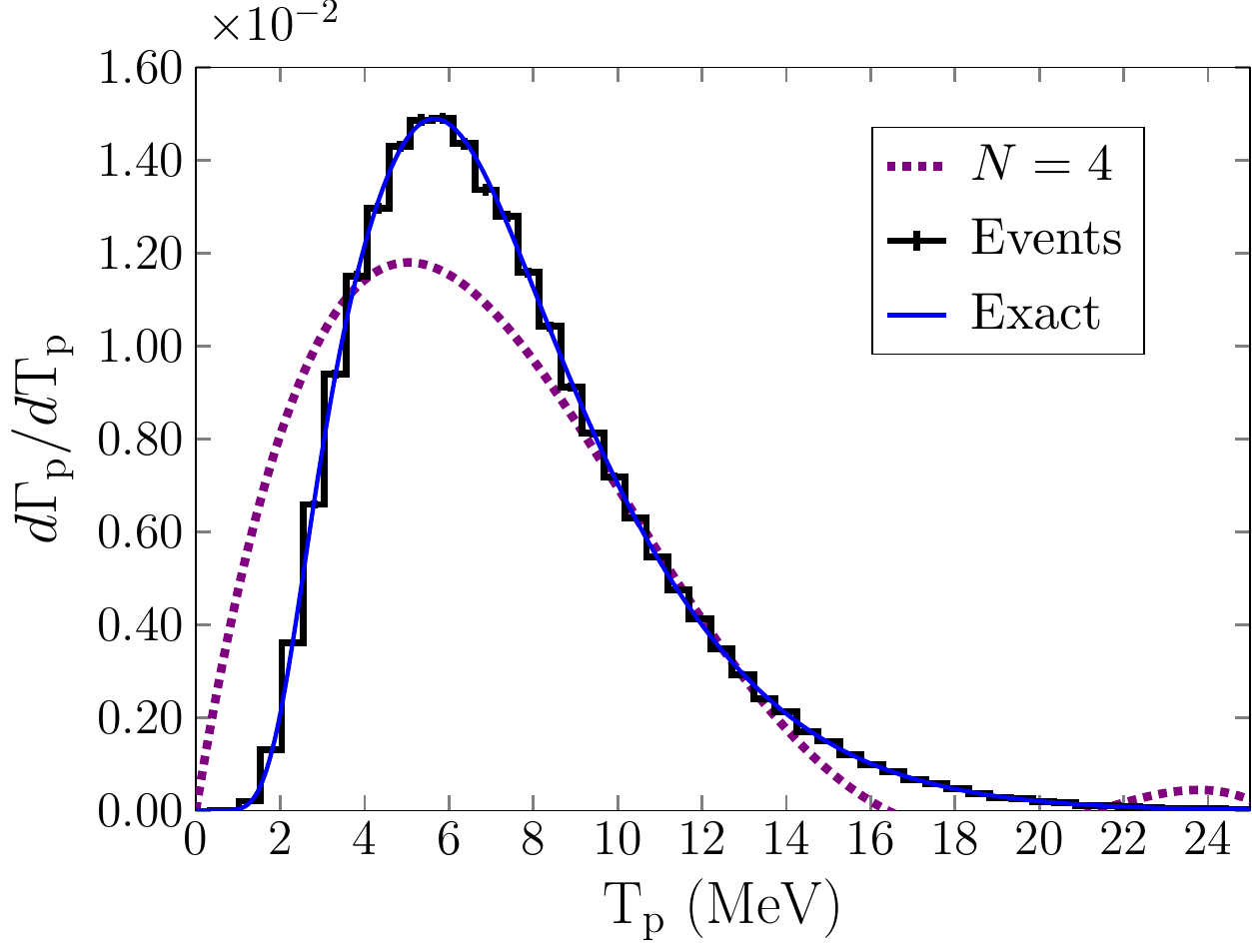}
\hfill
\includegraphics[height=0.28\textheight]{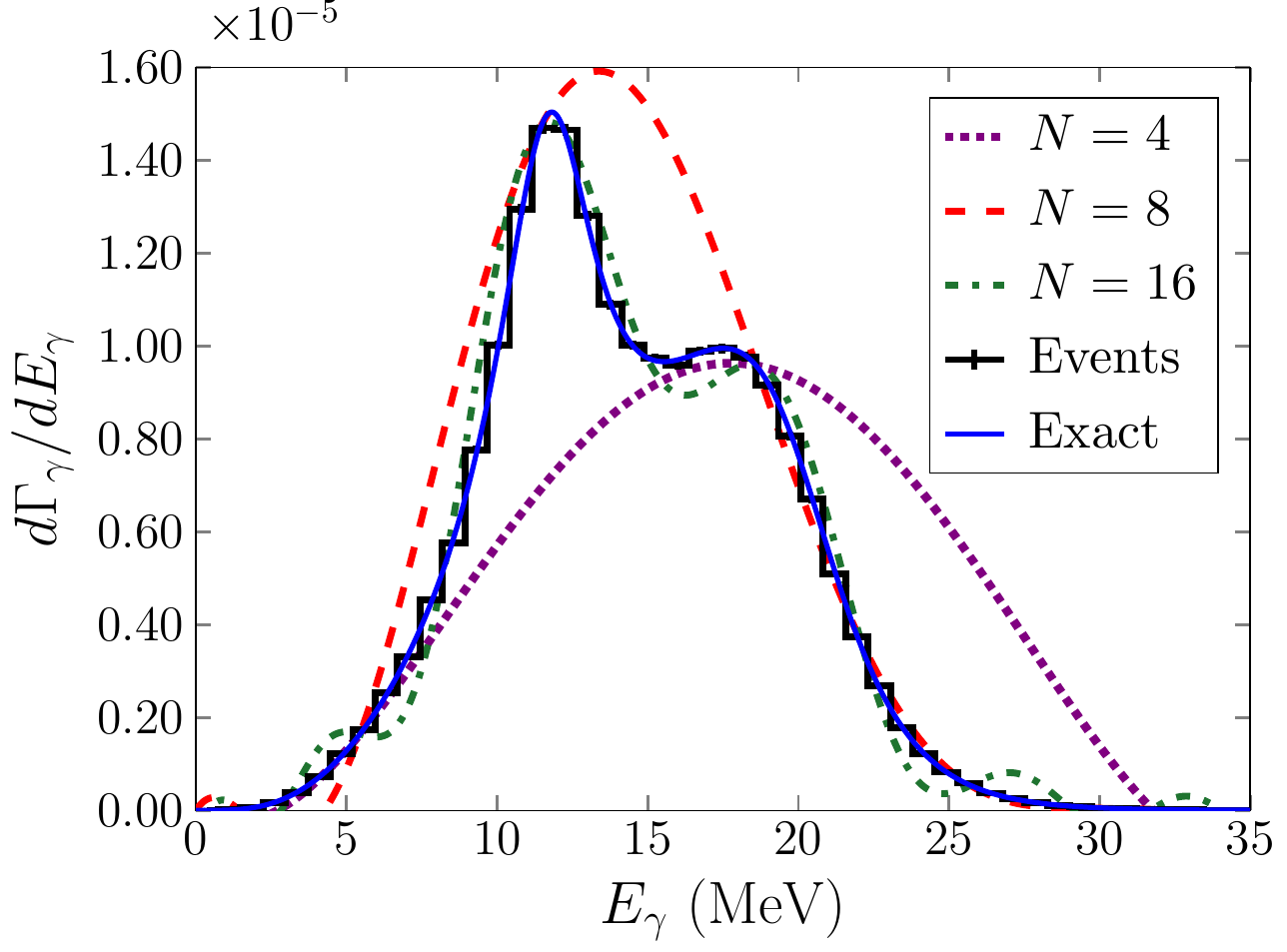}
\caption{Validation of \marley's inverse transform sampling algorithm
using simulations of proton (left) and $\gamma$-ray (right)
emission to the continuum from a highly-excited \isotope[40]{K}
state.}
\label{fig:chebyshev_demo}
\end{figure}

\Cref{fig:chebyshev_demo} shows an application of this sampling technique to
the modeling of compound nuclear decays in \marley. In the left-hand plot, the
black histogram shows the spectrum of proton kinetic energies $\mathrm{T_{p}}$
obtained from a simulation of \num[scientific-notation=fixed,
fixed-exponent=0]{2e5} decays $\isotope[40]{K} \to \mathrm{p} +
\isotope[39]{Ar}$ to the excitation energy continuum of the daughter
\isotope[39]{Ar} nucleus. The initial \isotope[40]{K} state had excitation
energy $E_x = \SI{45}{\MeV}$ and spin-parity $J^\pi = 4^{+}$. Decay modes other
than proton emission to the continuum were switched off for simplicity, and the
kinetic energy $\mathrm{T_{p}}$ is reported in the rest frame of the mother
\isotope[40]{K} nucleus. The contents of each histogram bin are normalized to
provide a calculation of the differential decay width $d\Gamma_\mathrm{p} /
d\mathrm{T_{p}}$, where the average value of this quantity in the $w$th bin is
approximated by the Monte Carlo estimator
\begin{equation}
\left< \frac{ d\Gamma_\mathrm{p} }{ d\mathrm{T_{p}} } \right>_{\!w}
\approx \frac{ n_w \, \Gamma_\mathrm{p} }{ \mathrm{N}_\text{events}
\, \Delta\mathrm{T}_\mathrm{p}^w }\,.
\end{equation}
Here $\mathrm{N}_\text{events} = \num[scientific-notation=fixed,
fixed-exponent=0]{2e5}$ is the total number of simulated decay events, $n_w$ is
the number of these that fall within the $w$th bin,
$\Delta\mathrm{T}_\mathrm{p}^w$ is the width of the $w$th bin, and
$\Gamma_\mathrm{p}$ is the total width for proton emission to the continuum.

The blue curve shows a direct calculation of the differential decay width via
\begin{equation}
\frac{ d\Gamma_\mathrm{p} }{ d\mathrm{T_{p}} }
= \frac{ M }{ M^\prime } \,
\frac{ d\Gamma_\mathrm{p} }{ dE_x^\prime }
\end{equation}
where $M$ ($M^\prime$) is the mass of the initial (final) nucleus and
$d\Gamma_\mathrm{p} / dE_x^\prime$ is evaluated according to
\cref{eq:fragment_diff_decay_width} with the fragment species $\fragment =
\mathrm{p}$. The agreement seen between the exact calculation and
the simulated events is achieved by sampling the latter from
a CDF constructed using a Chebyshev polynomial approximant
to $d\Gamma_\mathrm{p} / dE_x^\prime$ with grid size
$N = 64$. The polynomial approximant with $N = 64$ and the exact calculation
are indistinguishable on the scale of the plot. For reference, a lower-order
($N = 4$) Chebyshev polynomial approximation to the distribution is also
shown by the dotted purple line.

The right-hand plot in \cref{fig:chebyshev_demo} shows simulation results
obtained using an identical procedure, except that the decay process
$\isotope[40]{K} \rightarrow \gamma + \isotope[40]{K}$ is considered using the
differential width from \cref{eq:gamma_diff_decay_width}. Three lower-order
Chebyshev approximants are shown which provide improved agreement with the
exact calculation as the grid size $N$ grows.

\subsubsection{Implementation details}

Although the barycentric interpolation scheme described here is used in \marley\
solely for inverse transform sampling, the \cpp\ implementation is very general and
may find useful applications elsewhere. The \texttt{marley\allowbreak
::\allowbreak Chebyshev\allowbreak Interpola\allowbreak ting\allowbreak Function} class
constructs the polynomial interpolant $\tilde{f}$ for an arbitrary input
function $f(x)$, represented by a \texttt{std::\allowbreak function\allowbreak
<double\allowbreak (double)>}.
If the grid size $N$ is not specified in the constructor, then an adaptive
technique is used to choose a grid size sufficiently large to represent $f(x)$
at close to machine precision. Starting with $N = 2$, the value of $N$ is
doubled (and the Chebyshev expansion coefficients are recomputed) until the
stopping criterion
\begin{align}
\alpha_m &< 2 \, \varepsilon \max(\alpha_0, \alpha_1, \dots, \alpha_N)
& m &= N, N - 1
\end{align}
(with machine epsilon $\varepsilon$)
is satisfied or $N$ reaches a large maximum value.
The
\texttt{evaluate(double x)} member function returns the value of $\tilde{f}(x)$,
and the \texttt{cdf()} member function returns a new \texttt{marley\allowbreak
::\allowbreak Chebyshev\allowbreak Interpolating\allowbreak Function}
representing $\tilde{F}(x)$. The DCT-I calculations in
\cref{eq:DCTI-1,eq:DCTI-2} are carried out using the fast Fourier transform C
library \texttt{FFTPACK4} \cite{fftpack4Code}, which is included in the \marley\
source code distribution. This library is based on the original \texttt{FFTPACK}
Fortran code developed by P. Swarztrauber \cite{fftpackArticle,fftpackCode}.

No simultaneous sampling of multiple variables from a joint probability
distribution is needed to implement the physics models in the current version
of \marley.

\section{Event generation workflow}
\label{sec:implementation}

\begin{sidewaysfigure}
\rotatesidewayslabel
\centering
\includegraphics[width=0.7\paperheight]{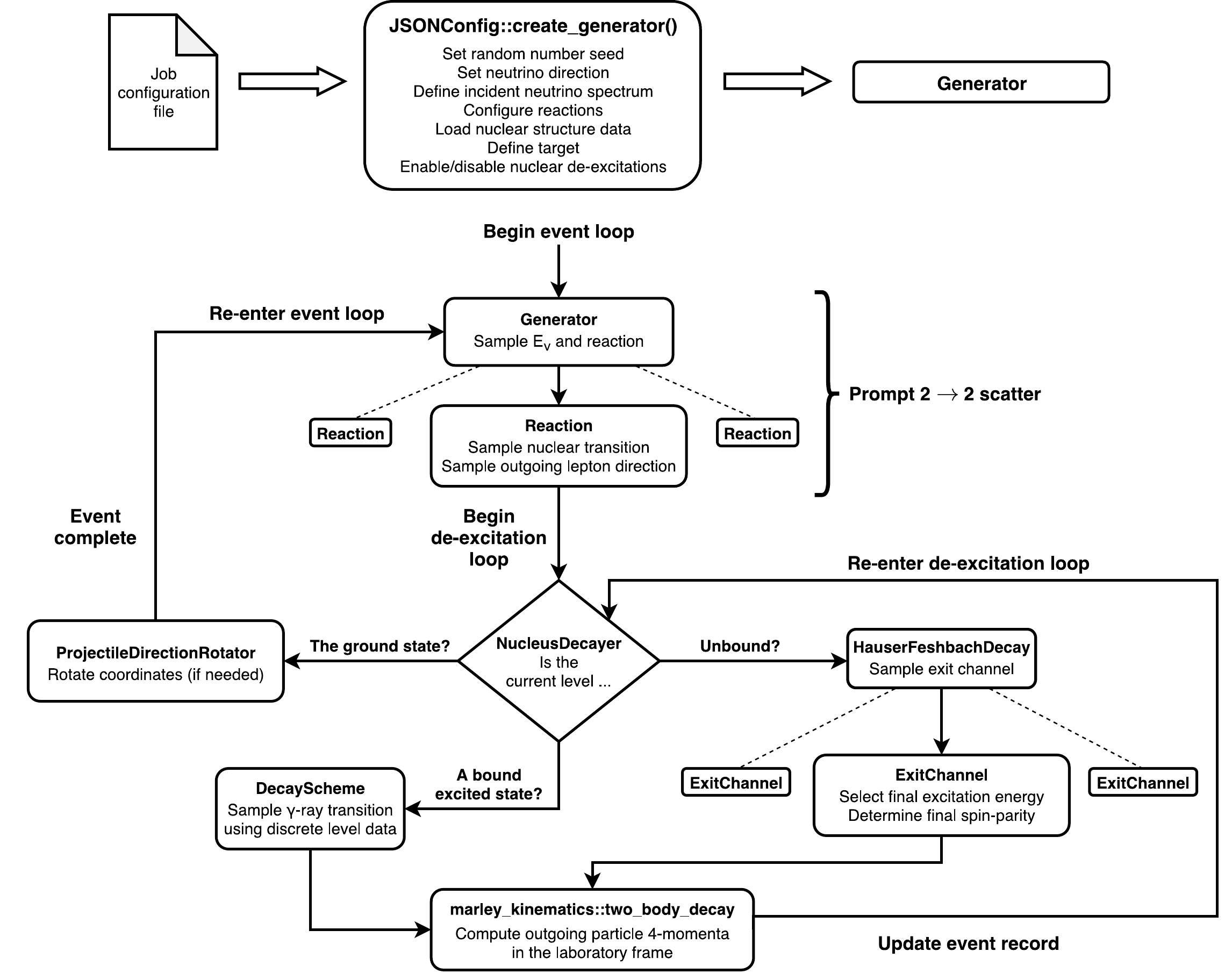}
\caption{Illustration of the workflow used by \marley\ to generate
neutrino-nucleus scattering events.}
\label{fig:flowchart}
\end{sidewaysfigure}

The flowchart in \cref{fig:flowchart} illustrates the procedure used by
\marley\ to generate events. In the following paragraphs, each stage in the
process will be described. Unless otherwise noted by providing an explicit
namespace specifier (e.g., \texttt{std::}), all \cpp\ classes referred to using
\texttt{typewriter font} in this section are defined within the \texttt{marley}
namespace.

\subsection{Generator initialization}
\label{sec:generator_init}

At the beginning of an event generation job, a \texttt{JSONConfig}
object is used to parse and interpret the settings stored in a configuration
file (whose format is described in \cref{sec:gen_config}). The
\texttt{create\_generator} member function is then used to construct a
\texttt{Generator} object which handles the actual simulation of events.
The steps below are followed to initialize the \texttt{Generator}.

\subsubsection{Set random number seed}

If the user has specified an integer value for the random number seed in the
configuration file (see \cref{sec:rng_seed_config}), then this value is used to
initialize a \texttt{std::mt19937\_\allowbreak 64} object owned by the
\texttt{Generator}. If not, then the system time since the Unix epoch is used
as a seed.

\subsubsection{Set neutrino direction}

By default, \marley\ generates events in a reference frame in which the
incident neutrino is traveling in the positive $z$ direction. If the user has
specified a different neutrino direction (see \cref{sec:neutrino_dir_config}),
then this information is passed to a \texttt{Projectile\allowbreak
Direction\allowbreak Rotator} object owned by the \texttt{Generator}. A
rotation matrix is precalculated which will allow the coordinate system to be
appropriately transformed at the end of each iteration of the event loop.

\subsubsection{Define incident neutrino spectrum}
\label{sec:define_spectrum}

Based on the user's description of the neutrino energy spectrum (see
\cref{sec:nu_source_config}), one of several derived classes of the
\texttt{NeutrinoSource} abstract base class is instantiated and stored as a
member of the \texttt{Generator} object. In typical \marley\ use cases, the
\texttt{NeutrinoSource} describes the energy distribution of \textit{incident}
neutrinos. This behavior is desirable so that the energy spectrum $P(E_\nu)$ of
\textit{reacting} neutrinos
\begin{align}
\label{eq:reacting_nu_spectrum}
P(E_\nu) &= \frac{ \phi(E_\nu) \, \sigma(E_\nu) }
{ {\scaleobj{1.17}{\int}}_{\!\!E_\nu^\text{min}}^{E_\nu^\text{max}} \phi(E_\nu)
\, \sigma(E_\nu) \, dE_\nu }
& E_\nu^\text{min} \leq E_\nu \leq E_\nu^\text{max}
\end{align}
where $\phi(E_\nu)$ is the \texttt{NeutrinoSource} spectrum and $\sigma(E_\nu)$
is the total cross section, is fully consistent with the \marley\
physics models. In unusual situations where cross section weighting is not
wanted (e.g., event generation with a uniform distribution of reacting neutrino
energies), it may be disabled in the job configuration file as described in
\cref{sec:weight_flux}. This corresponds to the substitution $\sigma(E_\nu)
\rightarrow 1$ in \cref{eq:reacting_nu_spectrum}.

\subsubsection{Configure reactions}
\label{sec:configure_reactions}

Each distinct $2\to2$ scattering mode that may be simulated during a \marley\
job is represented by an object that implements the abstract base class
\texttt{Reaction}. This class includes member functions called
\texttt{total\_xs} and \texttt{diff\_xs}, which respectively compute
(in units of $\si{\MeV\tothe{-2}}$ per target atom) the
reaction total cross section $\sigma_r(E_\nu)$ and differential cross section
$d\,\sigma_r(E_\nu) / d\cos\theta_\ell$, where $\theta_\ell$ is the lepton
scattering angle in the CM frame. The \texttt{create\_event} member function
simulates the scattering process using these quantities
(see \cref{sec:sample_level,sec:sample_lepton_direction})
and returns the results in a newly-created \texttt{Event} object.

Based on the set of one or more scattering modes enabled by the user in the
configuration file (see \cref{sec:reaction_config}), a vector of pointers to
\texttt{Reaction} objects is initialized during job startup and stored as the
\texttt{reactions\_} member of the \texttt{Generator}. This vector is populated
by calls to the factory method \texttt{Reaction\allowbreak::\allowbreak
load\_from\allowbreak\_file}, which processes a reaction specification given in
an input file (see \cref{sec:reaction_config,sec:reaction_file_format}). After
all reaction input files have been fully processed, the abundance-weighted sum
$\sigma(E_\nu)$ of the total cross sections $\sigma_r(E_\nu)$ for all of the
enabled reactions is used together with the \texttt{NeutrinoSource} spectrum
$\phi(E_\nu)$ to construct the probability density function for reacting
neutrino energies $P(E_\nu)$ shown in \cref{eq:reacting_nu_spectrum}. The
weights needed to compute $\sigma(E_\nu)$ are the nuclide fractions in the
neutrino target, as described in \cref{sec:target_definition}.

In \marley\ \version, two concrete derived classes of \texttt{Reaction} are
implemented. The \texttt{Nuclear\allowbreak Reaction} class implements the
neutrino-nucleus scattering model described in \cref{sec:xsec_theory}, while
the \texttt{Electron\allowbreak Reaction} class does the same for the
neutrino-electron elastic scattering model from \cref{sec:nu_electron_xsec}.

While simulating a neutrino-nucleus scattering event, \marley\ represents the
nuclear state in terms of the following quantities: the proton number $Z$,
nucleon number $A$, excitation energy above the ground state $E_x$, total spin
$J$, and parity $\Pi$. The values of these variables are tracked throughout the
event loop, both during $2 \to 2$ scattering (\cref{sec:loop_two_two_scatter})
and during simulation of nuclear de-excitations (\cref{sec:loop_deex}).

A distinction is made in \marley\ between bound versus unbound nuclear states.
A nuclear state is considered to be unbound if the excitation energy $E_x$
exceeds the separation energy for at least one nuclear fragment with mass
number $A \leq 4$. Separation energies are computed using atomic and particle
mass data from refs.~\cite{CODATA2010,AME2012} tabulated in the file
\texttt{data/mass\_table.js}. The singleton class \texttt{Mass\allowbreak
Table} provides a \cpp\ API for accessing and manipulating these data. If an
atomic mass value is not tabulated for a particular nuclide, the
\texttt{Mass\allowbreak Table} class computes an estimate using a formula for
the liquid-drop model mass excess developed by Myers and Swiatecki
\cite{Myers1966,Koning2008}.

A set of transition matrix elements to the final nuclear states that may be
populated by means of a $2 \to 2$ neutrino-nucleus scatter are represented in
\marley\ by a vector of \texttt{Matrix\allowbreak Element} objects. Each of
these objects is labeled as either a Fermi or Gamow-Teller matrix element (see
\cref{sec:xsec_theory}) using the \texttt{type\_} member variable, while the
\texttt{strength\_} member stores the corresponding $B(\mathrm{F})$ or
$B(\mathrm{GT})$ value. These quantities are given in the reaction input file
together with the final nuclear excitation energy, which is stored as the
\texttt{Matrix\allowbreak Element} member \texttt{level\_energy\_}. A
\texttt{Nuclear\allowbreak Reaction} is constructed using a
\texttt{std::shared\_ptr} to a vector of \texttt{Matrix\allowbreak Element}
objects, allowing shared ownership between multiple reactions as appropriate.

While parsing an input file repesenting a neutrino-nucleus reaction, the static
method \texttt{Reaction\allowbreak ::\allowbreak load\_from\allowbreak\_file}
constructs one \texttt{Matrix\allowbreak Element} object for each listed
nuclear transition. After the full list of transitions has been processed, a
\texttt{StructureDatabase} object owned by the \texttt{Generator} is consulted
to determine whether discrete level data are available for the final-state
nucleus of interest. The first such query will trigger loading of
these data in the form of \texttt{Decay\allowbreak Scheme} objects as
described in \cref{sec:load_nuclear_structure}. If a suitable
\texttt{DecayScheme} is available, then every \texttt{Matrix\allowbreak
Element} which represents a transition to a bound nuclear final state is
matched to a \texttt{Level} object owned by the \texttt{DecayScheme}. This
matching is performed by selecting the \texttt{Level} whose excitation energy
most closely matches the value of the \texttt{Matrix\allowbreak Element}'s
member variable \texttt{level\_energy\_}. Duplicate matches occurring for two
\texttt{Matrix\allowbreak Element} objects belonging to the same
\texttt{Nuclear\allowbreak Reaction} will lead to an exception being thrown. A
successful match will result in the \texttt{Matrix\allowbreak Element}'s member
variable \texttt{final\_level\_} being loaded with a pointer to the matched
\texttt{Level}. If the spin-parity of the matched \texttt{Level} is not
compatible with the relevant selection rules (from either
\cref{eq:Fermi_sel_rule} or \cref{eq:GT_sel_rule}), then a warning message will
be printed to the screen.

Matching of this kind is not performed for \texttt{Matrix\allowbreak Element}
objects representing transitions to unbound nuclear states. In such cases (or
when an appropriate \texttt{Decay\allowbreak Scheme} is not available), a null
pointer is assigned to the \texttt{final\_level\_} member.

\subsubsection{Load nuclear structure data}
\label{sec:load_nuclear_structure}

To simulate nuclear transitions at low excitation energies, \marley\ relies on
tabulated nuclear structure data files. These files contain listings of
discrete nuclear energy levels and branching ratios for $\gamma$-ray
transitions between them. The file format is documented in
\cref{sec:nuc_struct}.

For \marley\ \version, the recommended structure data files are largely taken
(with reformatting) from version 1.6 of the TALYS nuclear reaction code
\cite{Talys1,Talys2012}. The data are organized by element in the folder
\texttt{data/structure/}. This folder also includes a text file named
\texttt{nuclide\_index.txt} which serves as an index for the entire dataset:
each line of the text file has the format
\begin{center}
\begin{tabular}{c}
\begin{lstlisting}[backgroundcolor=\color{white}, numbers=none]
NucPDG  DataFileName
\end{lstlisting}
\end{tabular}
\end{center}
where \texttt{NucPDG} is an integer PDG code (see \cref{sec:PDG_codes})
representing a nuclear species and \texttt{Data\allowbreak File\allowbreak
Name} is the name of the corresponding file (assumed to appear in
\texttt{data/structure/}) in which its discrete level data are given.

A \texttt{Structure\allowbreak Database} object owned by the \texttt{Generator}
provides an interface for accessing and manipulating nuclear structure data
within \marley. The \texttt{Structure\allowbreak Database} manages a lookup
table indexed by nuclear PDG code that stores pointers to
\texttt{Decay\allowbreak Scheme} objects. Each \texttt{Decay\allowbreak Scheme}
represents a table of nuclear levels and $\gamma$-ray branching ratios for a
particular nuclide. A fully-initialized \texttt{DecayScheme} owns one
\texttt{Level} object for every discrete level listed for the nuclide of
interest in the associated structure data file. Each \texttt{Level} itself owns
one \texttt{Gamma} object for each listed $\gamma$-ray transition that may
originate from it.

External access to the \texttt{Decay\allowbreak Scheme} objects is provided by
the \texttt{Structure\allowbreak Database} member function
\texttt{get\_decay\_scheme}. This function takes a nuclear PDG code or a ($Z$,
$A$) pair as input. If a corresponding \texttt{Decay\allowbreak Scheme} already
exists in the lookup table, a pointer to it is immediately returned. If a
suitable \texttt{Decay\allowbreak Scheme} has not been constructed yet, the
\texttt{Structure\allowbreak Database} consults the nuclide index (which is
automatically loaded from \texttt{nuclide\_index.txt} when needed and cached
for repeated use) in an attempt to find a matching structure data file. If a
match is found, \texttt{Decay\allowbreak Scheme} objects are constructed and
added to the lookup table for every nuclide listed in the file. A pointer to
the requested \texttt{Decay\allowbreak Scheme} is then returned if it was
successfully loaded. A null pointer is stored in the lookup table and returned
when the attempt to load the requested \texttt{Decay\allowbreak Scheme} fails.

Additional information indexed by nuclear PDG code is also stored in the
\texttt{Structure\allowbreak Database}, including
\begin{enumerate*}[label=(\arabic*)]
\item ground-state nuclear spin-parities\footnote{Also obtained from
nuclear structure data files included with TALYS 1.6} loaded from the file
\texttt{data/\allowbreak structure/\allowbreak
gs\_spin\_\allowbreak parity\_table.txt} and
\item three varieties of objects (represented by the abstract base classes
\texttt{Level\allowbreak Density\allowbreak Model}, \texttt{Optical\allowbreak
Model}, and \texttt{Gamma\allowbreak Strength\allowbreak Function\allowbreak
Model}) used to compute quantities of interest for the Hauser-Feshbach
statistical model (see \cref{sec:HFSM,sec:simulation_unbound}).
\end{enumerate*}
In \marley\ \version, these three abstract base classes each have a single
concrete implementation which is used for calculations.\footnote{A second
\texttt{Gamma\allowbreak Strength\allowbreak Function\allowbreak Model}
implementation, \texttt{Weisskopf\allowbreak Single\allowbreak
Particle\allowbreak Model}, is preserved in the code base for historical
interest but has been deprecated.} The \texttt{Backshifted\allowbreak
Fermi\allowbreak Gas\allowbreak Model} class is derived from
\texttt{Level\allowbreak Density\allowbreak Model} and computes nuclear level
densities (e.g., $\rho_i$ and $\rho_f$ in \cref{eq:fragment_diff_decay_width})
as described in \cref{sec:nuc_fragments}. The \texttt{Koning\allowbreak
Delaroche\allowbreak Optical\allowbreak Model} class is derived from
\texttt{Optical\allowbreak Model} and computes nuclear fragment transmission
coefficients using the procedure outlined in \cref{sec:fragment_tr_coeff}.
Finally, the \texttt{Standard\allowbreak Lorentzian\allowbreak Model} class is
derived from \texttt{Gamma\allowbreak Strength\allowbreak Function\allowbreak
Model} and computes the $\gamma$-ray strength function from
\cref{eq:SLM_strength_function} as described in \cref{sec:gamma_tr_coeff}.

Beyond the nuclide-specific items managed by the \texttt{Structure\allowbreak
Database}, there are also two settings which are common to all nuclei: the
cutoff values $\ell_\text{max}$ and $\lambda_\text{max}$ used to truncate the
sums in \cref{eq:fragment_diff_decay_width} and
\cref{eq:gamma_diff_decay_width}, respectively, (see
\cref{sec:nuc_fragments,sec:gamma_emission,sec:l_max}) and a list (elements of
which are represented by the \texttt{Fragment} class) of the nuclear fragments
to consider when simulating decays using the Hauser-Feshbach model.

\subsubsection{Define target}
\label{sec:target_definition}

In a \marley\ simulation, the isotopic composition of the material illuminated
by the incident neutrinos is represented by a \texttt{Target} object owned by
the \texttt{Generator}. In the absence of an explicit list of abundances
specified by the user (see \cref{sec:target_config}), \marley\ assumes equal
amounts of each nuclide that appears in the initial state of at least one
configured \texttt{Reaction}. Using the information stored in the
\texttt{Target} object, the \texttt{Generator} computes the abundance-weighted
total cross section per target atom via
\begin{equation}
\label{eq:abundance_weighted_xsec}
\sigma(E_\nu) = \sum_r f_r \, \sigma_r(E_\nu)
\end{equation}
where $f_r$ is the nuclide fraction for the initial-state
atom involved in the $r$th \texttt{Reaction}.

\subsubsection{Enable/disable nuclear de-excitations}
\label{sec:nuc_deex_enable_disable}

By default, \marley\ simulates both the prompt $2\to2$ scattering reaction and
the subsequent nuclear de-excitations for every event. In applications where
only the prompt reaction is important, the user may disable de-excitations as
described in \cref{sec:disable_deex}. The choice of whether or not to simulate
de-excitations is represented as a boolean member variable owned by the
\texttt{Generator} object.

\subsection{Event loop: $\text{2} \to \text{2}$ scattering}
\label{sec:loop_two_two_scatter}

After the \texttt{Generator} object has been fully initialized, the
\texttt{marley} command-line executable enters an event loop. This loop
iterates until a fixed number of events requested by the user (see
\cref{sec:executable_settings}) has been reached or the loop is interrupted (by
an error condition or by the user pressing \texttt{ctrl+C}). A single event
loop iteration corresponds to a call to the \texttt{Generator} member function
\texttt{create\_event}. This function returns an \texttt{Event} object
constructed according to the following steps.

\subsubsection{Neutrino energy selection}

In the first step of the event loop, the \texttt{Generator} employs
rejection sampling (see \cref{sec:rejection_sampling}) to select a reacting
neutrino energy from the probability distribution $P(E_\nu)$ defined in
\cref{eq:reacting_nu_spectrum}. By default, Brent's method is used during the
first event loop iteration to obtain a numerical estimate $P_\text{max}$ of the
maximum value of the probability density function $P(E_\nu)$.  The value of
$P_\text{max}$ is cached and reused during neutrino energy sampling for
subsequent events.

In cases where Brent's method fails to converge to the global maximum of
$P(E_\nu)$, the resulting distribution of $E_\nu$ in the generated events will
be biased, with too few events produced in energy regions where $P(E_\nu)$ is
larger than $P_\text{max}$. At the start of each pass through the event loop,
\marley\ verifies that all values of $P(E_\nu)$ computed during rejection
sampling never exceed $P_\text{max}$. If the cached value of $P_\text{max}$ is
ever found to be an underestimate of the true global maximum, an error message
is printed (see \cref{lst:rejection_problem1} in \cref{sec:troubleshoot_reject}
for an example), and the value of $P_\text{max}$ is updated to match the
largest value of $P(E_\nu)$ encountered.

Users can override the default use of Brent's method by supplying their own
values of $P_\text{max}$ in the job configuration file. Instructions for doing
so can be found in \cref{sec:energy_pdf_max}. For convenience in
troubleshooting, the error message printed by \marley\ in response to a
rejection sampling problem contains a recommended value of $P_\text{max}$ to
use in this way.

\subsubsection{Reaction mode sampling}

Once a reacting neutrino energy $E_\nu$ has been chosen, the \texttt{Generator}
samples a specific reaction mode $r$ from the discrete probability distribution
\begin{equation}
\label{eq:reaction_pmf}
P(r) = \frac{ f_r \, \sigma_r(E_\nu) }{ \sigma(E_\nu) }.
\end{equation}
Control then passes to the \texttt{create\_event} member function of the $r$th
\texttt{Reaction} object owned by the \texttt{Generator}. This function handles
selection of $2 \to 2$ scattering kinematics and initializes an \texttt{Event}
object.

\subsubsection{Selection of a nuclear transition}
\label{sec:sample_level}

If the sampled reaction mode involves scattering on a nuclear target, then
simulation of the prompt $2 \to 2$ reaction is handled by the
\texttt{Nuclear\allowbreak Reaction} class, and the outgoing nucleus may be
left in one of potentially many final states. A specific final state is chosen
for a nuclear reaction by selecting a \texttt{Matrix\allowbreak Element}
according to the discrete distribution
\begin{equation}
\label{eq:level_probability}
P(\levelIndex) = \frac{ \sigma_r(E_\nu, \levelIndex) }{ \sigma_r(E_\nu) }
\end{equation}
where $\sigma_r(E_\nu, \levelIndex)$ is the partial reaction cross section
computed for the $\levelIndex$th \texttt{Matrix\allowbreak Element} owned by
the \texttt{Nuclear\allowbreak Reaction} of interest. That is, $\sigma_r(E_\nu,
\levelIndex)$ is given by the expression in \cref{eq:tot_xsec} with the
quantity $B(\mathrm{F}) + B(\mathrm{GT})$ set equal to the \texttt{strength\_}
member of the $\levelIndex$th \texttt{Matrix\allowbreak Element}. The total
cross section $\sigma_r(E_\nu)$ is obtained by summing over all of the owned
\texttt{Matrix\allowbreak Element} objects which represent
energetically-accessible nuclear transitions:
\begin{equation}
\sigma_r(E_\nu) = \sum_\levelIndex \sigma_r(E_\nu, \levelIndex)
\end{equation}

If the sampled \texttt{Matrix\allowbreak Element} has been matched to a known
discrete nuclear \texttt{Level} (i.e., if its member pointer
\texttt{final\_level\_} is non-null, see \cref{sec:configure_reactions}), then the
excitation energy, spin, and parity of the final nucleus are determined
directly from the \texttt{Level} object.

If a match is not available (e.g., because the associated nuclear state is
unbound), then the final nuclear excitation energy $E_x$ is set equal to the
sampled \texttt{Matrix\allowbreak Element}'s member variable
\texttt{level\_energy\_}. The final-state spin $J$ and parity $\Pi$ are
determined according to the selection rules given in either
\cref{eq:Fermi_sel_rule} or \cref{eq:GT_sel_rule} as appropriate for the
\texttt{Matrix\allowbreak Element} type. In the case of a Gamow-Teller
transition involving an initial nucleus with nonzero spin (as determined using
the table of ground-state spin-parities managed by the
\texttt{Structure\allowbreak Database}, see \cref{sec:load_nuclear_structure}),
multiple $J$ values will satisfy the spin selection rule in
\cref{eq:GT_sel_rule}. In such cases, \marley\ \version\ makes a rough
approximation: equipartition of spin is assumed, and a definite value of $J$ is
sampled from the discrete distribution
\begin{equation} \label{eq:dist_GT_J}
P(J) = \frac{ \rho(E_x, J, \Pi) }{ \sum_{K} \rho(E_x, K, \Pi) }
\end{equation}
where $\rho$ (which is calculated using the same level density treatment as in
\cref{sec:nuc_fragments}) is the density of final-state nuclear levels with
spin $J$ and parity $\Pi$ in the vicinity of excitation energy $E_x$. In
\cref{eq:dist_GT_J}, the sum in the denominator runs over all final spin values
$K$ that satisfy the Gamow-Teller spin selection rule from
\cref{eq:GT_sel_rule}.

\subsubsection{Outgoing lepton direction}
\label{sec:sample_lepton_direction}

Due to the simple kinematics of $2 \to 2$ scattering, the 4-momenta of the
outgoing particles are fully determined by specifying their masses (including
the final nuclear excitation energy) and the direction of the outgoing lepton.
This direction is represented in \marley\ using the lepton's azimuthal
scattering angle $\phi_\ell$ and polar scattering cosine $\cos\theta_\ell$,
both measured with respect to the incident neutrino direction in the
center-of-momentum (CM) frame.

For all reactions currently implemented in \marley\ \version, the differential
cross section is independent of $\phi_\ell$, and the value of this variable is
therefore sampled uniformly on the half-open interval $\left[0, 2\pi\right)$.

For neutrino-nucleus reactions, a value of $\cos\theta_\ell$ is obtained via
rejection sampling (see \cref{sec:rejection_sampling}) on the interval
$\left[-1, 1\right]$ from the probability density function
\begin{equation}
\label{eq:cos_dist}
P(\cos\theta_\ell) = \frac{1}{\sigma_r(E_\nu, \levelIndex)}
\frac{ d\sigma_r(E_\nu, \levelIndex) }
{d\cos\theta_\ell}
\end{equation}
where $d \, \sigma_r(E_\nu, \levelIndex) \, / \, d\cos\theta_\ell$ is the
CM-frame differential cross section for the reaction $r$ proceeding via a
nuclear transition described by the $\levelIndex$th \texttt{Matrix\allowbreak
Element}. This differential cross section is computed as in
\cref{eq:diff_xsec}, except that only one of the two matrix element terms is
used. For a Fermi transition, $B(\mathrm{F})$ is set equal to the
\texttt{strength\_} member variable of the \texttt{Matrix\allowbreak Element},
while $B(\mathrm{GT})$ is set to zero. For Gamow-Teller transitions, the
reverse is done. In the case of neutrino-electron elastic scattering, rejection
sampling is still performed using the distribution from \cref{eq:cos_dist}, but
the differential cross section is calculated as in \cref{eq:diff_xsec_nu_e}.

For nuclear transitions proceeding purely via a Fermi or Gamow-Teller operator,
the angular dependence in the CM frame reduces to (see \cref{eq:diff_xsec})
\begin{equation}
P(\cos\theta_\ell) \propto
\begin{cases}
1 + \beta_\ell \cos\theta_\ell & \text{Fermi} \\
1 - \frac{1}{3} \, \beta_\ell \cos\theta_\ell & \text{Gamow-Teller}
\end{cases}
\end{equation}
where $\beta_\ell \in [0, 1]$ is the speed of the outgoing lepton. The linear
expressions seen here for either case allow the maximum of the distribution
(needed for rejection sampling) to be found analytically. Nuclear transitions
that do not correspond to one of these two cases are neglected in the current
version of \marley. The maximum of the $\cos\theta_\ell$ distribution for
neutrino-electron elastic scattering is also found analytically.

After a CM frame direction is sampled for the outgoing lepton, the active
\texttt{Reaction} initializes \texttt{Particle} objects with the initial and
final 4-momenta for the $2 \to 2$ scattering event. The values of these in the
laboratory frame are stored in a newly-constructed \texttt{Event} object
together with the excitation energy and spin-parity of the final energy level
of the target.

At this point, program control passes back to the \texttt{Generator}, which
applies two additional operations to the \texttt{Event} object before the event
loop is complete. Once an \texttt{Event} object has been created by a
\texttt{Reaction}, further operations on it are handled by derived instances of
the \texttt{EventProcessor} abstract base class. These define a member function
called \texttt{process\_event} which takes references to the \texttt{Event}
and to the \texttt{Generator} as arguments.

\subsection{Event loop: Nuclear de-excitations}
\label{sec:loop_deex}

For reactions involving a transition to an excited state of a nuclear target,
the subsequent de-excitations are simulated using an \texttt{EventProcessor}
called \texttt{NucleusDecayer}. This class manages a de-excitation loop that is
initialized using the proton number ($Z$), nucleon number ($A$), excitation
energy ($E_x$), spin ($J$), and parity ($\Pi$) of the outgoing nuclear state
stored in the \texttt{Event} object. The values of these variables are updated
during each iteration of the de-excitation loop, which simulates a sequence of
binary decays until the nuclear ground state ($E_x = 0$) is reached.

Each binary decay step begins with a check to determine whether the current
nuclear state is bound or unbound (see \cref{sec:configure_reactions}).

\subsubsection{Unbound nuclear states}
\label{sec:simulation_unbound}

In the case of an unbound nuclear state, the \texttt{NucleusDecayer} constructs
a \texttt{HauserFeshbachDecay} object, which provides a Monte Carlo
implementation of the Hauser-Feshbach statistical model (see \cref{sec:HFSM})
for decays of a compound nucleus. Upon construction, the
\texttt{HauserFeshbachDecay} object uses the properties of the initial nuclear
state ($Z$, $A$, $E_x$, $J$, $\Pi$) and information from the
\texttt{Generator}'s owned \texttt{StructureDatabase} to initialize a member
vector of pointers to \texttt{ExitChannel} objects.

The abstract base class \texttt{ExitChannel} represents a nuclear de-excitation
via emission of a $\gamma$-ray or a particular nuclear fragment. Two pairs of
abstract derived classes virtually inherit from \texttt{ExitChannel}. The
classes in the first pair, \texttt{Discrete\allowbreak Exit\allowbreak Channel}
and \texttt{Continuum\allowbreak Exit\allowbreak Channel}, specialize to
handling transitions to a specific discrete nuclear level and to the continuum,
respectively. The classes in the second pair, \texttt{Fragment\allowbreak
Exit\allowbreak Channel} and \texttt{Gamma\allowbreak Exit\allowbreak Channel},
respectively represent de-excitations involving fragment and $\gamma$-ray
emission. Concrete \texttt{Exit\allowbreak Channel} objects belong to a class
that inherits from exactly one member of each pair, with the four possibilities
being \texttt{Fragment\allowbreak Discrete\allowbreak Exit\allowbreak Channel},
\texttt{Fragment\allowbreak Continuum\allowbreak Exit\allowbreak Channel},
\texttt{Gamma\allowbreak Discrete\allowbreak Exit\allowbreak Channel}, and
\texttt{Gamma\allowbreak Continuum\allowbreak Exit\allowbreak Channel}.

Every \texttt{Exit\allowbreak Channel} object owns a member variable called
\texttt{width\_}, which is initialized during construction with the partial
decay width of interest (in \si{\per\MeV}) via a call to the protected member
function \texttt{compute\_total\_width}. The
\texttt{Fragment\allowbreak Discrete\allowbreak Exit\allowbreak Channel} and
\texttt{Gamma\allowbreak Discrete\allowbreak Exit\allowbreak Channel}
classes implement this function using the expressions from
\cref{eq:fragment_discrete_decay_width,eq:gamma_partial_width}, respectively.
The \texttt{Fragment\allowbreak Continuum\allowbreak Exit\allowbreak Channel}
and \texttt{Gamma\allowbreak Continuum\allowbreak Exit\allowbreak Channel}
classes both share the \texttt{Continuum\allowbreak Exit\allowbreak
Channel\allowbreak ::\allowbreak compute\_total\_width} implementation,
which integrates the differential decay width over the energetically-accessible
continuum (see \cref{sec:nuc_fragments,sec:gamma_emission}) to obtain the
\texttt{width\_} value
\begin{equation}
\Gamma_\decayProductIndex = \int_{E_x^{ \prime \,
\text{min,c} }}^{E_x^{ \prime \, \text{max} }}
\frac{ d \Gamma_\decayProductIndex }{ d E_x^{\prime} } \,
d E_x^{\prime}
\end{equation}
where the emitted particle species
$\decayProductIndex \in \{ \mathrm{ p, \, n, \, d,\, t,\, h,}
\, \alpha, \gamma \}$. The numerical integration is performed using
Clenshaw-Curtis quadrature \cite{Clenshaw1960} by an instance of the
\texttt{Integrator} class.
Evaluation of the differential decay width
$d\Gamma_\decayProductIndex / dE_x^{\prime}$
is delegated to the pure virtual member function \texttt{differential\_width},
which is implemented in \texttt{Fragment\allowbreak Continuum\allowbreak
Exit\allowbreak Channel} and \texttt{Gamma\allowbreak Continuum\allowbreak
Exit\allowbreak Channel} according to the expressions given in
\cref{eq:fragment_diff_decay_width} and \cref{eq:gamma_diff_decay_width},
respectively.

When the \texttt{Hauser\allowbreak Feshbach\allowbreak Decay} object has been
fully initialized, its owned vector of \texttt{Exit\allowbreak Channel}
pointers will include an individual object derived from
\texttt{Discrete\allowbreak Exit\allowbreak Channel} for each accessible
transition to a discrete nuclear level present in the
\texttt{Structure\allowbreak Database}. It will also contain a pointer to a
single object derived from \texttt{Continuum\allowbreak Exit\allowbreak
Channel} for each particle species that may be emitted via a transition to the
continuum.

To simulate a binary decay step, the \texttt{NucleusDecayer} calls the
\texttt{do\_decay} method of the new \texttt{HauserFeshbachDecay} object. This
causes a particular \texttt{Exit\allowbreak Channel} $\exitChannelIndex$ to be
sampled with probability
\begin{equation}
P(\exitChannelIndex) = \frac{\Gamma_\exitChannelIndex}{ \sum_\beta \Gamma_\beta }.
\end{equation}
Here $\Gamma_\exitChannelIndex$ is the partial width for the
$\exitChannelIndex$th \texttt{Exit\allowbreak Channel} (given by the
\texttt{width\_} member variable) and the sum in the denominator runs over all
of the owned \texttt{Exit\allowbreak Channel} objects. Control then passes to
the \texttt{do\_decay} member function of the sampled \texttt{Exit\allowbreak
Channel} to determine the properties of the final nuclear state. If a
\texttt{Discrete\allowbreak Exit\allowbreak Channel} has been sampled, the
final nuclear excitation energy ($E_x^\prime$), spin ($J^\prime$), and parity
($\Pi^\prime$) are retrieved from the associated \texttt{Level} object.

For a \texttt{Continuum\allowbreak Exit\allowbreak Channel}, a
\texttt{Chebyshev\allowbreak Interpolating\allowbreak Function} approximation
to the differential decay width $d \Gamma_\decayProductIndex / d E_x^{\prime}$
is used to select a final excitation energy $E_x^\prime \in [ E_x^{ \prime \,
\text{min,c} }, E_x^{ \prime \, \text{max} }]$ via inverse transform sampling
(see \cref{sec:inverse_transform_sampling}). The differential decay width is
then evaluated (via a call to the \texttt{differential\_width} member function)
for the selected final excitation energy $E_x^\prime$. Each individual term
$\tau_\termIndex$ in the differential decay width sum (see either
\cref{eq:fragment_diff_decay_width} or \cref{eq:gamma_diff_decay_width} as
appropriate for the emitted particle species) is stored in a
\texttt{SpinParityWidth} object together with its corresponding $J^\prime$ and
$\Pi^\prime$ values. The final nuclear spin-parity is determined by choosing
the $\termIndex$th \texttt{SpinParityWidth} object with probability
\begin{equation} P(\termIndex) = \frac{ \tau_\termIndex }{ \sum_c \tau_c }
= \frac{ \tau_\termIndex }{ d \Gamma_\decayProductIndex / d E_x^{\prime} }
\end{equation} where the differential decay width in the denominator is
evaluated at the sampled value of $E_x^\prime$.

Once values of the variables defining the final nuclear state ($E_x^\prime$,
$J^\prime$, and $\Pi^\prime$) have been determined, the procedure required to
complete an iteration of the event loop (see \cref{sec:finish_deex}) is similar
to that used for decays of bound nuclear levels.

\subsubsection{Bound nuclear states}
\label{sec:simulation_bound}

If the nucleus is in a bound excited state, i.e., it has a nonzero
excitation energy $E_x$ which is below all of the nuclear fragment emission
thresholds, then the \texttt{NucleusDecayer} consults the
\texttt{StructureDatabase} to determine whether a \texttt{DecayScheme} has been
previously configured for the nuclide of interest (see
\cref{sec:load_nuclear_structure}). If this is the case, then simulation of the
remaining de-excitations is delegated to the \texttt{DecayScheme} member
function \texttt{do\_cascade}. During each binary decay step handled by this
function, a $\gamma$-ray transition $\gamma_j$ originating from the current
\texttt{Level} is chosen with probability
\begin{equation}
P(\gamma_j) = \frac{I_j}{\sum_k I_k}
\end{equation}
where $I_j$ is the relative intensity of the $j$th \texttt{Gamma} owned by the
current \texttt{Level} and the sum in the denominator runs over all of the
owned \texttt{Gamma} objects. The properties of the final nuclear state, known
immediately from the \texttt{Level} pointed to by the sampled \texttt{Gamma}
object (via its \texttt{end\_level\_} member variable) are used to complete
the de-excitation loop iteration as described in \cref{sec:finish_deex}.

If a suitable \texttt{DecayScheme} cannot be found in the
\texttt{StructureDatabase}, then \marley\ follows the same procedure as for
unbound nuclear levels (see \cref{sec:simulation_unbound}).

\subsubsection{Finishing a de-excitation step}
\label{sec:finish_deex}

At the end of each iteration of the de-excitation loop, a direction for the
emitted $\gamma$-ray or nuclear fragment is chosen by sampling a polar cosine
$\cos\theta$ and an azimuthal angle $\phi$ isotropically in the rest frame of
the decaying nucleus. This information is used together with the daughter
particle PDG codes and masses to initialize two new \texttt{Particle} objects
via a call to the utility function
\texttt{marley\allowbreak\_kinematics\allowbreak::\allowbreak
two\allowbreak\_body\allowbreak\_decay}. This function handles the elementary
kinematical calculations needed to obtain the outgoing particle 4-momenta in
the laboratory frame. The \texttt{Particle} representing the emitted
$\gamma$-ray or fragment is appended to the vector of final particles owned by
the \texttt{Event} object being processed. The \texttt{Particle} representing
the daughter nucleus, on the other hand, replaces the mother nucleus
\texttt{Particle} in the \texttt{Event}. After these adjustments have been made
to the \texttt{Event} object and the values of the variables tracked in the
de-excitation loop ($Z$, $A$, $E_x$, $J$, $\Pi$) have been updated,
simulation of the current binary decay step is complete.

The de-excitation loop will continue to simulate additional binary decay steps
until the nuclear ground state is reached, or, in cases where no discrete level
data are available for a particular final-state nucleus, until the excitation
energy falls below the cutoff $\ExCut = \SI{1}{\keV}$.

\subsection{Event loop: Rotation of coordinates}
\label{sec:rotation}

For convenience during internal calculations, all \marley\ events are initially
generated in a frame in which the incident neutrino direction lies along the
positive z axis. If the user has specified a different neutrino direction in
the job configuration file (see \cref{sec:neutrino_dir_config}), then an
instance of the \texttt{Projectile\allowbreak Direction\allowbreak Rotator}
class (which inherits from \texttt{Event\allowbreak Processor}) is used to
process the \texttt{Event} after the de-excitation loop terminates. All
particle 3-momenta in the event are rotated into a new reference frame in which
the neutrino is traveling along the direction specified by the user. After this
rotation is applied, generation of the new \marley\ event is complete.

\section{Installation and usage}
\label{sec:install}

The current release of \marley\ has been tested on both Linux and macOS
platforms and is expected to work in any Unix-like environment in which the
prerequisites are installed. Building and running \marley\ on Windows is not
currently supported. The installation instructions presented in this section
assume use of the Bash shell \cite{Bash} and the availability of several
standard command-line tools.

\subsection{Prerequisites}

The following prerequisites are required to build the \marley\ source code:
\begin{itemize}
\item A \cpp-14-compliant compiler. Two compilers are officially supported:
\begin{itemize}
  \item The GNU Compiler Collection (GCC) \cite{GCC}, version \minGCCversion\
or greater
  \item The Clang frontend for the LLVM compiler infrastructure \cite{Clang},
    version \minClangVersion\ or greater
\end{itemize}
\item GNU Make \cite{Make}
\item The GNU Scientific Library (GSL) \cite{Galassi2009,GSLWebsite}
\end{itemize}

On Linux architectures, all of these prerequisites will likely be available
for installation through the standard package manager. On macOS, the use of
Homebrew \cite{Homebrew} to install GSL is recommended. This may be done
by executing the terminal command
\begin{lstlisting}[backgroundcolor=\color{white}, numbers=none,
basicstyle=\normalsize\ttfamily]
brew install gsl
\end{lstlisting}
after Homebrew has been installed on the target system.

Although it is not required in order to build or use \marley, the popular
\rootcern\ data analysis framework \cite{Brun1997,ROOTWebsite} provides
convenient tools for plotting and analyzing simulation results. Users who wish
to use the optional interface between the two codes (see
\cref{sec:ROOT_format,sec:macros,sec:marsum}) should ensure that \rootcern\ is
installed before building \marley.

Other than GSL, \marley's only required external dependencies are the \cpp\
standard library and the symbols defined in the \texttt{dirent.h} and
\texttt{sys/stat.h} headers of the C POSIX library \cite{POSIX.1-2017}.

\subsection{Obtaining and building the code}
\label{sec:get_code}

The source code for \marley\ \version\ may be downloaded as a compressed
archive file from the releases webpage
(\url{https://github.com/MARLEY-MC/marley/releases}).
Both \texttt{zip} and \texttt{tar.gz} file formats are available. After
downloading the source code, the user should unpack the archive file in the
desired installation folder. For the \texttt{tar.gz} format, this may be done
via the command
\begin{lstlisting}[backgroundcolor=\color{white}, numbers=none,
basicstyle=\normalsize\ttfamily]
tar xvfz marley-1.2.0.tar.gz
\end{lstlisting}
After unpacking the source code, the user may build \marley\ by navigating to
the \texttt{build/} subdirectory and invoking GNU Make:
\begin{lstlisting}[backgroundcolor=\color{white}, numbers=none,
basicstyle=\normalsize\ttfamily]
cd marley-1.2.0/build
make
\end{lstlisting}

At build time, \marley\ verifies that GSL is installed by checking for the
presence of the \texttt{gsl-config} script on the system path. Similarly,
the optional \marley\ interface to \rootcern\ is automatically enabled or
disabled based on whether \texttt{root-config} script is present.
Users who have an existing installation of \rootcern\ but desire to build
\marley\ without \rootcern\ support may manually disable the interface
by setting the \texttt{IGNORE\_ROOT} variable while invoking \texttt{make},
e.g., via
\begin{lstlisting}[backgroundcolor=\color{white}, numbers=none,
basicstyle=\normalsize\ttfamily]
make IGNORE_ROOT=yes
\end{lstlisting}
If the build is successful, then executing
\begin{lstlisting}[backgroundcolor=\color{white}, numbers=none,
basicstyle=\normalsize\ttfamily, commentstyle=\normalsize\ttfamily]
./marley --version
\end{lstlisting}
from within the \texttt{build/} folder should yield the following console
output:
\begin{lstlisting}[backgroundcolor=\color{white}, numbers=none,
backgroundcolor=\color{lightgray},
basicstyle=\small\ttfamily\bfseries, commentstyle=\small\ttfamily\bfseries]
MARLEY (Model of Argon Reaction Low Energy Yields) 1.2.0
Copyright (C) 2016-2020 Steven Gardiner
License: GNU GPL version 3 <http://opensource.org/licenses/GPL-3.0>
This is free software: you are free to change and redistribute it.
\end{lstlisting}

Users wishing to contribute improvements to \marley\ may prefer to clone the
official source code repository instead of downloading a release archive file.
Instructions for doing so are available in the ``developer documentation''
section of the official website \cite{MARLEYWebsite}.

\subsection{Configuring the runtime environment}
\label{sec:setup_marley_script}

At runtime, the \texttt{marley} command-line executable relies on the system
environment variable \texttt{MARLEY} to store the path to the root folder of
the source code. If generation of events is attempted without this variable
being set, then the program will halt after printing the error message
\begin{lstlisting}[backgroundcolor=\color{white}, numbers=none,
backgroundcolor=\color{lightgray},
basicstyle=\small\ttfamily\bfseries, commentstyle=\small\ttfamily\bfseries]
[ERROR]: The MARLEY environment variable is not set. Please set it (e.g., by sourcing the setup_marley.sh script) and try again.
\end{lstlisting}
Although the user may manually set the value of the \texttt{MARLEY} variable,
use of the Bash shell script \texttt{setup\_marley.sh} to configure the system
environment is recommended. In addition to storing the path to the source code,
this script makes several other changes to environment variables for user
convenience, including adding the \texttt{build/} folder to the system search
paths for executables\footnote{The \texttt{PATH} environment variable} and
dynamic libraries.\footnote{Either the \texttt{LD\_LIBRARY\_PATH} (Linux) or
the \texttt{DYLD\_LIBRARY\_PATH} (macOS) environment variable} The
\texttt{setup\_marley.sh} script appears in the root source code folder and
should be executed (``sourced'') using the \texttt{source} command. From within
the \texttt{build/} folder, for example, one should source the script via
\begin{lstlisting}[backgroundcolor=\color{white}, numbers=none,
basicstyle=\normalsize\ttfamily, commentstyle=\normalsize\ttfamily]
source ../setup_marley.sh
\end{lstlisting}
Sourcing the setup script does not produce any console output. The instructions
given in the remainder of this section assume that the
\texttt{setup\_marley.sh} script has already been run in the current terminal
session.

\subsection{Running a simulation}
\label{sec:run_a_simulation}

The typical procedure for running a \marley\ simulation is to invoke the
executable via a command of the form\footnote{The \texttt{marley} executable
may also be invoked with the command-line option \texttt{--marley}.
This is best done using a terminal window set to display at least
80 columns and 53 rows in a small font.}
\begin{lstlisting}[backgroundcolor=\color{white}, numbers=none,
basicstyle=\normalsize\ttfamily, commentstyle=\normalsize\ttfamily]
marley CONFIG_FILE
\end{lstlisting}
where \texttt{CONFIG\_FILE} is the name (with any needed path specification) of
a job configuration file. \Cref{sec:gen_config} gives a full description of the
file format and configuration parameters. Three example job configuration files
are included with \marley\ in the \texttt{examples/config/} folder:
\begin{description}
\item[\texttt{annotated.js}] A heavily-commented example that provides
  documentation of the configuration file format similar to the contents
  of \cref{sec:gen_config}
\item[\texttt{COPY\_ME.js}] An example intended to serve as the basis for
  new job configuration files written by users
\item[\texttt{minimal.js}] An example of the simplest possible job
  configuration: default settings are used for all parameters except those
  that must be explicitly specified
\end{description}

As it executes, the \texttt{marley} program will print a variety of logging
messages to the screen. Several rows at the bottom of the screen are reserved
for a status display that tracks the progress of the simulation.
\Cref{lst:status_display} shows the format of the status display. Following the
first two rows, which report the current event count and the elapsed time since
the simulation began, zero or more rows record the total amount of data written
to each output file. The final row displays an estimate of the time at which
the simulation job will be completed.
\begin{lstlisting}[backgroundcolor=\color{white}, numbers=none,
backgroundcolor=\color{lightgray}, basicstyle=\small\ttfamily\bfseries,
caption={Example status display printed by the \texttt{marley} executable},
captionpos=b, label={lst:status_display}]
Event Count = 4001/10000 (40.0% complete, 300.4 events / s)
Elapsed time: 00:00:13 (Estimated total run time: 00:00:33)
Data written to events.ascii 3.80 MB
MARLEY is estimated to terminate on Mon Jun  1 23:54:52 2020 CDT
\end{lstlisting}

If it becomes necessary to end the simulation before all requested events have
been generated, the user may interrupt program execution by pressing ctrl+C. In
response, the \texttt{marley} executable will terminate gracefully after
writing the current event to any open output files. As discussed in
\cref{sec:executable_settings}, two of the available output file formats allow
for a simulation job interrupted in this way to be resumed from where it left
off.

\subsubsection{Troubleshooting rejection sampling problems}
\label{sec:troubleshoot_reject}

To select the energy $E_\nu$ of the reacting neutrino in each event, \marley\
employs the rejection sampling technique discussed in
\cref{sec:rejection_sampling} and the probability density function $P(E_\nu)$
given in \cref{sec:define_spectrum}. The validity of the technique depends on
obtaining an accurate estimate of the global maximum $P_\text{max}$ of
$P(E_\nu)$ within the sampling region of interest. Underestimates of
$P_\text{max}$ will lead to bias in the neutrino energy distribution of the
simulated events, while significant overestimates will adversely impact
sampling efficiency.

Although numerical estimation of $P_\text{max}$ in \marley\ is reasonably
robust for a variety of realistic neutrino spectra, it is not foolproof: there
exist pathological energy distributions (e.g., those with multiple sharp peaks)
for which automatic detection of the global maximum often fails. To protect
against this problem, \marley\ checks that each calculation of $P(E_\nu)$
performed during neutrino energy sampling yields a value that does not exceed
$P_\text{max}$. If a value larger than $P_\text{max}$ is ever encountered, a
set of warning and error messages similar to those in
\cref{lst:rejection_problem1} will be printed to the screen. A new estimate of
$P_\text{max}$, given by the problematic value of $P(E_\nu)$ increased by a
small ``safety factor,'' will be adopted in subsequent event generation.

\begin{lstlisting}[backgroundcolor=\color{white},
backgroundcolor=\color{lightgray}, basicstyle=\small\ttfamily\bfseries,
caption={Example warning and error messages printed in response to a rejection
sampling problem encountered when selecting reacting neutrino energies},
captionpos=b, label={lst:rejection_problem1}, escapechar=|]
[WARNING]: PDF value f(x) = 0.00133991 at x = 48.9882 exceeded the estimated maximum fmax = 0.001 during rejection sampling
[WARNING]: A new estimate fmax = 0.00135331 will now be adopted.
[ERROR]: Estimation of the maximum PDF value failed when using a rejection method to sample reacting neutrino energies.
This may occur when, e.g., an incident neutrino flux is used that includes multiple sharp peaks.
To avoid biasing the energy distribution, please rerun the simulation after adding the following line to the MARLEY job configuration file:
    energy_pdf_max: 0.00135331,|\label{line:epdfmax}|
If this error message persists after raising energy_pdf_max to a relatively high value, please contact the MARLEY developers for troubleshooting help.
\end{lstlisting}

While this adaptive approach to improving estimation of $P_\text{max}$ may
resolve rejection sampling problems in the later part of the simulation, it
cannot correct for bias in the energy distribution of the events that have
already been generated. A simple strategy for overcoming this difficulty is for
the user to restart the simulation from the beginning. By adding a line to the
job configuration file containing the \texttt{energy\_pdf\_max} key (see
\cref{sec:energy_pdf_max}) and the improved estimate of $P_\text{max}$
recommended by the rejection sampling error message (see
line~\ref{line:epdfmax} of \cref{lst:rejection_problem1}), automatic estimation
of $P_\text{max}$ will be disabled in favor of adopting the user-specified
value.

For severe underestimations of $P_\text{max}$, several repetitions of this
``interrupt and rerun'' strategy (each with a larger value of
\texttt{energy\_pdf\_max} than the one before) may be needed before correct
behavior is achieved. If rejection sampling errors persist across multiple fix
attempts, or if the corrected value of $P_\text{max}$ results in prohibitively
slow performance, a different approach to running the simulation (e.g.,
splitting a problematic spectrum into energy regions which are handled by
separate \marley\ jobs) should be pursued.

\subsection{The \texttt{marley-config} utility}
\label{sec:marley_config}

For a variety of applications, e.g., analyzing the standard \marley\ output
files (\cref{sec:example_analysis_scripts}) and interfacing \marley\ with
external codes (\cref{sec:gen_and_config}), it may be desirable to make use of
\marley\ classes within an external \cpp\ program. To facilitate compilation of
such programs, a Bash script named \texttt{marley-config} is placed in the
\texttt{build/} folder whenever \marley\ is successfully built.\footnote{The
\texttt{marley-config} script is loosely patterned after \texttt{pkg-config}
\cite{pkgconfigWebsite}. Similar scripts are used by \rootcern, GENIE, GSL, and
Geant4.} This script may be queried to obtain various pieces of information
about the build.

Each \texttt{marley-config} query is executed by providing the script with one
or more command-line options, each beginning with a prefix of two hyphens
(\texttt{--}). The command-line options may appear in any order and may be
repeated. If the \texttt{--help} option is present on the command line, then
all other arguments are ignored, and a multi-line usage message is printed. All
other recognized options produce console output that will be combined into a
single line and printed in the order that the corresponding options appeared on
the command line.

\begin{table}
\centering
\begin{tabular}{clc}
\toprule
Option             & Directory contents & Relative path \\
\midrule
\texttt{--bindir}  & executables & \texttt{build/} \\
\texttt{--datadir} & input data & \texttt{data/} \\
\texttt{--incdir}  & header files & \texttt{include} \\
\texttt{--libdir}  & libraries & \texttt{build/} \\
\texttt{--srcdir}  & source files & \texttt{src/} \\
\texttt{--topdir}  & top-level \marley\ directory & \texttt{./} \\
\bottomrule
\end{tabular}
\caption{Directory options recognized by the \texttt{marley-config} script}
\label{tab:dir_options}
\end{table}

Three categories of command-line options are recognized by the
\texttt{marley-config} script. The \textit{compilation options} are used to
retrieve compiler flags (compatible with both GCC and Clang) and other helpful
information for building external programs that link to the \marley\ shared
libraries (see \cref{sec:marley_config_compile}). Brief descriptions of the
four compilation options currently recognized by \texttt{marley-config} are
given below.

\begin{description}
\item[\texttt{--cflags}] Prints the \cpp\ compiler flags needed to build code
that includes \marley\ header files. If \marley\ has been built with \rootcern\
support, then the compiler flags will include \texttt{-DUSE\_ROOT}, which
defines the preprocessor macro \texttt{USE\_ROOT}. This macro may be used
together with the preprocessor directives \texttt{\#ifdef} or \texttt{\#ifndef}
to test for \rootcern-dependent \marley\ features in compiled code. An example
is given in \cref{sec:marley_config_compile}.

\item[\texttt{--cxx-std}]  Prints the version of the \cpp\ Standard used by the
compiler when \marley\ was built. The format used is the same as for parameters
passed via the \texttt{-std} flag to the compiler. For example, if \marley\ was
built using the \texttt{-std=c++14} flag, then \texttt{marley-config} will
print \texttt{c++14} for this option.

\item[\texttt{--libs}] Prints the compiler flags needed for linking to the
  \marley\ shared libraries

\item[\texttt{--use-root}] Prints the string \texttt{yes} if \marley\ was built
  with \rootcern\ support and \texttt{no} if it was not
\end{description}

The \textit{directory options} are used to print
the full paths to various subfolders of the \marley\ installation.
\Cref{tab:dir_options} lists the available directory options in the first
column. The second and third columns list, respectively, a description of the
directory's contents and its path relative to the root \marley\ folder.
For an installation of \marley\ in which the top-level directory
of the source code is \texttt{/home/\allowbreak marley}, executing
\begin{lstlisting}[backgroundcolor=\color{white}, numbers=none,
basicstyle=\normalsize\ttfamily, commentstyle=\normalsize\ttfamily]
marley-config --topdir --incdir --bindir
\end{lstlisting}
will produce the output
\begin{lstlisting}[backgroundcolor=\color{white}, numbers=none,
backgroundcolor=\color{lightgray}, basicstyle=\small\ttfamily\bfseries]
/home/marley /home/marley/include /home/marley/build
\end{lstlisting}

A final \texttt{marley-config} option category, the \textit{version options},
may be used to identify the installed version of \marley. Two options in this
category are currently recognized by the \texttt{marley-config} script:
\texttt{--git-revision} and \texttt{--version}.

The \texttt{--git-revision} option is used to print a unique hash value
reported by Git at build time. If \marley\ was built using a cloned Git
repository (see \cref{sec:get_code}), and the \texttt{git} executable is
present on the system path when GNU Make is invoked, then the hash identifier
for the latest commit will be retrieved\footnote{Via the command \texttt{git
rev-parse --short HEAD}} and saved in the \texttt{marley-config} script. If
uncommitted changes exist in the Git index or working tree when \marley\ is
built, then the string \texttt{-dirty} will be appended to the hash value. In
cases where a hash value is unavailable (e.g., because \marley\ was installed
without cloning a Git repository), the \texttt{marley-config} script will print
the string \texttt{unknown version} in response to this option.

If a tagged release of \marley\ was built,\footnote{This is determined by the
presence of a file named \texttt{.VERSION} in the top-level \marley\ folder at
build time.} then the \texttt{--version} option may be used to print the
corresponding version number (e.g., \texttt{1.2.0}). Otherwise, the output
produced by \texttt{marley-config} in response to \texttt{--version} is the
same as for \texttt{--git-revision}.

\subsubsection{Compiling programs using \texttt{marley-config}}
\label{sec:marley_config_compile}

To illustrate usage of the \texttt{marley-config} utility, three simple \cpp\
programs that employ \marley\ classes are provided in the folder
\texttt{examples/\allowbreak executables/\allowbreak minimal/}. Each one of
these may be compiled using GCC by executing a command of the form
\begin{lstlisting}[backgroundcolor=\color{white}, numbers=none,
basicstyle=\normalsize\ttfamily]
g++ -o EXECUTABLE_NAME $(marley-config --cflags --libs) SOURCE_NAME
\end{lstlisting}
from within \texttt{examples/\allowbreak executables/\allowbreak minimal/}.
Here \texttt{EXECUTABLE\_NAME} is the desired file name for the compiled
executable and \texttt{SOURCE\_NAME} is the name of one of the example \cpp\
source files. The \texttt{marley-config} script may also be used in a
\texttt{Makefile} as part of a more complicated build. Interested users should
consult the files \texttt{examples/\allowbreak executables/\allowbreak
build/\allowbreak Makefile} and \texttt{examples/\allowbreak marg4/\allowbreak
build/\allowbreak Makefile} for concrete examples.

\begin{lstlisting}[linewidth=\textwidth, caption={The \cpp\ source code for
\texttt{examples/\allowbreak executables/\allowbreak minimal/\allowbreak
mass\_40Ar.cc}, a trivial example program that uses the
\texttt{marley\allowbreak::\allowbreak Mass\allowbreak Table} class},
captionpos=b, label={lst:mass_table}, alsoletter=\#, emph={\#ifndef, \#endif},
emphstyle={\color{purple}}]
// Standard library includes
#include <iostream>

// MARLEY includes
#include "marley/MassTable.hh"

constexpr int Z = 18; // Ar|\label{line:Z}|
constexpr int A = 40;|\label{line:A}|

int main() {

  const marley::MassTable& mt = marley::MassTable::Instance();|\label{line:MassTable}|
  double mass_Ar40 = mt.get_atomic_mass( Z, A );|\label{line:atomic_mass}|

  std::cout << "MARLEY was";|\label{line:message_start}|
  #ifndef USE_ROOT
    std::cout << " not";|\label{line:message_not}|
  #endif
  std::cout << " built with ROOT support.\n";|\label{line:message_end}|

  std::cout << "The atomic mass of 40Ar is " << mass_Ar40 << " MeV/c^2\n";|\label{line:print_mass}|

  return 0;
}
\end{lstlisting}

\Cref{lst:mass_table} shows the source code for \texttt{mass\_40Ar.cc}, one of
the three example programs included in \texttt{examples/\allowbreak
executables/\allowbreak minimal/}. The other two, \texttt{efr.cc} and
\texttt{evgen.cc}, are discussed in \cref{sec:efr_class,sec:gen_and_config},
respectively. On lines~\ref{line:MassTable}--\ref{line:atomic_mass} of
\cref{lst:mass_table}, a constant reference to the singleton instance of the
\texttt{marley\allowbreak::\allowbreak MassTable} class is retrieved and used
to compute the atomic mass of \isotope[40]{Ar}.
Lines~\ref{line:message_start}--\ref{line:message_end} print a message to
standard output indicating whether \marley\ was built with \rootcern\ support.
The \texttt{USE\_ROOT} preprocessor macro mentioned in \cref{sec:marley_config}
is used to determine whether line~\ref{line:message_not} should be included in
the compiled program. A final message containing the atomic mass of
\isotope[40]{Ar} is printed to standard output on line~\ref{line:print_mass}.

\section{Generator configuration}
\label{sec:gen_config}

Before entering the event loop, the \marley\ executable reads in various
settings (see \cref{sec:generator_init}) from a job configuration file. The
format of the job configuration file is based on JavaScript Object Notation
(JSON) \cite{Bray2017}, which represents data structures as a nested
hierarchy of key-value pairs. A JSON \textit{object} is an unordered set of
key-value pairs surrounded by curly braces (\texttt{\{\}}). Each key is an
arbitrary string delimited by double quotes (\texttt{""}) and separated from
its corresponding value by a colon. Each value may be a JSON object itself, a
double-quoted string, a number, an array, or one of the words \texttt{true},
\texttt{false}, or \texttt{null} (without surrounding quotes). A JSON
\textit{array} is an ordered collection of values surrounded by square brackets
(\texttt{[]}). Elements of an array need not have the same data type, e.g., the
values \texttt{true} and \texttt{1.05} may be members of the same array. Array
elements are separated by commas, as are object key-value pairs. A valid JSON
document is itself an object and is thus delimited by a pair of curly braces.

For the convenience of users, the job configuration file format used by
\marley\ permits three minor deviations from the JSON standard:
\begin{itemize}
\item Single-word keys (no whitespace) may be given without surrounding double quotes.
\item \cpp-style comments (\verb|//| and \verb|/* */|) are allowed anywhere in the file.
\item A trailing comma is allowed after the final element in objects and arrays.
\end{itemize}
A filename extension of \texttt{.js} is recommended for \marley\ job
configuration files because the default JavaScript syntax highlighting used by
many text editors is well-suited to the JSON-like format.\footnote{Some email
services do not allow the attachment of files with the \texttt{.js} extension.
To enable sharing of a MARLEY job configuration file via email, it is usually
sufficient to change the filename extension to \texttt{.txt} before attaching
the file.}

To parse job configuration files, \marley\ relies on a modified version of the
SimpleJSON library \cite{SimpleJSON}, which was incorporated into the code base
as the class \texttt{marley::JSON}. An example \marley\ job configuration file
is shown in \cref{lst:config1}. Explanations of each parameter needed to fully
configure the generator are given in the following subsections.

\begin{lstlisting}[linewidth=\textwidth,caption={Example \marley\ job
configuration file},captionpos=b,label={lst:config1}]
// The full configuration is enclosed in a set of curly braces
{
  // Random number seed
  seed: 123456,

  // Incident neutrino direction
  direction: { x: 0.0, y: 0.0, z: 1.0 },|\label{line:direction}|

  // Target specification
  target: {|\label{line:start_target}|
    nuclides: [ 1000180400 ],
    atom_fractions: [ 1.0 ],
  },|\label{line:end_target}|

  // Reaction input files
  reactions: [ "ve40ArCC_Bhattacharya2009.react", "ES.react" ],|\label{line:reactions}|

  // Neutrino source specification
  source: {|\label{line:source_start}|
    neutrino: "ve", // The source produces electron neutrinos
    type: "monoenergetic",
    energy: 15.0,   // MeV
  },|\label{line:source_end}|

  // Settings for MARLEY command-line executable
  executable_settings: {|\label{line:exe_start}|
    // The number of events to generate
    events: 10000,

    // Event output configuration
    output: [ { file: "events.ascii", format: "ascii", mode: "overwrite" } ],
  },|\label{line:exe_end}|
}
\end{lstlisting}

\subsection{Random number seed}
\label{sec:rng_seed_config}

The optional \texttt{seed} key may be used to specify an integer seed for the
random number generator. Any value corresponding to a \cpp\ \texttt{long int}
(guaranteed by the standard to be at least 32 bits long) may be used as a seed.
This includes negative integers, although users should note that these will be
reinterpreted as unsigned integers by \marley. If the \texttt{seed} key is
omitted from the job configuration file, then the current Unix time will be
used as the random number seed.

\subsection{Neutrino direction}
\label{sec:neutrino_dir_config}

For simplicity, \marley\ initially generates all events in a frame where the
incident neutrino is traveling along the positive $z$ direction (see
\cref{sec:rotation}). However, it is often useful to simulate events for
neutrinos traveling in an arbitrary direction in the lab frame. This
functionality is enabled via use of the \texttt{direction} key to define a
lab-frame direction 3-vector for the incident neutrinos. The Cartesian
components of this vector are specified as shown on line \ref{line:direction}
of \cref{lst:config1}. They need not have any particular normalization, but at
least one component must be nonzero. If the \texttt{direction} key is omitted,
then the positive $z$ direction is assumed.

In certain special cases, e.g., studies of the Diffuse Supernova Neutrino
Background (DSNB) \cite{DSNB}, it may be desirable to generate \marley\ events
with an isotropic distribution of initial neutrino directions. This behavior
may be enabled by providing the string \texttt{"isotropic"}
as the value for the \texttt{direction} key instead of a direction 3-vector.

\subsection{Target composition}
\label{sec:target_config}

The nuclidic composition of the neutrino target material in a \marley\
simulation may be specified by defining a JSON object in the job configuration
file labeled with the \texttt{target} key. The JSON object must define two
arrays of equal size. The first of these, which is labeled with the
\texttt{nuclides} key, includes one or more nuclear PDG codes (see
\cref{sec:PDG_codes}) with one code per species. The second array,
which is labeled with the \texttt{atom\_fractions} key, contains the
corresponding nuclide fractions in the target material. After being
automatically normalized to sum to unity, these will be used as the abundance
weights $f_r$ that appear in \cref{eq:abundance_weighted_xsec,eq:reaction_pmf}.
If any of the elements of the \texttt{atom\_fractions} array is negative,
\marley\ will halt and print an error message.

Lines \ref{line:start_target}--\ref{line:end_target} of \cref{lst:config1}
provide an example configuration for a neutrino target composed of pure
\isotope[40]{Ar}. If the \texttt{target} key is omitted from the job
configuration file, then the default behavior mentioned in
\cref{sec:target_definition} will be used: every distinct nuclide that appears
in the initial state of at least one configured reaction (see
\cref{sec:reaction_config}) will be included in the neutrino target with equal
abundance.

\subsection{Reactions}
\label{sec:reaction_config}

To define the set of neutrino scattering processes that should be considered in
a \marley\ simulation, the user must specify a list of reaction input files as
a JSON string array labeled by the \texttt{reactions} key. This key is required
to be present in all \marley\ job configuration files. For neutrino-nucleus
reactions, each reaction input file provides the values of the allowed nuclear
matrix elements $B(\mathrm{F})$ and $B(\mathrm{GT})$ needed to compute the
differential cross section in \cref{eq:diff_xsec} for a particular target
nucleus and interaction mode. For neutrino-electron elastic scattering, the
reaction input file provides a list of nuclides for which this process should
be enabled. Details of the file format are discussed in
\cref{sec:reaction_file_format}.

In \marley\ \version, the code is capable of simulating reactions belonging to
four distinct interaction modes: charged-current neutrino-nucleus scattering
($\nu$ CC), charged-current antineutrino-nucleus scattering ($\bar{\nu}$ CC),
neutral-current (anti)neutrino-nucleus scattering (NC), and
(anti)neutrino-electron elastic scattering (ES). As shown in
\cref{eq:F_and_GT_operators}, the nuclear scattering modes may be distinguished
by the isospin operator that appears in the nuclear matrix elements (e.g.,
$t_{-}$ for $\nu$ CC).

\subsubsection{Example reaction input files}
\label{sec:example_reaction_input_files}

Three example reaction input files for $\nu$ CC on \isotope[40]{Ar}, intended
for use in simulations of the process
$\isotope[40]{Ar}(\nu_e,e^{-})\isotope[40]{K}^*$, are included in the
\texttt{data/react} folder as part of the standard \marley\ distribution. At
high excitation energies, all three files include the same set of theoretical
matrix elements taken from a QRPA\footnote{Quasiparticle Random Phase
Approximation} calculation by Cheoun, Ha, and Kajino \cite{Cheoun2012a}. At low
excitation energies, the files contain different sets of nuclear matrix
elements extracted from experimental data. The references used to obtain the
data-driven matrix elements for each file are as follows:
\begin{description}
\item[\texttt{ve40ArCC\_Bhattacharya1998.react}] Measurement of
\isotope[40]{Ti} $\beta^{+}$ decay reported in
ref.~\cite{Bhattacharya1998}
\item[\texttt{ve40ArCC\_Liu1998.react}] Independent \isotope[40]{Ti} $\beta^{+}$
decay measurement reported in ref.~\cite{Liu1998}
\item[\texttt{ve40ArCC\_Bhattacharya2009.react}] Forward-angle
(p,n) scattering data reported in ref.~\cite{Bhattacharya2009}.
\end{description}
A description of the data evaluation procedure used to prepare these files is
available in ref.~\cite[sec.~III]{marleyPRC}.

Two other reaction input files are included in the \texttt{data/react}
folder. The file \texttt{ES.react} enables simulations of neutrino-electron
scattering on an atomic \isotope[40]{Ar} target. Simulation of this process
for other nuclides may be enabled in \texttt{ES.react} by appending additional
nuclear PDG codes.

The file \texttt{CEvNS40Ar.react} provides a simple example of an NC
configuration. Although \marley\ is capable of simulating inelastic NC
reactions if provided with suitable input, \texttt{CEvNS40Ar.react} includes
only one $B(\mathrm{F})$ matrix element appropriate for simulating coherent
elastic neutrino nucleus scattering on \isotope[40]{Ar} (see \cref{sec:CEvNS}).
Altering the nuclear PDG code given in this file (see
\cref{sec:reaction_file_format}) will immediately enable this process to be
simulated for any spin-zero target nucleus.

\subsubsection{Configuration example}

Line \ref{line:reactions} of \cref{lst:config1} provides an example
\texttt{reactions} setting that enables $\nu$ CC and ES reactions on
\isotope[40]{Ar} to be simulated simultaneously. As shown in the example, a
full path specification is not needed for files that appear in the
\texttt{\$\{MARLEY\}/data/react/} folder, where \texttt{\$\{MARLEY\}} is the
value of the \marley\ environment variable at runtime (see
\cref{sec:setup_marley_script}). If multiple files representing the same
reaction mode and the same target nuclide are listed for the \texttt{reactions}
key, a warning message will be printed and only the configuration from the
first such file will be used in the simulation.

Every element of the \texttt{reactions} array must be a simple string literal
containing a single file name. The use of wildcard characters, regular
expressions, and environment variables is not currently allowed by the
\marley\ job configuration file format.

\subsection{Neutrino source}
\label{sec:nu_source_config}

The required \texttt{source} key allows the user to provide a description of
the incident neutrino energy spectrum using a JSON object. This object must
always contain at least two keys: \texttt{neutrino} and \texttt{type}.

The \texttt{neutrino} key represents the neutrino species to be used as the
projectile. This may be specified using either an integer PDG code or a string,
as shown in \cref{tab:source_neutrino}.

\begin{table}
\centering
\begin{tabular}{cSc}
\toprule
Neutrino & {PDG code} & string \\
\midrule
$\nu_e$ & 12 & \verb|"ve"| \\
$\bar{\nu}_e$ & -12 & \verb|"vebar"| \\
$\nu_\mu$ & 14 & \verb|"vu"| \\
$\bar{\nu}_\mu$ & -14 & \verb|"vubar"| \\
$\nu_\tau$ & 16 & \verb|"vt"| \\
$\bar{\nu}_\tau$ & -16 & \verb|"vtbar"| \\
\bottomrule
\end{tabular}
\caption{Allowed values for the \texttt{neutrino} key in the neutrino source
specification}
\label{tab:source_neutrino}
\end{table}

The \texttt{type} key may be used to request one of several built-in neutrino
spectra or to indicate to \marley\ that a user-defined spectrum should be used.
In the latter case, users should typically not apply any cross section
weighting themselves to the neutrino spectrum, as \marley\ will weight the
incident spectrum using the appropriate reaction cross section at runtime (see
\cref{sec:define_spectrum}). In unusual cases where this behavior is not
desirable, the user may disable cross section weighting by following the
procedure described in \cref{sec:weight_flux}.

The following paragraphs describe each of the allowed options
for the \texttt{type} key, together with any additional parameters
needed for each source type.

\subsubsection{Monoenergetic}
A neutrino source whose \texttt{type} is \texttt{mono} or \texttt{monoenergetic}
always produces a neutrino with a fixed energy $E_\nu$. This energy is specified
in \si{\MeV} using the \texttt{energy} key.
Lines \ref{line:source_start}--\ref{line:source_end} of \cref{lst:config1}
provide an example configuration for a monoenergetic $\nu_e$ source with $E_\nu
= \SI{15}{\MeV}.$

\subsubsection{Fermi-Dirac distribution}
The \texttt{fermi-dirac} or \texttt{fd} source type
emits neutrinos whose energies are drawn from a Fermi-Dirac
distribution with temperature $T$ (\si{\MeV}) and pinching parameter $\eta$.
The incident neutrino spectrum is given by
\begin{align} \phi(E_\nu) &=
\frac{C\,E_\nu^2}
{T^4\left[1+\exp\left(\frac{E_\nu}{T} - \eta\right)\right]}
& E_\nu^\text{min} \leq E_\nu \leq E_\nu^\text{max}
\end{align}
where $C$ is a normalization constant chosen so that
\begin{equation}
\label{eq:C_norm}
\int_{E_\nu^\text{min}}^{E_\nu^\text{max}} \! \phi(E_\nu) \, dE_\nu = 1.
\end{equation}
\vspace{-1\baselineskip}
\begin{lstlisting}[linewidth=\textwidth,caption={Example Fermi-Dirac source
specification},captionpos=b,label={lst:fd_source}]
source: {
  type: "fermi-dirac",
  neutrino: "ve",
  Emin: 0,           // Minimum neutrino energy (MeV)
  Emax: 60,          // Maximum neutrino energy (MeV)
  temperature: 3.5,  // Temperature (MeV)
  eta: 4             // Pinching parameter (default 0)
},
\end{lstlisting}
The values of the parameters $E_\nu^\text{min}$, $E_\nu^\text{max}$, $T$, and
$\eta$ are specified using the keys \texttt{Emin}, \texttt{Emax},
\texttt{temperature}, and \texttt{eta}, respectively. Omitting the \texttt{eta}
key will yield the default value $\eta = 0$. The other three parameters
must be given explicitly. \Cref{lst:fd_source} shows an example Fermi-Dirac
source specification in which the parameter values $E_\nu^\text{min} =
\SI{0}{\MeV}$, $E_\nu^\text{max} = \SI{60}{\MeV}$,
$T = \SI{3.5}{\MeV}$, and $\eta = 4$ have been chosen.

The form for the pinched Fermi-Dirac distribution used by \marley\ is
based on that of the ``Livermore model'' of the supernova neutrino energy
spectrum \cite{Totani1998}.

\subsubsection{``Beta fit'' spectrum}
The \texttt{beta-fit} or \texttt{bf} source type produces neutrinos
with energies drawn from the spectrum
\begin{align} \phi(E_\nu) &=
\label{eq:beta_fit_dist}
C\left(E_\nu \big/ \big<E_\nu\big>\right)^{\beta - 1}
\exp\left(-\beta\,E_\nu \big/ \big<E_\nu\big>\right)
& E_\nu^\text{min} \leq E_\nu \leq E_\nu^\text{max}
\end{align}
\vspace{-1.25\baselineskip}
\begin{lstlisting}[linewidth=\textwidth,caption={Example ``beta fit'' source
specification},captionpos=b,label={lst:beta_fit_source}]
source: {
  type: "beta-fit",
  neutrino: "ve",
  Emin: 0,   // Minimum neutrino energy (MeV)
  Emax: 60,  // Maximum neutrino energy (MeV)
  Emean: 15, // Mean neutrino energy (MeV)
  beta: 3.0, // Pinching parameter (default 4.5)
},
\end{lstlisting}
with mean neutrino energy
$\left<E_\nu\right>$ (\si{\MeV}) and dimensionless pinching parameter $\beta \geq 0$.
The normalization constant $C$ is chosen to satisfy \cref{eq:C_norm}.

The values of the parameters $E_\nu^\text{min}$, $E_\nu^\text{max}$,
$\left<E_\nu\right>$, and $\beta$ are specified using the keys \texttt{Emin},
\texttt{Emax}, \texttt{Emean}, and \texttt{beta}, respectively. Omitting the
\texttt{beta} key will yield the default value $\beta = 4.5$.
The other three parameters must be given explicitly.
Negative $\beta$ values are unphysical and will cause \marley\ to halt and
issue an error message. \Cref{lst:beta_fit_source} shows an example ``beta
fit'' source specification in which the parameter values $E_\nu^\text{min} =
\SI{0}{\MeV}$, $E_\nu^\text{max} = \SI{60}{\MeV}$,
$\left<E_\nu\right> = \SI{15}{\MeV}$, and $\beta = 3$ have been chosen.

The ``beta fit'' distribution has been employed in a recent theoretical study
\cite{Huang2016} of supernova neutrino detection and is equivalent to the
widely-used ``Garching model'' parameterization of supernova
neutrino spectra \cite{Keil2003}. The sole difference between the two
parameterizations is in the convention for the pinching parameter.
The Garching model employs a parameter $\alpha$ and makes
the substitution $\beta \rightarrow \alpha + 1$ in
\cref{eq:beta_fit_dist}. For $\alpha = \beta - 1$, the distributions
used by the two approaches are identical.

\subsubsection{Muon decay-at-rest}
\label{sec:muon_decay_at_rest_source}

The \texttt{decay-at-rest} or \texttt{dar} source type emits neutrinos
according to a Michel spectrum for a muon decaying at rest in the lab frame.
For a $\nu_e$ source (corresponding to the decay
$\mu^{+} \rightarrow e^{+} + \nu_e + \bar{\nu}_\mu$),
the energy spectrum is given by \cite{Kosmas2017}
\begin{align} \phi(E_\nu) &= 96 \, E_\nu^2 \, m_\mu^{-4} \, (m_\mu - 2E_\nu)
& 0 < E_\nu < m_\mu/2.
\end{align}
where $m_\mu$ is the muon mass. A $\bar{\nu}_\mu$ source has the
spectrum
\begin{align}
\phi(E_\nu) &= 16 \, E_\nu^2 \, m_\mu^{-4} \, (3 m_\mu - 4E_\nu)
& 0 < E_\nu < m_\mu/2.
\end{align}
\vspace{-1\baselineskip}
\begin{lstlisting}[linewidth=\textwidth,caption={Example muon decay-at-rest
source specification},captionpos=b,label={lst:dar_source}]
source: {
  type: "dar",
  neutrino: "ve",
},
\end{lstlisting}
Production of $\bar{\nu}_e$ and $\nu_\mu$ (from $\mu^{-} \rightarrow e^{-} +
{\bar{\nu}}_e + \nu_\mu$) may also be simulated using a \texttt{dar} neutrino
source. Because the decay kinematics fully determine the shape of the Michel
spectrum, no additional input parameters are needed for this source type.
\Cref{lst:dar_source} shows an example configuration for a muon decay-at-rest
$\nu_e$ source.

Neutrinos from muon decay-at-rest produced at facilities like the Oak Ridge
Spallation Neutron Source (SNS) \cite{Bolozdynya2012} and the Fermilab Neutrinos
at the Main Injector (NuMI) beamline \cite{Grant2015} offer a valuable
opportunity to study neutrino-nucleus scattering at tens-of-\si{\MeV} energies
with a terrestrial source.

\subsubsection{Histogram}
The \texttt{histogram} or \texttt{hist} source type allows the user to define
the incident neutrino spectrum as a histogram with one or more energy bins.
\begin{lstlisting}[linewidth=\textwidth,caption={Example histogram source
specification},captionpos=b,label={lst:histogram_source}]
source: {
  type: "histogram",
  neutrino: "ve",
  E_bin_lefts: [ 7., 8., 9. ], // Low edges of energy bins (MeV)
  weights: [ 0.2, 0.5, 0.3 ],  // Bin weights (dimensionless)
  Emax: 10.,                   // Upper edge of the final bin (MeV)
},
\end{lstlisting}

The $n$ bins are specified by their lower edges $E_\nu(k)$ and weights $W(k)$
for $k \in \{1,2,\dots,n\}$. Within bin $k$, incident neutrino energies are
distributed uniformly on the half-closed interval $[E_\nu(k), E_\nu(k+1))$.
Weights that do not sum to unity will be automatically renormalized by the
code. The keys \texttt{E\_bin\_lefts} and \texttt{weights} are used to provide
JSON arrays of $E_\nu(k)$ and $W(k)$ values, respectively. The sizes of these
arrays must be equal and are used by \marley\ to determine the bin count $n$.
The energy bin edges $E_\nu(k)$ must be given in increasing order. The upper
edge of the final energy bin $E_\nu(n+1)$ must also be specified using the key
\texttt{Emax}.

\Cref{lst:histogram_source} shows an example histogram
source specification in which three equally-spaced
bins are defined between $7$ and \SI{10}{\MeV}.

\subsubsection{Grid}
The \texttt{grid} source type emits neutrinos with energies drawn from a
tabulated probability density function $\phi(E_\nu)$. To define this function,
a grid of at least two monotonically increasing $E_\nu$ values is specified as
a JSON array via the \texttt{energies} key. The corresponding $\phi(E_\nu)$
values (which need not be normalized by the user to integrate to unity) must
also be given in an array of the same size using the \texttt{prob\_densities}
key.
\begin{lstlisting}[linewidth=\textwidth,caption={Example grid source
specification},captionpos=b,label={lst:grid_source}]
source: {
  type: "grid",
  neutrino: "ve",
  energies: [ 10., 15., 20. ],     // Energy grid points (MeV)
  prob_densities: [ 0., 1., 0. ],  // Probability densities
  rule: "lin", // Interpolation rule ("lin" default)
},
\end{lstlisting}

\begin{table}
\begin{center}
\begin{tabular}{ll}
\toprule
Rule &  Interpretation \\
\midrule
\texttt{"const"}   & histogram-like \\
\texttt{"lin"}     & linear in both $E_\nu$ and $\phi(E_\nu)$ \\
\texttt{"linlog"}  & linear in $E_\nu$, logarithmic in $\phi(E_\nu)$ \\
\texttt{"loglin"}  & logarithmic in $E_\nu$, linear in $\phi(E_\nu)$ \\
\texttt{"log"}     & logarithmic in both $E_\nu$ and $\phi(E_\nu)$ \\
\bottomrule
\end{tabular}
\caption{Allowed values of the \texttt{rule} key for the \texttt{grid} neutrino
source type.}
\label{tab:interp_rules}
\end{center}
\end{table}

To evaluate $\phi(E_\nu)$ at energies between the grid points,
\marley\ employs an interpolation rule specified by the \texttt{rule}
key. The default \texttt{"lin"} rule (linear in both $E_\nu$ and $\phi$)
is typically most useful, but \marley\ is also capable of interpolating
logarithmically along one or both axes. A \texttt{"const"} interpolation
rule is also available which treats $\phi(E_\nu)$ as a step function at each
energy grid point. \Cref{tab:interp_rules} lists all allowed values of
the \texttt{rule} key together with their interpretations.

\Cref{lst:grid_source} shows an example grid
source specification which defines a symmetric triangular
distribution between $10$ and \SI{20}{\MeV}.

\subsubsection{TH1 and TGraph}
\label{sec:th1_and_tgraph}

If \marley's optional interface to the \rootcern\ \cite{Brun1997} data analysis
framework is enabled (see \cref{sec:install}), then two additional neutrino
source types, \texttt{th1} and \texttt{tgraph}, are allowed.
\begin{lstlisting}[linewidth=\textwidth,caption={Example th1 source
specification},captionpos=b,label={lst:th1_source}]
source: {
  type: "th1",
  neutrino: "ve",
  tfile: "my_root_file.root",  // Name of the ROOT file
                               // containing the TH1 object

  namecycle: "MyHist",         // Name under which the TH1 object appears
                               // in the file (used to retrieve it)
},
\end{lstlisting}
The \texttt{th1} type takes a one-dimensional \rootcern\ histogram object (of
\cpp\ type \texttt{TH1}) as input and converts it into a \texttt{histogram}
neutrino source using \marley's native representation of the
distribution.\footnote{Both histogram and grid neutrino sources are implemented
using the \texttt{marley::InterpolationGrid} class.} The \texttt{tgraph} source
type is similar, except that it converts a \rootcern\ \texttt{TGraph} object
into a \texttt{grid} neutrino source.

To use either of the \texttt{th1} or \texttt{tgraph} source types, the user must
provide the name of an input \rootcern\ file (including any needed path
specification) using the \texttt{tfile} key. The string value given for the
\texttt{namecycle} key, which must also be provided, will be used by \marley\ in
a call to \texttt{TDirectoryFile::Get(const char*\allowbreak\ namecycle)} in
order to retrieve the object of interest from the input \rootcern\ file. Units
of \si{\MeV} should always be used for neutrino energies when preparing a
\texttt{TH1} or \texttt{TGraph} object for use with \marley.

\Cref{lst:th1_source} shows an example th1 source specification which uses a
neutrino energy histogram named \texttt{"MyHist"} stored in a \rootcern\ file in
the working directory called \texttt{my\_root\_file.root}.

\subsection{Executable settings}
\label{sec:executable_settings}

The optional \texttt{executable\_settings} key is used to provide a JSON
object that controls the \texttt{marley} command-line executable. However, the
entirety of the executable settings will be ignored if the configuration file
is used to initialize \marley\ in a different context (e.g., from within a
Geant4 application that links to the \marley\ shared libraries, see
\cref{sec:external_interfaces}).

Within the executable settings object, the \texttt{events} key is used to
specify the number of events that the \texttt{marley} executable should generate
before terminating. Because the \marley\ JSON parser expects the associated
value to be an integer literal, using scientific notation to specify the number
of events (e.g., \texttt{1e4} instead of \texttt{10000}) is not currently
allowed. If the \texttt{events} key is omitted, a default value of 1000 will
be assumed.

The \texttt{output} key in the executable settings object is used to
describe zero or more output files that will receive the generated events.
The corresponding value should be a JSON array containing one JSON object
per output file. Each object in the array should define the following keys:
\begin{description}
\item[file] The name of the output file (with any needed path specification).
\item[format] The format to use when storing events in the output file.
Valid values for this key are \texttt{"ascii"}, \texttt{"hepevt"},
\texttt{"json"}, and (if \marley's \rootcern\ interface is active)
\texttt{"root"}. Descriptions of each of the allowed output file formats
are given in \cref{sec:file_formats}.
\item[mode] The approach that the \texttt{marley} executable should use if the
output file is not initially empty. For the \ascii\ and \hepevt\ formats, valid
values are \texttt{"overwrite"} (erase any previously existing file contents)
and \texttt{"append"} (continue output immediately after any existing file
contents). For the JSON and \rootcern\ formats, valid values are
\texttt{"overwrite"} and \texttt{"resume"}. If the \texttt{"resume"} mode is
chosen, the generator will restore its previous state from an incomplete run
(e.g., a run that was interrupted by the user by pressing ctrl+C) that was
saved to the output file and continue from where it left off. If the
\texttt{"resume"} mode is selected for more than one output file, then
\marley\ will halt and print an error message.
\end{description}
Two additional keys may be used in specific contexts:
\begin{description}
\item[force] Boolean value used only for the \texttt{"overwrite"} mode. If it is
\texttt{true}, then the marley executable will not prompt the user before
overwriting existing data. If this key is omitted, a value of \texttt{false} is
assumed.
\item[indent] Nonnegative integer value used only for the JSON output format.
It represents the number of spaces that should be used as a tab stop when
pretty-printing the JSON output. If this key is omitted, all unnecessary
whitespace will be suppressed. This default behavior results in the most
compact JSON-format output files.
\end{description}
If the \texttt{output} key in the \texttt{executable\_settings} object is
omitted entirely, then the default configuration
\begin{center}
\begin{tabular}{c}
\begin{lstlisting}[backgroundcolor=\color{white}, numbers=none,
  linewidth=\paperwidth, basicstyle=\small\ttfamily]
output: [ { file: "events.ascii", format: "ascii", mode: "overwrite" } ]
\end{lstlisting}
\end{tabular}
\end{center}
will be used.

Lines \ref{line:exe_start}--\ref{line:exe_end} of \cref{lst:config1} define an
example \texttt{executable\_settings} object that includes the default settings
described in this section.

\subsection{Less commonly-used parameters}

Although the job configuration file shown in \cref{lst:config1} provides usage
examples for all of the key-value pairs typically needed to define a \marley\
simulation, a number of additional parameters are recognized by the
configuration parser.\footnote{That is, the
\texttt{marley\allowbreak::\allowbreak JSON\allowbreak Config} class} These
parameters, which are intended for advanced users or for addressing unusual
situations, are documented in the remainder of this section.

\subsubsection{Logging}

The singleton \texttt{marley\allowbreak::\allowbreak Logger} class provides
rudimentary support for writing diagnostic messages to zero or more output
streams including standard output, standard error, and plaintext log files.
This is done in a prioritized way by configuring a \textit{logging level} for
each output stream. Messages are likewise categorized by a logging level. In
order of severity, the logging levels recognized by
\texttt{marley\allowbreak::\allowbreak Logger} are \texttt{debug},
\texttt{info}, \texttt{warning}, \texttt{error}, and \texttt{disabled}, with
the last available for output streams but not for messages. When a message with
logging level \texttt{MSG\_LEVEL} is passed by \marley\ to the \texttt{Logger},
the message will be ignored by all streams whose logging levels are more severe
than \texttt{MSG\_LEVEL}. The message will be written to all other configured
streams.

Nearly all output written to the terminal by the \texttt{marley} executable is
handled by the \texttt{Logger} class. The main exception is the status display
discussed in \cref{sec:run_a_simulation} (see \cref{lst:status_display}).

The default behavior of the \texttt{Logger} class may be adjusted via the
optional \texttt{log} key in the job configuration file. This key is used to
specify an array of JSON objects with one element per output stream. Each
object in the array may define the following keys:
\begin{description}
\item[file] A string giving the name of a text file that will receive the
  output logging messages. If the string value is \texttt{"stdout"}
  (\texttt{"stderr"}), then the messages will be written to standard
  output (standard error) rather than a text file.
\item[level] A string specifying the logging level that should be used as a
  threshold for writing messages to the stream. Valid values are
  \texttt{"debug"}, \texttt{"info"}, \texttt{"warning"}, \texttt{"error"}, and
  \texttt{"disabled"}.
\item[overwrite] A boolean value indicating whether new messages should be
  appended to the end of the output file (\texttt{false}) or whether any
  previously existing file contents should be erased (\texttt{true}). This key
  is ignored when writing to standard output and standard error. A default
  value of \texttt{false} is assumed when this key is not present.
\end{description}
If the \texttt{log} key is omitted from the job configuration file, then
\marley\ will use default settings equivalent to
\begin{center}
\begin{tabular}{c}
\begin{lstlisting}[backgroundcolor=\color{white}, numbers=none,
  linewidth=\paperwidth, basicstyle=\small\ttfamily]
log: [ { file: "stdout", level: "info" } ]
\end{lstlisting}
\end{tabular}
\end{center}

\subsubsection{Disabling simulation of nuclear de-excitations}
\label{sec:disable_deex}

As described in \cref{sec:theory_summary}, \marley\ models neutrino-nucleus
scattering as proceeding via a prompt $2 \to 2$ reaction and a subsequent
sequence of zero or more binary decays of the remnant nucleus. For certain
calculations,\footnote{For example, those in which only the final-state lepton
is of interest} it may be convenient to simulate events in which only the
primary $2 \to 2$ scattering process is considered. This behavior may be
enabled in the job configuration file by using the optional
\texttt{do\_deexcitations} key together with a boolean value. If the
accompanying value is \texttt{true}, then the \marley\ event loop will proceed
normally, and nuclear de-excitations will be simulated. If it is
\texttt{false}, then the de-excitation loop will be skipped.

\subsubsection{Disabling cross-section weighting of the neutrino spectrum}
\label{sec:weight_flux}

By default, \marley\ samples the reacting neutrino energy $E_\nu$ for each
event from a probability density function $P(E_\nu) \propto \sigma(E_\nu) \,
\phi(E_\nu)$, where $\sigma(E_\nu)$ is the abundance-weighted total cross
section and $\phi(E_\nu)$ is the incident neutrino spectrum (see
\cref{sec:define_spectrum}). For some use cases, it may be desirable to sample
$E_\nu$ directly from $\phi(E_\nu)$ without any cross-section weighting. This
may be achieved by adding the optional \texttt{weight\_flux} key, which takes a
boolean value, to the contents of the \texttt{source} JSON object
(\cref{sec:nu_source_config}). If the \texttt{weight\_flux} value is set to
\texttt{true}, then $E_\nu$ will be sampled from the usual probability density
function $P(E_\nu)$. If the \texttt{weight\_flux} value is set to
\texttt{false}, then $E_\nu$ will be drawn directly from $\phi(E_\nu)$.

\subsubsection{Orbital angular momentum and multipolarity cutoffs}
\label{sec:l_max}

When computing differential decay widths for nuclear fragment ($\gamma$-ray)
emission from a compound nucleus, \marley\ truncates the infinite sum over
orbital angular momenta (multipolarities) that appears in
\cref{eq:fragment_diff_decay_width} (\cref{eq:gamma_diff_decay_width}) by
imposing a cutoff value $\ell_\text{max}$ ($\lambda_\text{max}$). The user may
modify the default value ($\ell_\text{max} = \lambda_\text{max} = 5$) by
including two keys in the job configuration file. The \texttt{fragment\_lmax}
key is used to specify a nonnegative integer which will be used as the maximum
orbital angular momentum $\ell_\text{max}$ to consider in nuclear fragment
emission. Similarly, the \texttt{gamma\_lmax} key is used to specify a positive
integer which will serve as the maximum multipolarity $\lambda_\text{max}$
considered in $\gamma$-ray emission.

\subsubsection{Coulomb correction factor}
\label{sec:coulomb_mode}

As described in \cref{sec:coulomb_corrections}, \marley\ accounts for
final-state Coulomb effects in charged-current neutrino-nucleus interactions by
the inclusion of a Coulomb correction factor $F_C$ in the expression for the
differential cross section. By default, this factor is computed according to
\cref{eq:FC}, which uses the smaller of the two corrections given by the Fermi
function (\cref{eq:Lorentz_invariant_Fermi_function}) and the modified
effective momentum approximation (MEMA, \cref{eq:FMEMA}). The \marley\
prescription for the Coulomb correction factor may be controlled in the
job configuration file by use of the \texttt{coulomb\_mode} key. The
following string values are allowed for this key:

\begin{description}
\item[\texttt{"none"}] No Coulomb corrections will be applied
($F_C = 1$ in all cases).
\item[\texttt{"Fermi"}] The Fermi function defined in
\cref{eq:Lorentz_invariant_Fermi_function} will always be used to compute
the Coulomb correction factor ($F_C = F_\text{Fermi}$).

\item[\texttt{"EMA"}] The effective momentum approximation (EMA) will be used
for Coulomb corrections ($F_C = F_\text{EMA}$, see \cref{eq:EMA}).

\item[\texttt{"MEMA"}] The modified effective momentum approximation
(MEMA) will be used for Coulomb corrections
($F_C = F_\text{MEMA}$, see \cref{eq:FMEMA}).

\item[\texttt{"Fermi-EMA"}] The Fermi function will be compared to the EMA, and
the approach that yields the smaller correction (i.e., the $F_C$ value closer
to unity) will be used. The Coulomb correction factor $F_C$ is calculated as in
\cref{eq:FC} with the substitution $F_\text{MEMA} \to F_\text{EMA}$.

\item[\texttt{"Fermi-MEMA"}] Similar to \texttt{"Fermi-EMA"}, except that the
MEMA is used instead of the EMA. The Coulomb correction factor $F_C$ is
calculated as in \cref{eq:FC}. This is \marley's default approach.
\end{description}

\subsubsection{Status update interval}

As discussed in \cref{sec:run_a_simulation}, the \texttt{marley} executable
prints a text-based status display at the bottom of the screen. By default,
this display is updated after each set of $n_\text{update} = 100$ events has
been generated. The user may modify the value of $n_\text{update}$ by adding
the \texttt{status\_update\_interval} key to the \texttt{executable\_settings}
object in the job configuration file (see \cref{sec:executable_settings}). Only
positive integer values of $n_\text{update}$ may be specified for this key.

\subsubsection{Manually specifying a maximum value of the neutrino energy PDF}
\label{sec:energy_pdf_max}

To select a reacting neutrino energy for each event, \marley\ relies on
rejection sampling from the probability density function given in
\cref{eq:reacting_nu_spectrum}. In situations where automatic estimation of the
global maximum $P_\text{max}$ of this PDF is found to be inadequate, \marley\
will print a set of messages to alert the user (see
\cref{sec:troubleshoot_reject}). These messages will include an improved
estimate of $P_\text{max}$ (see line~\ref{line:epdfmax} of
\cref{lst:rejection_problem1}) which may be manually supplied by the user in
the job configuration file as a floating-point value associated with the
\texttt{energy\_pdf\_max} key. Doing so will disable numerical estimation of
$P_\text{max}$ in favor of using the specified value, which may help to resolve
some rejection sampling problems.

\subsubsection{Keys specific to \texttt{mardumpxs}}
\label{sec:mardumpxs_keys}

Several job configuration file keys are available to modify the default
behavior of the \texttt{mardumpxs} utility described in \cref{sec:mardumpxs}.
They are otherwise ignored by \marley. The \texttt{mardumpxs}-specific
keys are listed below together with their interpretations:

\begin{description}
\item[\texttt{xsec\_dump\_pdg}] Labels an integer value that specifies the PDG
code (see the second column of \cref{tab:source_neutrino}) for the projectile
of interest. If this key is not present, then a $\nu_e$ (PDG code \num{12}) is
assumed by \texttt{mardumpxs}.
\item[\texttt{xsec\_dump\_KEmin}] Used to set the minimum projectile kinetic
energy (\si{MeV}) that should be included in the output. By default,
\texttt{mardumpxs} sets the minimum kinetic energy to \SI{0}{\MeV}.
\item[\texttt{xsec\_dump\_KEmax}] Used to set the maximum projectile kinetic
energy (\si{MeV}) that should be included in the output. By default,
\texttt{mardumpxs} sets the maximum kinetic energy to \SI{100}{\MeV}.
\item[\texttt{xsec\_dump\_steps}] Used to set the number of equally-spaced
points at which the total cross section should be evaluated as a function of
projectile kinetic energy. Only positive integer values are allowed. If this
key is absent, then \texttt{mardumpxs} will use a default value of \num{10000}.
\end{description}

\section{Interpreting the output}
\label{sec:output}

The neutrino scattering events generated by the \texttt{marley} command-line
executable may be saved to disk in four distinct output formats.
\Cref{sec:executable_settings} explains how one or more of these output formats
may be selected by the user in the job configuration file. Following a brief
discussion of the particle numbering scheme used by \marley\ in
\cref{sec:PDG_codes}, documentation for each of the output file formats is
provided in \cref{sec:file_formats}. \Cref{sec:example_analysis_scripts}
presents a \cpp-based API which may be used to parse all of the standard output
formats via a unified interface. \Cref{sec:xsec_normalization} provides
guidance on how histograms of quantities of interest from the generated events
may be converted into easily interpretable physics distributions, e.g.,
differential cross sections and event rates.

\subsection{Particle numbering scheme}
\label{sec:PDG_codes}

Like many other modern physics event generators, \marley\ identifies particle
species using a standard set of integer codes defined by the Particle Data
Group (PDG) \cite[sec.~44, pp. 661--664]{PDG2020}. Each kind of particle is
assigned a unique positive integer, and the corresponding antiparticle is
assigned a negative integer with the same absolute value. These \textit{PDG
codes} are used by \marley\ both internally and in output files.

\Cref{tab:PDG_codes} provides the interpretation of the most common PDG
codes that appear in \marley\ events. In general, a nucleus with proton
number $Z$ and mass number $A$ is labeled with the PDG code
\begin{equation}
\label{eq:nuclear_PDG_codes}
\pdgCodeVar = \num{10000}\,Z + \num{10}\,A + \num{1000000000}\,.
\end{equation}
The only two exceptions to the rule given in \cref{eq:nuclear_PDG_codes} are
for a free neutron and \isotope[1]{H}, which are represented by the codes
\num{2112} and \num{2212}, respectively.

\begin{table}
\renewcommand{\arraystretch}{1.4}
\centering
\begin{tabular}{cc}
\toprule
PDG code & Particle \\
\midrule
11 & $e^{-}$ \\
12 & $\nu_e$ \\
13 & $\mu^{-}$ \\
14 & $\nu_\mu$ \\
13 & $\tau^{-}$ \\
14 & $\nu_\tau$ \\
22 & $\gamma$ \\
2112 & $\mathrm{n}$ \\
2212 & $\mathrm{p}$ \\
1000010020 & $\mathrm{d}$ \\
1000010030 & $\mathrm{t}$ \\
1000020030 & $\mathrm{h}$ \\
1000020040 & $\alpha$ \\
\bottomrule
\end{tabular}
\caption{The most commonly encountered PDG codes used by \marley\ to
identify kinds of particles.}
\label{tab:PDG_codes}
\end{table}

\subsection{Output file formats}
\label{sec:file_formats}

In the current version of \marley, generated events may be saved to disk
using four possible file formats:
\begin{description}
\item[\ascii] The native text-based format for \marley\ events.
\item[\hepevt] A legacy format for interfacing event generators with each other
and with other software. Despite its relative age and emergence in the specific
context of QCD event generation for Large Electron-Positron Collider
experiments \cite{hepevt}, compatibility with this format is maintained in many
modern high energy physics software tools. Examples include the HepMC3
\cite{Buckley2019} event record library and (via its \texttt{TextFileGen}
module) the LArSoft toolkit discussed in \cref{sec:larsoft}.
\item[JSON] In addition to serving as the input language for \marley\ job
configuration files (with slight extensions to the standard grammar, see
\cref{sec:gen_config} for details), JSON is also available as an output file format.
\item[\rootcern] If \marley's interface to \rootcern\ is active, a binary
representation of the full \texttt{marley::Event} objects may be saved to disk
in an output \texttt{TTree}. A simplified ``flat'' form of the \rootcern\
output that can be read without recourse to the \marley\ shared libraries may
also be produced using the \texttt{marsum} command-line tool (see
\cref{sec:marsum}).
\end{description}

\subsubsection{\ascii\ file format}

An \ascii-format output file begins with the line
\vspace{-0.25\baselineskip}
\begin{center}
\begin{tabular}{c}
\begin{lstlisting}[backgroundcolor=\color{white}, numbers=none]
FluxAvgXsec
\end{lstlisting}
\end{tabular}
\end{center}
in which \texttt{FluxAvgXsec} is the flux-averaged total reaction cross section
in natural units (\si[per-mode=symbol] {\MeV\tothe{-2}\per\atom}, see
\cref{sec:xsec_normalization}). This line is followed by one or more event
records, each of which begins with the header
\begin{center}
\begin{tabular}{c}
\begin{lstlisting}[backgroundcolor=\color{white}, numbers=none]
Ni Nf Ex twoJ P
\end{lstlisting}
\end{tabular}
\end{center}
where \texttt{Ni} (\texttt{Nf}) is the number of particles in the initial (final)
state. The three remaining fields in the event header report
properties of the recoiling nucleus immediately following
the prompt $2 \rightarrow 2$ scattering reaction.
The quantity \texttt{Ex} is the excitation energy (\si{\MeV}),
\texttt{twoJ} is an integer equal to two times
the total spin,\footnote{This allows representation of half-integer
spins using a \cpp\ \texttt{int}} and \texttt{P} is a single
character representing either positive (\texttt{+}) or negative
(\texttt{-}) parity.

Following the event header, each of the \texttt{Ni} initial-state particles is
described by a single line of the form
\begin{center}
\begin{tabular}{c}
\begin{lstlisting}[backgroundcolor=\color{white}, numbers=none]
PDG Etot Px Py Pz M Q
\end{lstlisting}
\end{tabular}
\end{center}
where \texttt{PDG} is the PDG code (see \cref{sec:PDG_codes}) identifying the
particle species and \texttt{Etot}, \texttt{Px}, \texttt{Py}, and \texttt{Pz}
are the Cartesian components of the particle 4-momentum (in \si{\MeV}). The
particle mass \texttt{M} (\si{MeV}) and (net) electric charge \texttt{Q} (in
units of the elementary charge) are also included in each line. To complete the
event record, \texttt{Nf} lines describing the final-state particles appear in
the same format used for the initial-state particles.

To preserve full numerical precision when converting back and forth between
\texttt{marley::Event} objects held in memory and \ascii-format event files, all
floating point numbers are output in scientific notation with the required
number of base-10 digits needed to uniquely represent all distinct values of
the \cpp\ type \texttt{double}.\footnote{This number is called
\texttt{std::\allowbreak numeric\_limits<double>\allowbreak ::max\_digits10} in
the \cpp\ standard library.} However, this level of precision is not enforced
by the code when reading events as input from an \ascii\ file.
\begin{lstlisting}[linewidth=\textwidth,caption={Example \ascii-format
output file},captionpos=b,label={lst:ascii_events},
basicstyle=\footnotesize\ttfamily]
5.98368867447267264e-19
2 4 3.79748000000000019e+00 2 +
12 1.00000000000000000e+01 0.00000000000000000e+00 0.00000000000000000e+00 1.00000000000000000e+01 0.00000000000000000e+00 0
1000180400 3.72247225431518091e+04 0.00000000000000000e+00 0.00000000000000000e+00 0.00000000000000000e+00 3.72247225431518091e+04 0
11 5.20690886815266740e+00 -4.63535385338761508e+00 1.35706546730579314e+00 -1.87687187323011373e+00 5.10998927645907708e-01 -1
1000190400 3.72257175068044089e+04 4.98066184879143847e+00 -2.27058729207187593e+00 9.28456410767741858e+00 3.72257159465162476e+04 1
22 1.53694352434620662e+00 -1.11400145523226901e+00 9.50751758100756628e-01 4.66119350851738223e-01 0.00000000000000000e+00 0
22 2.26118395490673985e+00 7.68693459828445835e-01 -3.72299333346743089e-02 2.12618841470095354e+00 0.00000000000000000e+00 0
2 3 1.03964200000000009e+01 2 +
12 2.99304885549511290e+01 0.00000000000000000e+00 0.00000000000000000e+00 2.99304885549511290e+01 0.00000000000000000e+00 0
1000180400 3.72247225431518091e+04 0.00000000000000000e+00 0.00000000000000000e+00 0.00000000000000000e+00 3.72247225431518091e+04 0
11 1.85223477954320721e+01 -9.86795370224160173e+00 -1.55256312663347771e+01 -2.09630900945355991e+00 5.10998927645907708e-01 -1
1000190390 3.62940260743929030e+04 -1.74322423197586254e+01 -4.21863040207651636e+01 5.85225771236273147e+01 3.62939501877290750e+04 1
2112 9.42104609518426400e+02 2.73001960220002289e+01 5.77119352870999407e+01 -2.64957795592226226e+01 9.39565378653339735e+02 0
\end{lstlisting}

\Cref{lst:ascii_events} shows an example \marley\ output file in \ascii\
format. Line 1 gives the value of $\SI{5.984e-19}{\MeV\tothe{-2}} =
\SI{2.330e-40}{\centi\meter\squared}$ for the flux-averaged total cross
section. Line 2 begins the record for the first event, which involves a
transition to the \isotope[40]{K} nuclear level with an excitation energy of
\SI{3.797}{\MeV} above the ground state and spin-parity $1^+$. Lines 3--4
describe the initial state: a \SI{10}{\MeV} $\nu_e$ traveling in the $+z$
direction toward a \isotope[40]{Ar} atom at rest. Lines 5--8 describe the
final-state particles: a \SI{5.2}{\MeV} electron, the recoiling \isotope[40]{K}
ion, and two de-excitation $\gamma$-rays with energies of \num{1.54} and
\SI{2.26}{\MeV}. The second and final event in the file, which appears on lines
9--14, involves a $\nu_e$-\isotope[40]{Ar} collision which produces an
electron, a \isotope[39]{K} ion, and a neutron.

\subsubsection{\hepevt\ file format}

For the sake of brevity, only those aspects of the \hepevt\ format needed to
interpret \marley\ output are discussed in this section. A full description of
the \hepevt\ standard is available in ref.~\cite[pp. 327--330]{hepevt}.

A \hepevt-format output file consists of one or more text-based event records.
Each of these records begins with the header
\vspace{-0.6\baselineskip}
\begin{center}
\begin{tabular}{c}
\begin{lstlisting}[backgroundcolor=\color{white}, numbers=none]
NEVHEP NHEP
\end{lstlisting}
\end{tabular}
\end{center}
where \texttt{NEVHEP} is the event number (untracked by \marley\ and thus
always set to zero) and \texttt{NHEP} is the number of particles in the event.
The header is followed by \texttt{NHEP} lines, each representing a single
particle. These have the format
\vspace{-1\baselineskip}
\begin{center}
\begin{tabular}{c}
\begin{lstlisting}[backgroundcolor=\color{white}, numbers=none]
ISTHEP IDHEP JMOHEP1 JMOHEP2 JDAHEP1 JDAHEP2 PHEP1 PHEP2 PHEP3 PHEP4 PHEP5 VHEP1 VHEP2 VHEP3 VHEP4
\end{lstlisting}
\end{tabular}
\end{center}
where \texttt{ISTHEP} is an integer code identifying the particle status and
\texttt{IDHEP} is the PDG particle ID code. In agreement with the \hepevt\
standard, \marley\ uses status code 1 for the final-state particles and 3 for
the initial-state particles, the latter of which are not considered part of the
event history \cite{hepevt}. The \texttt{JMOHEP1}, \texttt{JMOHEP2},
\texttt{JDAHEP1}, and \texttt{JDAHEP2} entries record the indices $j$ ($1 \leq
j \leq \text{\texttt{NHEP}}$) of particles in the event record that correspond
to the first mother, second mother, first daughter, and last daughter of the
current particle, respectively. These indices are set to zero in cases where
they do not apply (e.g., a particle which has not decayed will have
$\texttt{JDAHEP1} = \texttt{JDAHEP2} = 0$). Entries \texttt{PHEP1} through
\texttt{PHEP3} record the $x$, $y$, and $z$ components of the particle
3-momentum, while \texttt{PHEP4} gives the total energy and \texttt{PHEP5}
gives the particle mass (all in \si{GeV}). Entries \texttt{VHEP1} through
\texttt{VHEP3} store the $x$, $y$, and $z$ positions of the particle production
vertex (\si{\milli\meter}), and \texttt{VHEP4} gives the production time
(\si[per-mode=symbol]{\milli\meter\per\clight}).

Because \marley\ currently treats all nuclear de-excitations as instantaneous
and does not perform any particle tracking, \texttt{VHEP1} through
\texttt{VHEP4} are always identically zero in \hepevt\ output files.
Intermediate de-excitation steps are also not currently stored in the event
record, so \texttt{JMOHEP1}, \texttt{JMOHEP2}, \texttt{JDAHEP1}, and
\texttt{JDAHEP2} are also identically zero in most cases.

In addition to the initial- and final-state particles, \marley\ adds a dummy
particle with $\text{\tt ISTHEP} = 11$ to each \hepevt\ event record. All data
fields are zero for this particle except for
\begin{enumerate*}[label=(\arabic*)] \item \texttt{JMOHEP1}, which
contains the nuclear spin multiplied by two, \item \texttt{JMOHEP2},
which reports the parity of the nucleus as an integer,
\item \texttt{PHEP4}, which gives the
excitation energy of the nucleus (\si{\MeV}), and
\item \texttt{PHEP5}, which records the
flux-averaged total cross section (see \cref{sec:xsec_normalization})
in units of \si[per-mode=symbol, per-symbol=\,/\,]{\MeV\tothe{-2}\per\atom}.
\end{enumerate*}
As is the case for the \ascii\ format, the excitation energy, spin, and parity
values refer to the nuclear state that existed immediately after the initial
$2 \to 2$ scattering reaction.
\begin{lstlisting}[linewidth=\textwidth,caption={Example \hepevt-format output
file},captionpos=b,label={lst:hepevt_events},
basicstyle=\scriptsize\ttfamily]
0 7
3 12 0 0 0 0 0.00000000000000000e+00 0.00000000000000000e+00 1.00000000000000002e-02 1.00000000000000002e-02 0.00000000000000000e+00 0. 0. 0. 0.
3 1000180400 0 0 0 0 0.00000000000000000e+00 0.00000000000000000e+00 0.00000000000000000e+00 3.72247225431518061e+01 3.72247225431518061e+01 0. 0. 0. 0.
11 0 2 1 0 0 0.00000000000000000e+00 0.00000000000000000e+00 0.00000000000000000e+00 3.79748000000000019e+00 5.98368867447267264e-19 0. 0. 0. 0.
1 11 0 0 0 0 -4.63535385338761496e-03 1.35706546730579320e-03 -1.87687187323011366e-03 5.20690886815266749e-03 5.10998927645907710e-04 0. 0. 0. 0.
1 1000190400 0 0 0 0 4.98066184879143812e-03 -2.27058729207187593e-03 9.28456410767741942e-03 3.72257175068044077e+01 3.72257159465162459e+01 0. 0. 0. 0.
1 22 0 0 0 0 -1.11400145523226908e-03 9.50751758100756655e-04 4.66119350851738206e-04 1.53694352434620668e-03 0.00000000000000000e+00 0. 0. 0. 0.
1 22 0 0 0 0 7.68693459828445808e-04 -3.72299333346743093e-05 2.12618841470095347e-03 2.26118395490674000e-03 0.00000000000000000e+00 0. 0. 0. 0.
0 6
3 12 0 0 0 0 0.00000000000000000e+00 0.00000000000000000e+00 2.99304885549511283e-02 2.99304885549511283e-02 0.00000000000000000e+00 0. 0. 0. 0.
3 1000180400 0 0 0 0 0.00000000000000000e+00 0.00000000000000000e+00 0.00000000000000000e+00 3.72247225431518061e+01 3.72247225431518061e+01 0. 0. 0. 0.
11 0 2 1 0 0 0.00000000000000000e+00 0.00000000000000000e+00 0.00000000000000000e+00 1.03964200000000009e+01 5.98368867447267264e-19 0. 0. 0. 0.
1 11 0 0 0 0 -9.86795370224160216e-03 -1.55256312663347770e-02 -2.09630900945356009e-03 1.85223477954320724e-02 5.10998927645907710e-04 0. 0. 0. 0.
1 1000190390 0 0 0 0 -1.74322423197586264e-02 -4.21863040207651613e-02 5.85225771236273160e-02 3.62940260743929031e+01 3.62939501877290738e+01 0. 0. 0. 0.
1 2112 0 0 0 0 2.73001960220002303e-02 5.77119352870999400e-02 -2.64957795592226236e-02 9.42104609518426450e-01 9.39565378653339778e-01 0. 0. 0. 0.
\end{lstlisting}

\Cref{lst:hepevt_events} shows an example \marley\ output file in \hepevt\
format. The same two events from the \ascii-format example file (see
\cref{lst:ascii_events}) are used for easy comparison of the formats.

\subsubsection{JSON file format}

Unlike \marley\ job configuration files, which allow a few non-standard JSON
language extensions (see \cref{sec:gen_config} for details), the JSON-format
output files fully conform to the standard grammar. Each output file includes
two top-level keys: \texttt{events}, which is associated with an array of event
objects, and \texttt{gen\_state}, which stores a JSON object representation of
the generator state at the moment that the file was created.

Each element of the \texttt{events} array is a JSON object containing five
key-value pairs. The first three of these, \texttt{Ex}, \texttt{twoJ},
and \texttt{parity}, provide the excitation
energy (\si{\MeV}), two times the total spin, and the parity
of the final nucleus after the initial $2 \rightarrow 2$
scattering reaction but before any de-excitations have taken place. The other
two keys, \texttt{initial\_particles} and \texttt{final\_\allowbreak
particles}, are used store arrays of particles represented as JSON objects.
Each particle object defines the following keys:
\begin{enumerate*}[label=(\arabic*)]
\item{\tt charge:} the (net) electric charge in units of the elementary charge,
\item{\tt pdg:} the PDG code identifying the particle type,
\item{\tt E:} the total energy,
\item{\tt px:} the $x$ momentum component,
\item{\tt py:} the $y$ momentum component,
\item{\tt pz:} the $z$ momentum component, and
\item{\tt mass:} the particle mass.
\end{enumerate*}
The particle 4-momentum components and mass are all given in
natural units (\si{\MeV}).

The \texttt{gen\_state} JSON object includes several key-value pairs. The
\texttt{config} key refers to a JSON object which reproduces the full contents
(except for comments) of the job configuration file used to generate the
events. The \texttt{event\_count}, \texttt{flux\_avg\_xsec}, and \texttt{seed}
keys label the total number of events generated at the time the file was
written, the flux-averaged total cross section (\si[per-mode=symbol,
per-symbol=\,/\,]{\MeV\tothe{-2}\per\atom}, see \cref{sec:xsec_normalization}),
and the integer random number seed used to initialize the event generator. A
final key, \texttt{generator\_state\_string}, records a string representation
of the internal state of the \texttt{std::mt19937\_\allowbreak 64} object used
by \marley\ to obtain pseudorandom numbers.

An example \marley\ output file in JSON format (\texttt{example.json}) is
included in the supplemental materials.

\subsubsection{\rootcern\ file format}
\label{sec:ROOT_format}

If \marley\ has been built against the appropriate shared libraries from the
\rootcern\ data analysis toolkit (see \cref{sec:install}), then the
\texttt{marley} command-line executable may also produce output files in the
standard \rootcern\ compressed binary format.

Within a \rootcern-format output file, access to the generated events is
managed by an instance of the \rootcern\ \texttt{TTree} class, which provides
a hierarchical data structure for storing many objects belonging to the
same \cpp\ type. In general, a \texttt{TTree} may own one or more branches
(represented by the \texttt{TBranch} class), each of which owns one or
more leaves (represented by \texttt{TLeaf}). Branches may be read
from a file independently of one another, allowing for efficient access to
elements of a large dataset stored in a suitably-organized \texttt{TTree}
\cite{TTrees}.

The \rootcern-format output files generated by \marley\ contain a single
\texttt{TTree} called \texttt{MARLEY\allowbreak\_event\allowbreak\_tree}. This
\texttt{TTree} contains a single branch called \texttt{event}, which stores one
\texttt{marley\allowbreak::\allowbreak Event} object per tree entry. Although
direct access to the events is possible by manipulating the
\texttt{MARLEY\allowbreak\_event\allowbreak\_tree}, use of the simplified \cpp\
API described in \cref{sec:example_analysis_scripts} is recommended instead.

In addition to the \marley\ events themselves, four pieces of metadata are
stored in a \rootcern-format output file:
\newcommand{\fakeunderscore}{\hspace{-0.1em}\rule{0.5em}{0.6pt}\hspace{0.1em}}
\newcommand{\secondfakeunderscore}{\rule{0.5em}{0.6pt}\hspace{0.1em}}
\begin{description}[font=\normalfont\normalsize]
\item[\textbf{MARLEY\fakeunderscore config} (\textit{std::string})]
  A JSON-format string which stores the contents (except for comments) of the
job configuration file used to generate the events (see \cref{sec:gen_config})
\item[\textbf{MARLEY\fakeunderscore state} (\textit{std::string})]
  Serialized internal state (obtained using the stream insertion operator
\texttt{<<}) of the \texttt{std::mt19937\allowbreak\_64} object (see
\cref{sec:random_sampling}) owned by the \texttt{marley\allowbreak::\allowbreak
Generator} object used to create the events
\item[\textbf{MARLEY\fakeunderscore seed} (\textit{std::string})]
  A string representation of the integer random number seed used to initialize
the simulation
\item[\textbf{MARLEY\secondfakeunderscore flux\secondfakeunderscore
avg\secondfakeunderscore xsec} (\textit{TParameter\textless
double\textgreater})]
  The flux-averaged total cross section (\si[per-mode=symbol,
  per-symbol=\,/\,]{\MeV\tothe{-2}\per\atom}) needed to normalize physics
  distributions computed from the events (see \cref{sec:xsec_normalization})
\end{description}
In \cref{lst:print_meta}, an example \cpp\ function is shown which retrieves
all of these metadata objects from a file called \texttt{events.root} and
prints their contents to standard output.
\begin{lstlisting}[language=C++, linewidth=\textwidth,
caption={Example access to \marley\ metadata stored in a \rootcern-format
output file}, captionpos=b, label={lst:print_meta},
basicstyle=\small\ttfamily, escapechar=|]
// Standard library includes
#include <iostream>
#include <string>

// ROOT includes
#include "TFile.h"
#include "TParameter.h"

void print_metadata() {

  TFile event_file( "events.root", "read" );

  std::string* config = NULL;
  std::string* state = NULL;
  std::string* seed = NULL;
  TParameter<double>* xsec = NULL;

  event_file.GetObject( "MARLEY_config", config );
  event_file.GetObject( "MARLEY_state", state );
  event_file.GetObject( "MARLEY_seed", seed );
  event_file.GetObject( "MARLEY_flux_avg_xsec", xsec );

  std::cout << "MARLEY configuration: " << *config << '\n';
  std::cout << "Generator internal state: " << *state << '\n';
  std::cout << "Random number seed: " << *seed << '\n';
  std::cout << "Flux-averaged total cross section: " << xsec->GetVal()
    << " MeV^{-2} / atom\n";

}
\end{lstlisting}

The \rootcern\ output files described above may be converted into an
alternative ``flat'' format which is readable by \rootcern\ without a need for
the \marley\ shared libraries. This is done by running the \texttt{marsum}
command-line tool described in \cref{sec:marsum}.

\subsection{\cpp\ analysis API}
\label{sec:example_analysis_scripts}

Although the file format descriptions given in \cref{sec:file_formats} should
be sufficient to enable processing of \marley\ events using any programming
language, a \cpp\ API allowing manipulation of events stored in any of the
standard output formats has been developed for the convenience of users.
The API is usable within compiled code as well as via the interactive \cpp\
interpreters included with \rootcern.\footnote{The CINT \cite{CINT} interpreter
is distributed with version 5 of \rootcern, while version 6
uses Cling \cite{Cling}.}

\subsubsection{The EventFileReader class}
\label{sec:efr_class}

Programmatic access to \marley\ event records stored in a file is provided by
the \texttt{marley\allowbreak::\allowbreak Event\allowbreak File\allowbreak
Reader} class. The constructor of this class takes as its sole argument a
\texttt{std::string} containing the name (including any needed path
specification) of the file to be parsed. Events may be read one-by-one from the
file using the stream extraction operator \texttt{>>}, which returns a boolean
value indicating whether a new event was successfully loaded.

\Cref{lst:efr_program} shows the source code for \texttt{examples/\allowbreak
executables/\allowbreak minimal/\allowbreak efr.cc}, an example \cpp\ program
that illustrates the recommended usage of the \texttt{Event\allowbreak
File\allowbreak Reader} class. Line \ref{line:efr_constructor} creates a new
\texttt{Event\allowbreak File\allowbreak Reader} object that will read \marley\
events from a file whose name is given as the first command-line argument when
the program is run. The while loop on lines
\ref{line:event_start}--\ref{line:event_end} streams all events in the file to
standard output, where they will be printed in ASCII format. Line
\ref{line:xsec_start} contains a call to the \texttt{Event\allowbreak
File\allowbreak Reader} member function \texttt{flux\_averaged\_xsec}, which
returns the flux-averaged total cross section (see
\cref{sec:xsec_normalization}) used to generate the events in units of
\SI[per-mode=symbol, per-symbol=\,/\,]{e-42}{\centi\meter \squared\per\atom}.
Passing a boolean value of \texttt{true} to this function will cause it to
return the cross section in natural units (\si[per-mode=symbol,
per-symbol=\,/\,]{\MeV\tothe{-2}\per\atom}) instead.

\begin{lstlisting}[language=C++, linewidth=\textwidth, caption={The source code
for \texttt{examples/\allowbreak executables/\allowbreak minimal/\allowbreak
efr.cc}, an example \cpp\ program that uses the
\texttt{marley\allowbreak::\allowbreak Event\allowbreak File\allowbreak Reader}
class}, captionpos=b, label={lst:efr_program}, basicstyle=\small\ttfamily,
escapechar=|]
#include <iostream>

#include "marley/Event.hh"
#include "marley/EventFileReader.hh"|\label{line:efr_header}|

int main(int argc, char** argv) {

  if ( argc < 2 ) return 1;

  std::string input_file_name( argv[1] );

  marley::EventFileReader efr( input_file_name );|\label{line:efr_constructor}|
  marley::Event event;

  double avg_xsec = efr.flux_averaged_xsec();|\label{line:xsec_start}|
  std::cout << "flux-averaged total cross section = "
    << avg_xsec << " * 10^{-42} cm^2 / atom\n";|\label{line:xsec_end}|

  while ( efr >> event ) {|\label{line:event_start}|
    std::cout << event << '\n';
  }|\label{line:event_end}|

  return 0;
}
\end{lstlisting}

To build the \texttt{efr.cc} example program, it is necessary to link the
executable to the \marley\ shared library. This is most easily done using the
\texttt{marley-config} script with the syntax described in
\cref{sec:marley_config_compile}.

The \texttt{Event\allowbreak File\allowbreak Reader} class can parse \marley\
output files written in the ASCII, HEPEVT, and JSON formats. If \marley\ has
been built against \rootcern\ (see \cref{sec:install}), then support for the
\rootcern\ output format in addition to the others is available via the
\texttt{Root\allowbreak Event\allowbreak File\allowbreak Reader} class. This
class is derived from \texttt{Event\allowbreak File\allowbreak Reader} and
implements an identical user interface. Making the replacement
\texttt{Event\allowbreak File\allowbreak Reader} $\to$ \texttt{Root\allowbreak
Event\allowbreak File\allowbreak Reader} on lines \ref{line:efr_header} and
\ref{line:efr_constructor} of \cref{lst:efr_program} will yield a program that
is capable of reading events from any \marley\ output file. Compiling this
modified program requires linking to the shared library containing the \marley\
interface to \rootcern. The build command given in
\cref{sec:marley_config_compile} will automatically include the necessary
compiler flags for linking to \rootcern\ if \marley\ was successfully built
with \rootcern\ support.

\subsubsection{Accessing information from an event record}
\label{sec:event_access}

As discussed in \cref{sec:theory_summary}, \marley\ conceives of each
scattering event as consisting of a primary $2 \to 2$ reaction possibly
followed by nuclear de-excitations. The particle content of the primary
reaction may be written in the general form
\begin{equation}
a + b \to c + d
\end{equation}
where particles $a$, $b$, $c$, and $d$, are respectively termed the
\textit{projectile}, \textit{target}, \textit{ejectile}, and \textit{residue}.
Where a mass difference exists, \marley\ chooses the projectile (ejectile) to
be the lighter of the two particles in the initial (final) state. Otherwise,
the choice is arbitrary. All four-vector components stored in a \marley\ event
record are expressed in the laboratory frame, i.e., the rest frame of the
target.

A full \marley\ event record is represented in the code itself by the
\texttt{marley\allowbreak::\allowbreak Event} class. Instances of this class
own two vectors of pointers to \texttt{marley\allowbreak::\allowbreak Particle}
objects, with one (\texttt{initial\allowbreak\_particles\_}) describing the
initial state and the other (\texttt{final\_particles\_}) describing the final
state. Within the \texttt{initial\_particles\_} vector, the first and second
elements always correspond to the projectile and target, respectively. A
similar ordering (ejectile followed by residue) is enforced for the first two
elements of the \texttt{final\_particles\_} vector. If present, elements of
\texttt{final\_particles\_} beyond the second correspond to nuclear
de-excitation products, which are listed in the order that they were emitted.
If nuclear de-excitations were enabled in the simulation (as they are by
default, see \cref{sec:nuc_deex_enable_disable}), then the \texttt{Particle}
object for the residue stores its properties after it has reached the ground
state.

The \texttt{Event} class provides the following member functions for accessing
\texttt{Particle} objects stored in the event record:
\begin{description}[font=\normalfont\itshape\normalsize]
\item[projectile(), target(), ejectile(), residue()] Returns a reference to
  a specific particle involved in the primary $2 \to 2$ reaction
\item[get\_initial\_particles()] Returns a reference to the owned
  vector of initial particles
\item[get\_final\_particles()] Returns a reference to the owned
  vector of final particles
\item[initial\_particle\_count()] Returns the number of initial-state particles
  in the event
\item[final\_particle\_count()] Returns the number of final-state particles
  in the event
\item[initial\_particle(size\_t idx)] Returns a constant reference to the
  initial-state particle at position \texttt{idx}
\item[final\_particle(size\_t idx)] Returns a constant reference to the
  final-state particle at position \texttt{idx}
\end{description}

The \texttt{Particle} class stores a 4-momentum (represented as a C-style array
of type \texttt{double[4]}) together with variables representing particle
properties. It implements the following member functions
for data retrieval:
\begin{description}[font=\normalfont\itshape\normalsize]
\item[charge()] Electric charge (in units of the proton charge)
\item[mass()] Mass (\si{MeV})
\item[pdg\_code()] PDG code \cite[sec.~44, pp. 661--664]{PDG2020}
  representing the particle species
\item[px()] 3-momentum x-component (\si{\MeV})
\item[py()] 3-momentum y-component (\si{\MeV})
\item[pz()] 3-momentum z-component (\si{\MeV})
\item[kinetic\_energy()] Kinetic energy (\si{\MeV})
\item[total\_energy()] Total energy (\si{\MeV})
\end{description}
All of these functions return a value of type \texttt{double} except for
\texttt{pdg\_code()}, which returns an \texttt{int}.

In addition to \texttt{Particle} objects, the \texttt{Event} class also
stores information about the state of the residue immediately after its
creation by the prompt $2 \to 2$ reaction. Access to this information
is provided by these member functions:
\begin{description}[font=\normalfont\itshape\normalsize]
\item[Ex()] Initial residue excitation energy (\texttt{double}, \si{\MeV}),
  measured with respect to its own ground state
\item[twoJ()] Two times the initial residue spin (\texttt{int})
\item[parity()] Initial residue parity (\texttt{marley\allowbreak::\allowbreak
Parity})
\end{description}
The \texttt{Parity} object returned by \texttt{Event\allowbreak::\allowbreak
parity()} provides a type-safe representation of a parity value $\pm1$. It may
be converted to a boolean (\texttt{true} $\leftrightarrow +1$, \texttt{false}
$\leftrightarrow -1$) or integer representation via an explicit cast. A
\texttt{Parity} object may also be converted to a \texttt{char} value
(\texttt{+} $\leftrightarrow +1$, \texttt{-} $\leftrightarrow -1$) via the
member function \texttt{to\_char()}.

Full technical documentation for the \texttt{Event}, \texttt{Particle}, and
\texttt{Parity} classes is available on the official \marley\ website
\cite{MARLEYWebsite}. Similar HTML documentation may be automatically generated
for offline use from the \marley\ source code using Doxygen \cite{Doxygen}.
After Doxygen has been installed, one may build the \marley\ API documentation
files by executing
\begin{lstlisting}[backgroundcolor=\color{white}, numbers=none,
basicstyle=\normalsize\ttfamily, commentstyle=\normalsize\ttfamily]
make doxygen
\end{lstlisting}
from within the \texttt{build/} folder. The generated documentation may then be
viewed using any web browser by opening the file
\texttt{docs/\_build/html/doxygen/index.html}

\subsubsection{Executable example: \texttt{marprint}}

The \texttt{marprint} program (\texttt{examples/\allowbreak
executables/\allowbreak marprint.cc}) included with \marley\ provides a
detailed example of how the \cpp\ analysis API may be used in compiled code.
After instantiating either an \texttt{Event\allowbreak File\allowbreak Reader}
or a \texttt{Root\allowbreak Event\allowbreak File\allowbreak Reader} object
(depending on whether \marley\ was built against \rootcern), the program prints
all information stored in each event to standard output using a human-readable
format. Usage examples for many of the functions listed in
\cref{sec:event_access} are provided within the \texttt{marprint.cc} source
file. After sourcing the \texttt{setup\_marley.sh} script, one may build the
\texttt{marprint} example program via the commands
\begin{lstlisting}[backgroundcolor=\color{white}, numbers=none,
basicstyle=\normalsize\ttfamily, commentstyle=\normalsize\ttfamily]
cd ${MARLEY}/build
make marprint
\end{lstlisting}
The \texttt{marprint} executable takes the names of one or more files
containing \marley\ events as command-line arguments. For example, one may
print all of the events stored in the files \texttt{/home/\allowbreak
events1.ascii} and \texttt{/home/\allowbreak events2.json} by executing
\begin{lstlisting}[backgroundcolor=\color{white}, numbers=none,
basicstyle=\normalsize\ttfamily, commentstyle=\normalsize\ttfamily]
marprint /home/events1.ascii /home/events2.json
\end{lstlisting}
\Cref{lst:marprint_output} shows the printout generated by \texttt{marprint}
while parsing the first event given in the example \ascii-format output file
shown above (\cref{lst:ascii_events}).
\begin{lstlisting}[linewidth=\textwidth, numbers=none, caption={Example output
generated by the \texttt{marprint} program}, captionpos=b,
label={lst:marprint_output},
basicstyle=\footnotesize\ttfamily]
*** Event 0 has 2 initial particles and 4 final particles. ***
The residual nucleus initially had excitation energy 3.79748 MeV and spin-parity 1+
Initial particles
  particle with PDG code = 12 has total energy 10 MeV,
    3-momentum = (0 MeV, 0 MeV, 10 MeV),
    mass = 0 MeV, and charge = 0 times the proton charge.
  particle with PDG code = 1000180400 has total energy 37224.7 MeV,
    3-momentum = (0 MeV, 0 MeV, 0 MeV),
    mass = 37224.7 MeV, and charge = 0 times the proton charge.
Final particles
  particle with PDG code = 11 has total energy 5.20691 MeV,
    3-momentum = (-4.63535 MeV, 1.35707 MeV, -1.87687 MeV),
    mass = 0.510999 MeV, and charge = -1 times the proton charge.
  particle with PDG code = 1000190400 has total energy 37225.7 MeV,
    3-momentum = (4.98066 MeV, -2.27059 MeV, 9.28456 MeV),
    mass = 37225.7 MeV, and charge = 1 times the proton charge.
  particle with PDG code = 22 has total energy 1.53694 MeV,
    3-momentum = (-1.114 MeV, 0.950752 MeV, 0.466119 MeV),
    mass = 0 MeV, and charge = 0 times the proton charge.
  particle with PDG code = 22 has total energy 2.26118 MeV,
    3-momentum = (0.768693 MeV, -0.0372299 MeV, 2.12619 MeV),
    mass = 0 MeV, and charge = 0 times the proton charge.
\end{lstlisting}
The commands needed to build \texttt{marprint} (and a second example program,
\texttt{mardumpxs}, described in \cref{sec:mardumpxs}) against the \marley\
shared libraries are given in the \texttt{Makefile} located in the folder
\texttt{examples/executables/build/}. Users are encouraged to adapt this
\texttt{Makefile} for building their own \cpp\ programs that interface with
\marley.

\subsubsection{Example \rootcern\ macros}
\label{sec:macros}

As a second example of the \marley\ analysis API, several \textit{\rootcern\
macros}, i.e., programs intended to be run using the interactive \cpp\
interpreter distributed with \rootcern, are provided in the
\texttt{examples/\allowbreak macros/} folder. Although \texttt{Root\allowbreak
Event\allowbreak File\allowbreak Reader} and the other \marley\ classes are
fully compatible with version 6 of \rootcern, version 5 lacks support for \cpp
11 features, which are used extensively by \marley. To work around this
problem, \marley\ provides a class called \texttt{Macro\allowbreak
Event\allowbreak File\allowbreak Reader} which is usable within macros written
for both versions 5 and 6 of \rootcern. This class provides the same interface
to \marley\ events as \texttt{Root\allowbreak Event\allowbreak File\allowbreak
Reader}, but the use of \cpp\ syntax that is incompatible with \rootcern\ 5 has
been hidden from the interpreter.

In order to correctly interact with \marley\ classes, \rootcern\ requires
precompiled dictionaries to be loaded at runtime. These are stored in the
\texttt{MARLEY\_ROOT} shared library. To avoid the need for users to manually
load the \marley\ class dictionaries at the start of each \rootcern\
interpreter session, a startup script called \texttt{mroot} is placed in the
\texttt{build/} folder whenever \marley\ is successfully built against
\rootcern. After sourcing the \texttt{setup\_marley.sh} setup script (see
\cref{sec:setup_marley_script}), issuing the command
\begin{lstlisting}[backgroundcolor=\color{white}, numbers=none,
basicstyle=\normalsize\ttfamily, commentstyle=\normalsize\ttfamily]
mroot
\end{lstlisting}
will start the interactive \rootcern\ interpreter and automatically load the
\marley\ class dictionaries. For version 5 of \rootcern, only the
\texttt{Event}, \texttt{Particle}, \texttt{Parity}, and
\texttt{Macro\allowbreak Event\allowbreak File\allowbreak Reader} classes (all
defined within the \texttt{marley} namespace)  may be used within an
\texttt{mroot} session. For \rootcern\ 6, all \marley\ classes will be
available if the appropriate header files are loaded via \texttt{\#include}
statements.

The \texttt{README.md} file in the \texttt{examples/\allowbreak macros/} folder
gives brief descriptions of each of the example \rootcern\ macros together with
usage instructions. Users are encouraged to adopt these macros as a starting
point for implementing their own calculations.

\subsubsection{``Flat'' \rootcern\ files: the \texttt{marsum} utility}
\label{sec:marsum}

Although they are expected to be suitable for many applications, the
standard \rootcern-format output files generated by \marley\ (see
\cref{sec:ROOT_format}) have two important limitations:
\begin{enumerate*}[label=(\arabic*)]
\item They are only fully readable in environments in which both \rootcern\ and
\marley\ are installed, and
\item Each \texttt{Event} object to be analyzed must be loaded from disk in
its entirety.
\end{enumerate*}

To address these limitations, \marley\ provides a command-line utility called
\texttt{marsum}, which takes as input one or more files containing \marley\
events stored in any of the standard output formats. The \texttt{marsum}
program creates a new file in which many quantities of interest from the input
events have been saved as individual branches of a \rootcern\ \texttt{TTree}.
Following a successful build of \marley\ with \rootcern\ support (see
\cref{sec:install}), the \texttt{marsum} executable will be present in the
\texttt{build/} folder. Assuming that the \texttt{setup\_marley.sh} script has
already been sourced, one may execute the command
\begin{lstlisting}[backgroundcolor=\color{white}, numbers=none,
basicstyle=\normalsize\ttfamily, commentstyle=\normalsize\ttfamily]
marsum myout.root /home/events1.root /home/events2.ascii
\end{lstlisting}
to create a new file called \texttt{myout.root} in the working directory.
This file will contain a single \rootcern\ \texttt{TTree} named \texttt{mst}
(for ``\texttt{\underline{M}ARLEY} \underline{s}ummary
\underline{t}ree''\footnote{This output format was inspired by the similar
``gst'' \texttt{TTree} produced by the \texttt{gntpc} utility distributed with
GENIE \cite[sec.~7.6.2, pp. 112--115]{GENIEManual}.}) with one entry for each
event present in the two input files
(\texttt{/home/events1.root} and \texttt{/home/events2.ascii}). The
\texttt{mst} \texttt{TTree} will contain the following branches:
\begin{description}[font=\normalfont\normalsize]
\item[\textbf{pdgv} (\textit{int})] Projectile PDG code
\item[\textbf{Ev} (\textit{double})] Projectile total energy (\si{\MeV})
\item[\textbf{KEv} (\textit{double})] Projectile kinetic energy (\si{\MeV})
\item[\textbf{pxv} (\textit{double})]
  Projectile 3-momentum x-component (\si{\MeV})
\item[\textbf{pyv} (\textit{double})]
  Projectile 3-momentum y-component (\si{\MeV})
\item[\textbf{pzv} (\textit{double})]
  Projectile 3-momentum z-component (\si{\MeV})
\item[\textbf{pdgt} (\textit{int})] Target PDG code
\item[\textbf{Mt} (\textit{double})] Target mass (\si{\MeV})
\item[\textbf{pdgl} (\textit{int})] Ejectile PDG code
\item[\textbf{El} (\textit{double})] Ejectile total energy (\si{\MeV})
\item[\textbf{KEl} (\textit{double})] Ejectile kinetic energy (\si{\MeV})
\item[\textbf{pxl} (\textit{double})]
  Ejectile 3-momentum x-component (\si{\MeV})
\item[\textbf{pyl} (\textit{double})]
  Ejectile 3-momentum y-component (\si{\MeV})
\item[\textbf{pzl} (\textit{double})]
  Ejectile 3-momentum z-component (\si{\MeV})
\item[\textbf{pdgr} (\textit{int})] Residue PDG code
\item[\textbf{Er} (\textit{double})] Residue total energy (\si{\MeV})
\item[\textbf{KEr} (\textit{double})] Residue kinetic energy (\si{\MeV})
\item[\textbf{pxr} (\textit{double})]
  Residue 3-momentum x-component (\si{\MeV})
\item[\textbf{pyr} (\textit{double})]
  Residue 3-momentum y-component (\si{\MeV})
\item[\textbf{pzr} (\textit{double})]
  Residue 3-momentum z-component (\si{\MeV})
\item[\textbf{Ex} (\textit{double})]
  Initial residue excitation energy (\si{\MeV})
\item[\textbf{twoJ} (\textit{int})]
  Two times the initial residue spin
\item[\textbf{parity} (\textit{int})] Initial residue parity
\item[\textbf{np} (\textit{int})]
  Number of de-excitation products
\item[\textbf{pdgp} (\textit{int{[np]}})]
  De-excitation product PDG codes
\item[\textbf{Ep} (\textit{double{[np]}})]
  De-excitation product total energies (\si{\MeV})
\item[\textbf{KEp} (\textit{double{[np]}})]
  De-excitation product kinetic energies (\si{\MeV})
\item[\textbf{pxp} (\textit{double{[np]}})]
  De-excitation product 3-momentum x-components (\si{\MeV})
\item[\textbf{pyp} (\textit{double{[np]}})]
  De-excitation product 3-momentum y-components (\si{\MeV})
\item[\textbf{pzp} (\textit{double{[np]}})]
  De-excitation product 3-momentum z-components (\si{\MeV})
\item[\textbf{xsec} (\textit{double})]
  Flux-averaged total cross section (\SI[per-mode=symbol,
  per-symbol={\,/\,}]{e-42}{\centi\meter
  \tothe{2}\per\atom})
\end{description}
The ``flat'' file created in this way will be readable by \rootcern\ without
the need to load the \marley\ class dictionaries. Individual branches may also
be loaded and plotted (e.g., via the \texttt{TTree::Draw} member function)
without the need to manipulate the event as a whole.

\subsection{Converting event distributions to physics quantities}
\label{sec:xsec_normalization}

The theoretical predictions made by an event generator like \marley\ may most
usefully be compared to competing calculations and experimental data in the
form of cross sections and event rates. To see how these quantities may be
obtained from a set of simulated events, first note that the expression given
in \cref{eq:reacting_nu_spectrum} for the probability density $P(E_\nu)$ of
the energy $E_\nu$ of a reacting neutrino may be rewritten in the form
\begin{equation}
\label{eq:PE2}
P(E_\nu) = \frac{ \phi(E_\nu)\,\sigma(E_\nu) }
{ \Phi \FluxAvgTotXSec }
\end{equation}
where
\begin{equation}
\label{eq:total_flux}
\Phi \equiv
\int_{E_\nu^\text{min}}^{E_\nu^\text{max}} \! \! \phi(E_\nu) \, dE_\nu
\end{equation}
is the total neutrino flux and
\begin{equation}
\label{eq:flux_avg_xsec}
\FluxAvgTotXSec \equiv
\frac{ 1 }{ \Phi } \,
\int_{E_\nu^\text{min}}^{E_\nu^\text{max}} \! \! \phi(E_\nu)
\, \sigma(E_\nu) \, dE_\nu
\end{equation}
is the (abundance-weighted) flux-averaged total cross section.
Let $x$ denote an arbitrary, continuously-distributed observable that is
computable from a \marley\ event record. Then, for a single event,
the probability $P_j$ that $x$ falls within the $j$th bin
$x \in [x_j, x_{j+1})$ is given by
\begin{equation}
\label{eq:Pj}
P_j = \int_{E_\nu^\text{min}}^{E_\nu^\text{max}} \! \!
P(E_\nu) \int_{x_j}^{x_{j+1}} P(x | E_\nu) \, dx \, dE_\nu
= \frac{ 1 }{ \FluxAvgTotXSec }
\int_{x_j}^{x_{j+1}} \FluxAvgDiffXSec dx
\end{equation}
where
\begin{equation}
\label{eq:Px}
P(x | E_\nu) = \frac{ 1 }{ \sigma(E_\nu) }
\frac{ d\sigma(E_\nu) }{ dx }
\end{equation}
is the conditional probability density of $x$ at fixed neutrino energy,
$d\sigma(E_\nu)/dx$ is the total differential cross section with respect to
$x$, and
\begin{equation}
\label{eq:flux_avg_diff_xsec}
\FluxAvgDiffXSec \equiv
\frac{ 1 }{ \Phi } \int_{E_\nu^\text{min}}^{E_\nu^\text{max}} \! \!
\phi(E_\nu) \, \frac{ d\sigma(E_\nu) }{ dx } \, dE_\nu
\end{equation}
is the flux-averaged total differential cross section. A Monte Carlo estimator
for the average value of this quantity in the $j$th bin may be obtained via
\begin{equation}
\label{eq:diff_xsec_from_events}
\FluxAvgDiffXSec_{\!\! j}
\equiv \frac{ 1 }{ \Delta x_j } \int_{x_j}^{x_{j+1}} \FluxAvgDiffXSec dx
\approx \frac{ \FluxAvgTotXSec f_j }{ \Delta x_j }\,,
\end{equation}
where $\Delta x_j \equiv x_{j + 1} - x_j$ is the bin width
and $f_j = n_j / N$ is the ratio of the $n_j$ events that fall
within the $j$th bin to the total number $N$ of simulated events.
Recognizing that $n_j$ follows a binomial distribution allows for an estimate
of the associated Monte Carlo statistical uncertainty (standard deviation)
via
\begin{equation}
\label{eq:stat_error_diff_xsec}
\text{SD}\!\left(\FluxAvgDiffXSec_{\!\! j}\right)
\approx \frac{ \FluxAvgTotXSec }{ \Delta x_j  \, N }
\sqrt{ \frac{ (N - n_j) \, n_j }{ N } } \,.
\end{equation}
The results from \cref{eq:diff_xsec_from_events,eq:stat_error_diff_xsec} may
readily be extended to multiple dimensions. For a Monte Carlo calculation of an
$n$-dimensional differential cross section, simply let $\Delta x_j$ denote the
product of the $n$ widths of the $j$th bin in the $n$-dimensional phase space.

For a discrete observable, similar expressions may be used with the bin width
$\Delta x_j$ omitted. The flux-averaged partial cross section
$\FluxAvgTotXSec_{\!j}$ to produce events fulfilling a criterion $j$ (e.g.,
involving emission of a single neutron) is estimated from the simulation
results via $\FluxAvgTotXSec_{\!j} \approx \FluxAvgTotXSec f_j$, where $f_j =
n_j / N$ is the fraction of simulated events satisfying the criterion. The
statistical uncertainty of this estimator is approximated by the expression on
the right-hand side of \cref{eq:stat_error_diff_xsec} with the substitution
$\Delta x_j \to 1$.

\subsubsection{Examples}

\begin{figure}
\centering
\includegraphics[height=0.270\textheight]{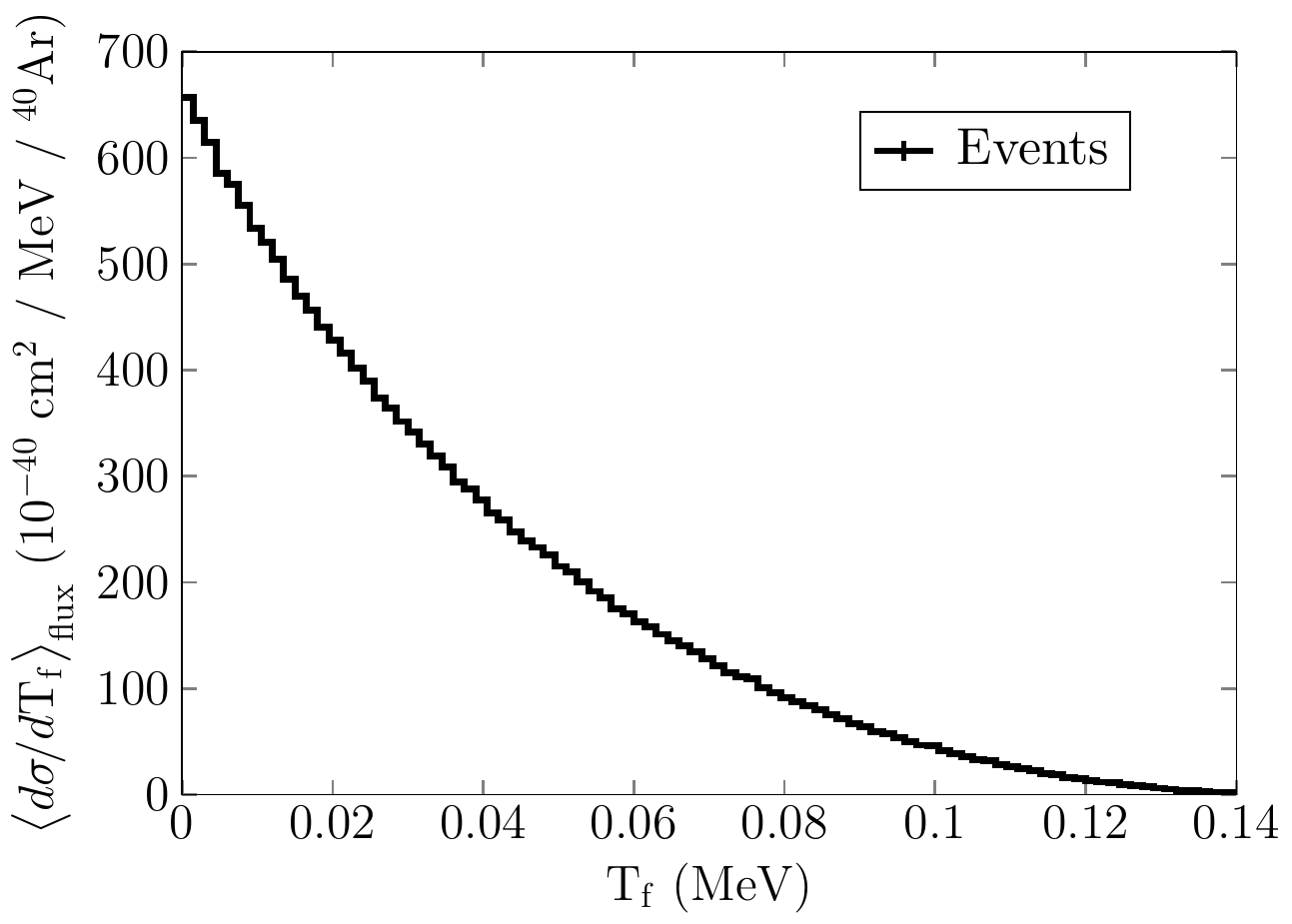}
\hfill
\includegraphics[height=0.270\textheight]{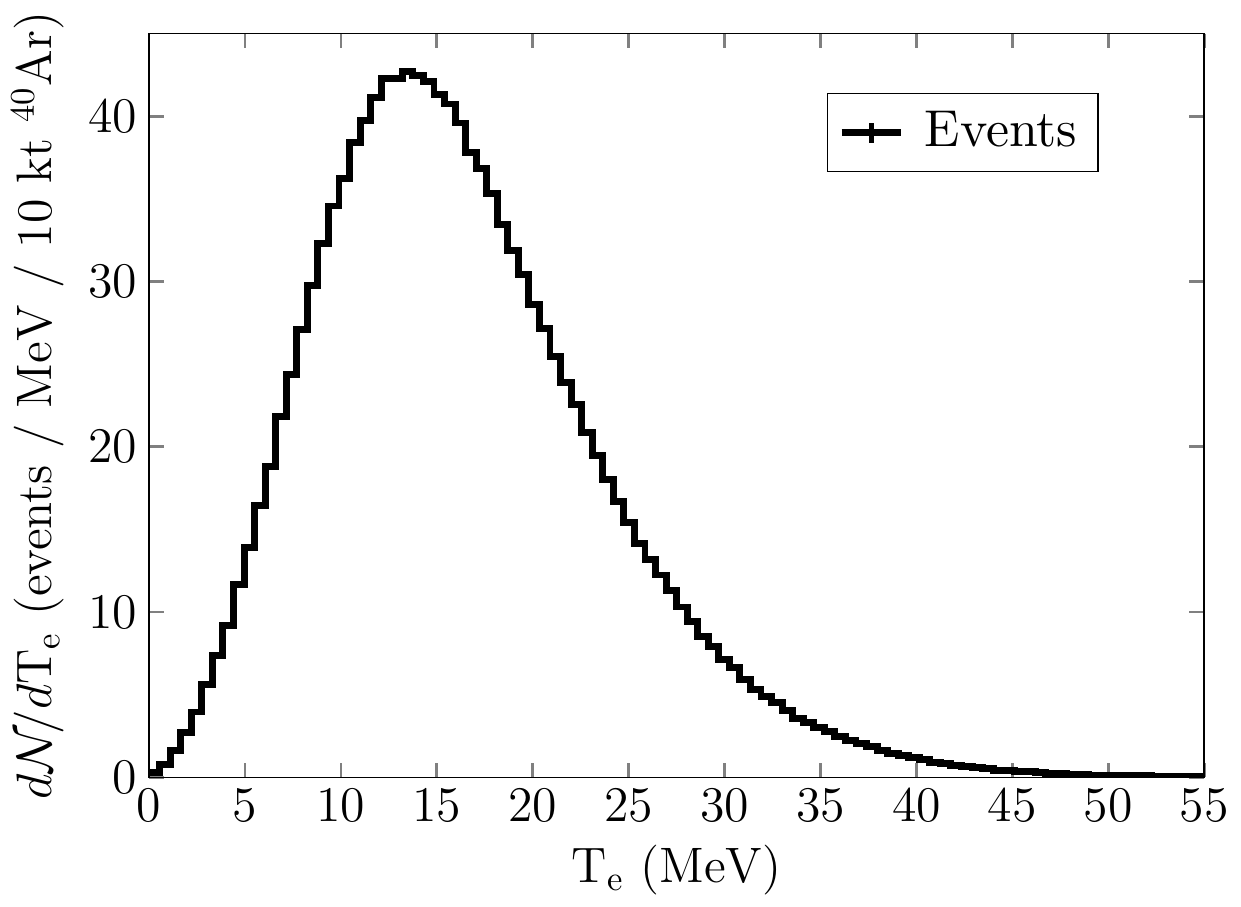}
\caption{Example \marley\ calculations of physics observables.
LEFT: The flux-averaged differential cross section for muon decay-at-rest
$\bar{\nu}_\mu$ undergoing coherent elastic neutrino-nucleus scattering on
\isotope[40]{Ar}. RIGHT: The differential event rate for
charged-current absorption of supernova $\nu_e$ on \SI{10}{\kilo\ton}
of pure \isotope[40]{Ar}. The Fermi-Dirac neutrino source
from \cref{lst:fd_source} was used as a toy model of the
time-integrated supernova $\nu_e$ spectrum.
}
\label{fig:xsec_demo}
\end{figure}

\Cref{fig:xsec_demo} shows the results of two example calculations of physics
observables performed using \marley\ events and the procedure outlined in
\cref{sec:xsec_normalization}. The histograms shown in the left and right
panels of \cref{fig:xsec_demo} were computed using independent simulations of
$N = \num{2e6}$ events each. To produce the left-hand plot, coherent elastic
neutrino-nucleus scattering (CEvNS) on \isotope[40]{Ar} was simulated for
$\bar{\nu}_\mu$ produced by $\mu^{+}$ decays at rest (see
\cref{sec:muon_decay_at_rest_source}). A histogram describing the distribution
of the kinetic energy $\mathrm{T_f}$ of the recoiling final-state nucleus was
prepared from the simulated events using a uniform bin width
$\Delta\mathrm{T_f} = \SI{1.5}{\keV}$. The event counts $n_j$ from each bin
were renormalized according to \cref{eq:diff_xsec_from_events} to yield a Monte
Carlo estimator for the mean value of the flux-averaged differential cross
section in the $j$th bin:
\begin{equation}
\label{eq:cevns_example_bin_content}
\left< \frac{ d\sigma }{ d\mathrm{T_f} } \right>_{\!\!j}
\approx \frac{ \FluxAvgTotXSec n_j }{ N \, \Delta\mathrm{T_f} } \,.
\end{equation}
Here $\FluxAvgTotXSec = \SI{26.69e-40}{\centi\meter\squared} / \,
\isotope[40]{Ar}$ is the \marley\ prediction for the CEvNS flux-averaged total
cross section. \Cref{eq:cevns_example_bin_content} gives the expression used to
obtain the content of each histogram bin shown in the left-hand plot of
\cref{fig:xsec_demo}. Based on \cref{eq:stat_error_diff_xsec}, an estimate of
the statistical uncertainty
\begin{equation}
\text{SD}\!\left( \left< \frac{ d\sigma }{ d\mathrm{T_f} }
\right>_{\!\!j} \right) \approx \frac{ \FluxAvgTotXSec }{ N \,
\Delta\mathrm{T_f} }
\sqrt{ \frac{ (N - n_j) \, n_j }{ N } }
\end{equation}
associated with each bin was also calculated but is small on the scale of the
plot.

A similar procedure was used to obtain the distribution shown in the right-hand
plot of \cref{fig:xsec_demo}, but the quantity of interest is $\mathrm{T_e}$,
the kinetic energy of the electron produced in the charged-current reaction
$\isotope[40]{Ar}(\nu_e, e^{-})\isotope[40]{K}^{*}$. The simulation of this
process was carried out using the reaction input file
\texttt{ve40ArCC\allowbreak\_\allowbreak Bhattacharya\allowbreak 1998.react}
(see \cref{sec:example_reaction_input_files}). The incident neutrino spectrum
was defined using the Fermi-Dirac source shown in \cref{lst:fd_source}, which
was treated as a toy model of the time-integrated $\nu_e$ flux at Earth
produced by a core-collapse supernova. The flux-averaged differential cross
section was converted into a differential event rate via
\begin{equation}
\label{eq:diff_event_rate_SN}
\frac{ d\mathcal{N} }{ d\mathrm{T_e} } = \Phi
\left< \frac{ d\sigma }{ d\mathrm{T_e} } \right>
n_\text{targets}
\end{equation}
where a total time-integrated flux\footnote{The quoted value is a rough
estimate for a core-collapse supernova at \SI{10}{\kilo\parsec} from Earth.} of
$\Phi = \num{1.0e11}\,\, \nu_e \, / \, \si{\cm\squared}$ was assumed. There are
$n_\text{targets} = \num{1.5e32}$ atoms in \SI{10}{\kilo\ton} of pure
\isotope[40]{Ar}.

\subsubsection{Energy-dependent total cross sections: \texttt{mardumpxs}}
\label{sec:mardumpxs}

One of the observables predicted by \marley\ that allows the most
straightforward comparison to other theoretical models is the
(abundance-weighted) total cross section as a function of neutrino energy
$\sigma(E_v)$. While a Monte Carlo estimate of this quantity may be computed
from generated events\footnote{Specifically, this involves generating events
using a uniform neutrino energy spectrum $\phi(E_\nu)$ and then constructing a
histogram of the neutrino energy distribution.} using the approach described in
\cref{sec:xsec_normalization}, $\sigma(E_\nu)$ is exactly calculable given only
the neutrino target composition (see \cref{sec:target_definition}) and the
information stored in the reaction input file(s) (see
\cref{sec:configure_reactions}). For the convenience of users, an example \cpp\
program called \texttt{mardumpxs} (\texttt{examples/\allowbreak
executables/\allowbreak mardumpxs.cc}) is provided with \marley\ that produces
tables of $\sigma(E_\nu)$. Assuming that the \texttt{setup\_marley.sh} script
has already been sourced (see \cref{sec:setup_marley_script}), one may build
\texttt{mardumpxs} by executing the commands
\begin{lstlisting}[backgroundcolor=\color{white}, numbers=none,
basicstyle=\normalsize\ttfamily, commentstyle=\normalsize\ttfamily]
cd ${MARLEY}/build
make mardumpxs
\end{lstlisting}
The \texttt{mardumpxs} executable takes the name of an output file followed by
the name of a \marley\ job configuration file (see \cref{sec:gen_config}) as
command-line arguments. For example, the command
\begin{lstlisting}[backgroundcolor=\color{white}, numbers=none,
basicstyle=\normalsize\ttfamily, commentstyle=\normalsize\ttfamily]
mardumpxs xsec_table.txt config.js
\end{lstlisting}
will write a table of $\sigma(E_\nu)$ values to the output file
\texttt{xsec\_table.txt} after configuring \marley\ according to the
settings given in \texttt{config.js}. Each line of the output file
will have the format
\begin{center}
\begin{tabular}{c}
\begin{lstlisting}[backgroundcolor=\color{white}, numbers=none]
KE XSec
\end{lstlisting}
\end{tabular}
\end{center}
where \texttt{KE} is the projectile kinetic energy\footnote{Since \marley\
treats neutrinos as massless, this is the same as the total energy $E_\nu$.
Kinetic energy is used by \texttt{mardumpxs} in anticipation of extensions to
\marley\ involving massive projectiles.} in \si{\MeV} and \texttt{XSec} is the
abundance-weighted total reaction cross section $\sigma(E_\nu)$ (see
\cref{sec:target_definition}) in \SI[per-mode=symbol,
per-symbol=\,/\,]{e-42}{\centi\meter\squared\per\atom}.

By default, \texttt{mardumpxs} assumes a $\nu_e$ projectile and produces a
table of \num{10000} cross section values using a regularly-spaced energy grid
between \SIrange[range-units=single, range-phrase=\text{~and~}]{0}{100}{\MeV}.
This behavior may be altered using \texttt{mardumpxs}-specific keys
in the job configuration file, as described in \cref{sec:mardumpxs_keys}.

\section{Interfacing MARLEY with external tools}
\label{sec:external_interfaces}

While it is hoped that \marley's capabilities as a standalone software package
will be beneficial to the low-energy neutrino physics community, interfacing
the generator with external codes has the potential to greatly extend its
usefulness. This is particularly true for neutrino detection experiments, which
typically rely on end-to-end simulations of neutrino production, neutrino
interactions in and around the detector, final-state particle propagation, and
the detector electronics response in order to interpret their measurements. For
some applications, simply passing \marley\ output files as input to the next
stage of a multi-step simulation may be satisfactory. In other contexts, direct
calls to \marley\ functions by a client code may be more appropriate.

\Cref{sec:gen_and_config} presents the recommended approach to generating
\marley\ events within an external \cpp\ application, which involves use of the
\texttt{marley\allowbreak::\allowbreak JSON\allowbreak Config} and
\texttt{marley\allowbreak::\allowbreak Generator} classes. An example
application of this kind, \texttt{marg4}, is discussed in \cref{sec:marg4}. The
\texttt{marg4} program produces \marley\ events and tracks the final-state
particles through a simple geometry using the popular Geant4
\cite{Geant4,Geant4Website} particle transport code. \Cref{sec:larsoft} then
briefly discusses the \marley\ interface included in the LArSoft
\cite{Church2013,Snider2017,Pordes2017,LArSoftWebsite} toolkit.

\subsection{\cpp\ event generation API}
\label{sec:gen_and_config}

The core functionality of \marley\ is encapsulated for use by external
applications in the form of the \texttt{marley\allowbreak::\allowbreak
Generator} class. While a \texttt{Generator} object may be instantiated and
configured without reference to a file, doing so is not recommended in most
situations. Instead, users are encouraged to construct \texttt{Generator}
objects indirectly by means of the \texttt{create\_generator} member function
of the \texttt{marley\allowbreak::\allowbreak JSON\allowbreak Config} class.
The constructor of \texttt{JSON\allowbreak Config} takes a single string
argument containing the name of a \marley\ job configuration file. The contents
of this file are parsed and used to initialize a \texttt{Generator} object
during a subsequent call to \texttt{JSON\allowbreak
Config\allowbreak::\allowbreak create\_\allowbreak generator}. With the
exception of the parameters described in \cref{sec:executable_settings} (which
are unique to the \texttt{marley} command-line executable), all configuration
options listed in \cref{sec:gen_config} will be recognized and respected by the
\texttt{JSON\allowbreak Config} class. Once a fully-initialized
\texttt{Generator} object has been created, a single neutrino scattering event
may be simulated and returned as a \texttt{marley\allowbreak::\allowbreak
Event} object by calling the \texttt{create\allowbreak\_event} member function.
A new event will be generated with each call to this function.

\Cref{lst:evgen_program} shows the source code for \texttt{examples/\allowbreak
executables/\allowbreak minimal/\allowbreak evgen.cc}, an example \cpp\ program
that uses the recommended event generation API. On line~\ref{line:evgen_cfg}, a
\texttt{JSON\allowbreak Config} object called \texttt{cfg} is constructed using
settings from the job configuration file \texttt{/home/config.js}. The settings
parsed by \texttt{cfg} are then used to create a \texttt{Generator} object
called \texttt{gen} on line~\ref{line:evgen_gen}. A simple event loop is
defined on lines~\ref{line:evgen_loop_begin}--\ref{line:evgen_loop_end} and
iterates until ten events have been generated (line~\ref{line:evgen_event}) and
printed in \ascii\ format to standard output
(line~\ref{line:evgen_loop_print}). The \texttt{evgen.cc} example program may
most easily be compiled by using the \texttt{marley-config} script as
described in \cref{sec:marley_config_compile}.

\begin{lstlisting}[language=C++, linewidth=\textwidth, caption={Minimal working
example of a \cpp\ program that generates \marley\ events}, captionpos=b,
label={lst:evgen_program}, basicstyle=\small\ttfamily, escapechar=|]
// Standard library includes
#include <iostream>

// MARLEY includes
#include "marley/Event.hh"|\label{line:evgen_event_header}|
#include "marley/Generator.hh"|\label{line:evgen_gen_header}|
#include "marley/JSONConfig.hh"|\label{line:evgen_config_header}|

constexpr int NUM_EVENTS = 10;

int main() {

  marley::JSONConfig cfg( "/home/config.js" );|\label{line:evgen_cfg}|
  marley::Generator gen = cfg.create_generator();|\label{line:evgen_gen}|

  for ( int j = 0; j < NUM_EVENTS; ++j ) {|\label{line:evgen_loop_begin}|
    marley::Event ev = gen.create_event();|\label{line:evgen_event}|
    std::cout << ev << '\n';|\label{line:evgen_loop_print}|
  }|\label{line:evgen_loop_end}|

}
\end{lstlisting}
In cases where handling of ROOT-dependent configuration file parameters (see,
e.g., \cref{sec:th1_and_tgraph}) is desired, the \texttt{Root\allowbreak
JSON\allowbreak Config} class should be used instead of \texttt{JSON\allowbreak
Config} . This amounts to making the replacement \texttt{JSON\allowbreak
Config} $\to$ \texttt{Root\allowbreak JSON\allowbreak Config} on
lines~\ref{line:evgen_config_header} and~\ref{line:evgen_cfg} of
\cref{lst:evgen_program}. Assuming that \marley\ has been built with \rootcern\
support, the compilation command given in \cref{sec:marley_config_compile} will
automatically link to the required \rootcern\ libraries.

\subsection{Interfacing MARLEY with Geant4}
\label{sec:marg4}

The \cpp-based Geant4 software framework provides a large suite of tools for
simulating particle propagation through matter. In the \texttt{main} function
of a typical Geant4 application, a \texttt{G4RunManager} object is constructed
and used to drive the simulation. Before the simulation can begin,
the \texttt{G4RunManager} must be initialized with pointers
to three objects, each of which is derived from a distinct abstract base
class defined by Geant4. The three required classes used to initialize the
run manager are
\begin{description}
\item[G4VUserDetectorConstruction] Defines the geometry and material
composition of the \textit{world volume} (and zero or more subvolumes) through
which the simulated particles will be tracked
\item[G4VUserPhysicsList] Defines the particle species and physics processes to
be included in the simulation
\item[G4VUserPrimaryGeneratorAction] Defines a member function,
\texttt{GeneratePrimaries}, which will be used to populate each new event with
the starting locations, momenta, etc.\ of all primary particles to be tracked
\end{description}
In addition to these required classes, pointers to objects that instantiate
optional \textit{user action} classes may also be registered with the run
manager. These allow user-defined functions to be executed at various stages of
the simulation. Further details about general Geant4 application development
are available in ref.~\cite{Geant4AppDevGuide}.

\subsubsection{The \texttt{marg4} Geant4 application}

The folder \texttt{examples/\allowbreak marg4/} contains the source code for
\texttt{marg4}, an example Geant4 application that uses \marley\ events as a
source of primary particles. The world volume is defined by the
\texttt{Detector\allowbreak Construction} class and consists of a single
uniform sphere of liquid argon with a radius of \SI{10}{\meter} and centered on
the origin. A built-in Geant4 physics list suitable for \si{\MeV}-scale
particle transport, \texttt{QGSP\_BIC\_HP}, is constructed using a factory
method in \texttt{examples/\allowbreak marg4/\allowbreak src/\allowbreak
marg4.cc}. As a trivial example of a user action, an \texttt{Event\allowbreak
Action} object is used to print the current event count to standard output at
the beginning of every hundredth event.

An example of a direct interface between \marley\ and Geant4 is provided by the
\texttt{Marley\allowbreak Primary\allowbreak Generator\allowbreak Action}
class, which is derived from \texttt{G4VUser\allowbreak Primary\allowbreak
Generator\allowbreak Action}. The constructor of this class takes a single
\texttt{std::string} argument containing the name of a \marley\ job
configuration file. This file name is used to create either a
\texttt{JSON\allowbreak Config} or a \texttt{Root\allowbreak JSON\allowbreak
Config} object, with the latter being chosen if the \texttt{USE\_ROOT}
preprocessor macro is defined (see \cref{sec:marley_config}). The member
variable \texttt{marley\_generator\_}, which is a \texttt{Generator} object, is
then initialized using the parameters from the configuration file.

\Cref{lst:marg4_excerpt} shows the \texttt{Generate\allowbreak Primaries}
member function defined by the \texttt{Marley\allowbreak Primary\allowbreak
Generator\allowbreak Action} class. This function is called once during
initialization of each Geant4 event. On line~\ref{line:G4PrimaryVertex}, a new
\texttt{G4PrimaryVertex} object is created at the spacetime origin. All primary
particles which are associated with it will begin their Geant4 trajectories at
the same 4-position. After a single \texttt{marley\allowbreak::\allowbreak
Event} object is created on \cref{line:G4Event}, each of its final-state
particles is converted into a new \texttt{G4\allowbreak Primary\allowbreak
Particle} by the loop defined on
lines~\ref{line:G4LoopStart}--\ref{line:G4LoopEnd}. Line~\ref{line:G4PP4}
associates each fully-initialized \texttt{G4\allowbreak Primary\allowbreak
Particle} with the primary vertex defined previously, and
line~\ref{line:G4AddVertex} adds the completed primary vertex to the current
\texttt{G4Event} object. Propagation of the primary particles obtained from the
\marley\ event is simulated by Geant4 after the \texttt{Generate\allowbreak
Primaries} function returns.

\begin{lstlisting}[linewidth=\textwidth,
caption={Definition of the \texttt{Generate\allowbreak Primaries} member
function of the \texttt{Marley\allowbreak Primary\allowbreak
Generator\allowbreak Action} class. This listing is an excerpt from the
file \texttt{examples/\allowbreak marg4/\allowbreak src/\allowbreak
Marley\allowbreak Primary\allowbreak Generator\allowbreak Action.cc}
included with \marley.}, captionpos=b,
label={lst:marg4_excerpt}, escapechar=|]
void MarleyPrimaryGeneratorAction::GeneratePrimaries(G4Event* anEvent)
{
  // Create a new primary vertex at the spacetime origin.
  G4PrimaryVertex* vertex = new G4PrimaryVertex(0., 0., 0., 0.); // x,y,z,t0|\label{line:G4PrimaryVertex}|

  // Generate a new MARLEY event using the owned marley::Generator object
  marley::Event ev = marley_generator_.create_event();|\label{line:G4Event}|

  // This line, if uncommented, will print the event in ASCII format
  // to standard output
  //std::cout << ev << '\n';|\label{line:G4ToStdOut}|

  // Loop over each of the final particles in the MARLEY event
  for ( const auto& fp : ev.get_final_particles() ) {|\label{line:G4LoopStart}|

    // Convert each one from a marley::Particle into a G4PrimaryParticle.
    // Do this by first setting the PDG code and the 4-momentum components.
    G4PrimaryParticle* particle = new G4PrimaryParticle( fp->pdg_code(),|\label{line:G4PP1}|
      fp->px(), fp->py(), fp->pz(), fp->total_energy() );|\label{line:G4PP2}|


    // Also set the charge of the G4PrimaryParticle appropriately
    particle->SetCharge( fp->charge() );|\label{line:G4PP3}|


    // Add the fully-initialized G4PrimaryParticle to the primary vertex
    vertex->SetPrimary( particle );|\label{line:G4PP4}|
  }|\label{line:G4LoopEnd}|

  // The primary vertex has been fully populated with all final-state particles
  // from the MARLEY event. Add it to the G4Event object so that Geant4 can
  // begin tracking the particles through the simulated geometry.
  anEvent->AddPrimaryVertex( vertex );|\label{line:G4AddVertex}|
}
\end{lstlisting}

\subsubsection{Building and running \texttt{marg4}}

Building the example \texttt{marg4} program requires the \texttt{geant4-config}
utility (included as part of a standard Geant4 installation) to be present on
the system's executable search path. Additionally, use of the
\texttt{QGSP\_BIC\_HP} physics list mentioned above requires installation of
the data files belonging to the Geant4 Neutron Data Library (G4NDL)
\cite{Mendoza2018,Mendoza2012}. These files, which are available for download
from the Geant4 website (\url{https://geant4.web.cern.ch/support/download}),
are required for high-precision (HP) tracking of low-energy neutrons by Geant4.
The use of Geant4's HP treatment of neutron transport is strongly recommended
for propagation of neutrino-induced neutrons from \marley\ events.

Assuming that the \texttt{setup\_marley.sh} script has already been sourced
(see \cref{sec:setup_marley_script}), the \texttt{marg4} executable may
be built against Geant4 via the commands
\begin{lstlisting}[backgroundcolor=\color{white}, numbers=none,
basicstyle=\normalsize\ttfamily, commentstyle=\normalsize\ttfamily]
cd ${MARLEY}/build
make marg4
\end{lstlisting}
The \texttt{marg4} executable takes the number of desired events
followed by the name of a \marley\ job configuration file as its
command-line arguments. For example, invoking the program via
\begin{lstlisting}[backgroundcolor=\color{white}, numbers=none,
basicstyle=\normalsize\ttfamily, commentstyle=\normalsize\ttfamily]
marg4 500 /home/myconfig.js
\end{lstlisting}
will simulate \num{500} Geant4 events. Each of these events will contain a
single primary vertex populated with the final-state particles from one
\marley\ event. The \marley\ events will be generated using the job
configuration given in the file \texttt{/home/\allowbreak myconfig.js}.

The \texttt{marg4} program is provided as a simple usage example for the API
described in \cref{sec:gen_and_config}. As such, it does not produce any output
other than logging messages from both \marley\ and Geant4. Interested users are
encouraged to copy and modify the \texttt{marg4} source code to meet their
needs. Information from the \texttt{marley\allowbreak::\allowbreak Event}
objects themselves may be accessed using the member functions described in
\cref{sec:event_access}. The functions needed to extract quantities of interest
from the Geant4 simulation are documented in ref.~\cite{Geant4AppDevGuide}.

\subsection{LArSoft interface}
\label{sec:larsoft}

The liquid argon software toolkit (LArSoft) provides a flexible simulation,
reconstruction, and analysis framework designed for liquid argon time
projection chamber (LArTPC) experiments. Based on the \texttt{art}
event-processing framework \cite{Green2012}, LArSoft has been adopted as a key
part of the software infrastructure for various experimental collaborations,
including ArgoNeuT \cite{Anderson2012}, LArIAT \cite{Acciarri2020}, MicroBooNE
\cite{Acciarri2017}, ICARUS \cite{Rubbia2011}, SBND \cite{McConkey2017}, and
DUNE \cite{DUNEtdrVol1}.

The LArSoft source code is hosted on GitHub and split into multiple
respositories, each of which provides a particular type of functionality. The
\texttt{larsim} respository (\url{https://github.com/LArSoft/larsim}) includes,
among other things, interfaces to external physics event generators for
neutrino interactions (e.g., GENIE) and for other processes (e.g., the
cosmic-ray generators CORSIKA \cite{CORSIKA} and CRY \cite{CRY}).

Version 6.04.00 of LArSoft was the first\footnote{Ref.~\cite{Gardiner2018}
incorrectly identifies the initial LArSoft version as 6.03.00. This number
refers instead to the corresponding version of the \texttt{larsim} subpackage.}
to include a direct \texttt{larsim} interface to \marley\ contributed by the
present author.\footnote{The first \marley\ version to be included in LArSoft
was 0.9.5, an August 2016 public beta release.} Ever since the initial version
of the interface was added, \marley\ has been included as part of the standard
LArSoft distribution. A detailed description of the LArSoft interface to
\marley, which is more sophisticated than the examples given in
\cref{sec:gen_and_config} and \cref{sec:marg4}, is beyond the present scope.
However, brief descriptions of the relevant \cpp\ classes defined in the
\texttt{larsim} source code (see the \texttt{larsim/\allowbreak
Event\allowbreak Generator/\allowbreak MARLEY/} subfolder of the
\texttt{larsim} repository) are given below. All four of these classes are
defined within the \texttt{evgen} namespace used by LArSoft for event
generation.

\begin{description}

\item[ActiveVolumeVertexSampler] Used to sample neutrino vertex
positions uniformly over the active volume(s) of a detector. The approach used
by this class is only suitable for simulations in which
\begin{enumerate*}[label=(\arabic*)]
\item the detector is uniformly illuminated by the incident neutrino flux, and
\item neutrino interactions occurring outside the detector active volume(s) are
not of interest.
\end{enumerate*}

\item[MarleyGen] Creates neutrino scattering events in the native
LArSoft format using the \texttt{Active\allowbreak Volume\allowbreak
Vertex\allowbreak Sampler} and \texttt{MARLEY\allowbreak Helper} classes

\item[MARLEYHelper] Implements the low-level interface between \marley\ and
LArSoft. At the beginning of a simulation job, this class initializes a
\texttt{marley\allowbreak::\allowbreak Generator} object by converting a
LArSoft configuration given in the Fermilab Hierarchical Configuration Language
(FHiCL) \cite{FHiCLSpec} to the JSON-like format (see \cref{sec:gen_config})
used by \marley. With the exception of the ``executable settings'' described in
\cref{sec:executable_settings}, all \marley\ job configuration file parameters
are available for use via FHiCL.

This class also provides a member function (\texttt{create\_MCTruth}) which
generates a \marley\ event and stores a representation of it in an instance of
the \texttt{simb\allowbreak::\allowbreak MCTruth} class. The
\texttt{simb\allowbreak::\allowbreak MCTruth} class is used by LArSoft as a
generator-agnostic event record.

\item[MarleyTimeGen] Similar to \texttt{MarleyGen}, but provides experimental
support for generating events using a time-dependent neutrino spectrum
\end{description}

Further technical details about these and other LArSoft classes are available
in the LArSoft Doxygen documentation
(\url{https://nusoft.fnal.gov/larsoft/doxsvn/html/index.html}).

\section{Prospects for future development}
\label{sec:prospects}

This work describes a new Monte Carlo event generator, \marley, suitable for
simulating tens-of-\si{\MeV} neutrino scattering on complex nuclei. The release
of \marley\ \version, the first version of the code to be documented with the
present level of detail, represents a significant milestone. However, active,
open-source development of \marley\ is ongoing, and contributions from the
community are encouraged. To coordinate development efforts,
those interested in contributing improvements to \marley\ are asked to contact
the author at the earliest opportunity.

Potential avenues for future development of \marley\ include the following:

\begin{itemize}

\item Preparation of additional reaction input files. In addition to new
scattering modes (inelastic NC and $\bar{\nu}_e$ CC) for \isotope[40]{Ar}, the
addition of input files for several other nuclides will likely be of immediate
interest. These include but are not limited to
\begin{enumerate}
\item \isotope[12]{C} and \isotope[16]{O}, both of which make
small but non-negligible contributions to the total event rate
in hydrocarbon (e.g., NOvA \cite{NOvAsn}) and water Cherenkov
(e.g., Super-Kamiokande \cite{Ikeda2007}) supernova neutrino detectors
\item The stable lead isotopes (especially \isotope[208]{Pb}), which
serve as the target material for the HALO \cite{Duba2008,Zuber2015} detector
\item The stable iron and copper isotopes (especially \isotope[56]{Fe} and
\isotope[63]{Cu}), which, together with lead, are under study by the COHERENT
\cite{COH2018} experiment in an effort to understand low-energy
neutrino-induced neutron production
\end{enumerate}

\item Inclusion of forbidden transitions in the \marley\ neutrino-nucleus cross
section model. These are currently neglected via the allowed approximation.
Recent theoretical calculations using a Continuum Random Phase Approximation
(CRPA) approach \cite{CRPA1,CRPA2} indicate that the forbidden contributions
can have a significant impact on both the total cross section and the kinematic
distributions of the outgoing lepton.

\item Addition of new keys to the job configuration file format to allow
variations of model parameters, e.g., those used by the nuclear optical
model. With some related enhancements to other parts of the code,
e.g., storage of the full de-excitation history in the \marley\ event
record, these variations could be used to assess theoretical uncertainties
on \marley\ predictions via event reweighting. An approach of this kind
is commonly used by accelerator neutrino experiments. Documentation of the
GENIE event reweighting framework, for example, is given in
ref.~\cite[ch. 9, pp. 129--154]{GENIEManual}.

\item Improvements to the \marley\ treatment of nuclear de-excitations. These
could include
  \begin{itemize}
    \item Realistic angular distributions for the de-excitation products
    (instead of the isotropic emission that is currently assumed)
    \item Storage of particle emission times in the event record. Related
    updates could also be made to the nuclear structure data format to
    allow measured discrete level lifetimes to be listed.
    \item Competition of internal conversion with $\gamma$-ray emission
    \item Pre-equilibrium emission of nuclear fragments
    \item Modeling of neutrino-induced fission
  \end{itemize}

\item Creation of new subclasses of \texttt{marley\allowbreak::\allowbreak
Reaction} to simulate events for projectiles other than neutrinos. Processes
likely to be of interest include electron-nucleus scattering and beyond the
Standard Model reactions like nuclear absorption of \si{\MeV}-scale dark matter
\cite{DMAbsorb}.

\item Implementation of a tool to facilitate comparisons of \marley\ model
predictions with cross section measurements from low-energy neutrino
experiments. This could potentially be accomplished by adding low-energy
datasets and a \marley\ interface to the NUISANCE \cite{Stowell2017} framework
used by the accelerator neutrino community.

\item Simulation of events within a detector geometry in which both the target
material and the incident neutrino flux are spatially non-uniform. This may
most easily be achieved by interfacing \marley\ with flux and geometry
navigation drivers present in an existing generator (see
ref.~\cite[sec.~6]{GENIEManual} for a description of GENIE's). A ``community
based'' version of these tools, independent of any particular generator, has
also been proposed \cite{Barrow2020}.

\end{itemize}

\section{Acknowledgements}

I am grateful to Myung-Ki Cheoun for providing the QRPA Gamow-Teller matrix
elements which are used to compute $\isotope[40]{Ar}(\nu_e,
e^{-})\isotope[40]{K}^{*}$ cross sections at high excitation energies (see
\cref{sec:example_reaction_input_files} and ref.~\cite{marleyPRC}). I also
thank the authors of the TALYS code for sharing their nuclear level data files
under the GNU General Public License.

Vishvas Pandey provided helpful feedback on a draft of this paper, and I thank
Robert Svoboda, Ramona Vogt, and Michael Mulhearn for their comments on the PhD
thesis \cite{Gardiner2018} which served as the first formal description of
\marley. I am also indebted to Sam Hedges and Erin Conley for their tests of
the \marley\ \version\ release candidate.

My work to develop \marley\ while at the University of California, Davis was
supported in part by the John Jungerman-Charles Soderquist Graduate Fellowship
and by the DOE National Nuclear Security Administration through the Nuclear
Science and Security Consortium under award number DE-NA0003180.

This manuscript has been authored by Fermi Research Alliance, LLC under
Contract No. DE-AC02-07CH11359 with the U.S. Department of Energy, Office of
Science, Office of High Energy Physics.

\appendix

\section{Nuclear structure data file format}
\label{sec:nuc_struct}

The \marley\ nuclear structure data files mentioned in
\cref{sec:load_nuclear_structure} contain tables of nuclear energy levels
and $\gamma$-ray branching ratios for one or more nuclides. Each table follows
a whitespace-delimited format that begins with the header
\begin{center}
\begin{tabular}{c}
\vspace{-1.5\baselineskip}
\begin{lstlisting}[backgroundcolor=\color{white}, numbers=none]
Z A num_levels
\end{lstlisting}
\end{tabular}
\end{center}
in which \verb|Z| is the proton number, \verb|A| is the mass number, and
\verb|num_levels| is the number of tabulated nuclear levels. The header is
followed by data blocks for each level in order of increasing
excitation energy. Each block begins with the line
\begin{center}
\begin{tabular}{c}
\begin{lstlisting}[backgroundcolor=\color{white}, numbers=none]
Ex twoJ Pi num_gammas
\end{lstlisting}
\end{tabular}
\end{center}
where \verb|Ex| is the level excitation energy (\si{MeV}), \verb|twoJ| is two
times the level spin (the factor of two allows half-integer spins to be
represented by an integer), and \verb|Pi| is the parity (denoted by the
character \verb|+| or \verb|-|). The data block for a level terminates with a
set of lines describing each available $\gamma$-ray transition (for a total of
\verb|num_gammas| transitions). Each of these lines has the format
\begin{center}
\begin{tabular}{c}
\begin{lstlisting}[backgroundcolor=\color{white}, numbers=none]
E_gamma RI Lf_index
\end{lstlisting}
\end{tabular}
\end{center}
where \verb|E_gamma| is the energy of the emitted $\gamma$-ray (\si{\MeV}) and
\verb|RI| is the relative intensity of the transition. Although the relative
intensities in the official \marley\ structure data files are normalized to sum
to unity, this is not required. The parameter \verb|Lf_index| gives the
position of the final nuclear level accessed by the transition, with
$\verb|Lf_index| = 0$ corresponding to the first listed level (presumably the
ground state).

\begin{lstlisting}[linewidth=\textwidth,caption={Example \marley\ nuclear
structure data file},captionpos=b,label={lst:structure1}]
17 43 5
  0 1 + 0
  0.329 3 + 1
    0.329 1 0
  0.944 5 + 1
    0.615 1 1
  1.34 5 + 1
    1.34 1 0
  1.829 7 + 1
    0.885 1 2
\end{lstlisting}

\Cref{lst:structure1} shows the table for \isotope[43]{Cl} which appears in the
structure data file \texttt{data/\allowbreak structure/\allowbreak Cl.dat}
included with \marley. Five nuclear levels appear in the table, each (apart
from the ground state) having one listed $\gamma$-ray transition. Although the
lines giving the nuclide header, level descriptions, and $\gamma$-ray
descriptions are shown here with different indentations to improve readability,
this is not required by the file format.

\section{Reaction input file format}
\label{sec:reaction_file_format}

The reaction input files mentioned in \cref{sec:reaction_config} provide
information needed for \marley\ to configure cross section calculations
according to \cref{eq:diff_xsec} (for neutrino-nucleus scattering) or
\cref{eq:diff_xsec_nu_e} (for neutrino-electron elastic scattering). Any line
of a reaction input file that begins with a \texttt{\#} character will be
treated as a comment and ignored by the parser. The first line of the file
that is not a comment contains a header of the form
\begin{center}
\begin{tabular}{c}
\vspace{-1.5\baselineskip}
\begin{lstlisting}[backgroundcolor=\color{white}, numbers=none]
ProcessType NucPDG
\end{lstlisting}
\end{tabular}
\end{center}
where \texttt{ProcessType} is an integer code representing the reaction mode
and \texttt{NucPDG} is a nuclear PDG code (see \cref{sec:PDG_codes})
identifying the target nuclide involved in the initial state.
\Cref{tab:process_types} lists the allowed values of \texttt{ProcessType} in
the second column together with their corresponding reaction modes and isospin
operators. In the first column, the symbol $A$ is used as a stand-in for any
target nucleus. The third column lists the elements of the enumerated type used
in the source code to represent each reaction mode.

\begin{table}
\centering
\begin{tabular}{cSrc}
\toprule
Reaction mode & {\texttt{ProcessType} integer} & \texttt{ProcessType} enum & Isospin operator \\
\midrule
$A(\nu, \ell^{-})$ & 0 & \texttt{NeutrinoCC} & $t_{-}$ \\
$A(\bar{\nu}, \ell^{+})$ & 1 & \texttt{AntiNeutrinoCC} & $t_{+}$ \\
$A(\nu, \nu)$ or $A(\bar{\nu}, \bar{\nu})$ & 2 & \texttt{NC} & $t_{3}$ or $1$ \\
$e^{-}(\nu, \nu)$ or $e^{-}(\bar{\nu}, \bar{\nu})$ & 3 & \texttt{NuElectronElastic}
& none \\
\bottomrule
\end{tabular}
\caption{Reaction mode labeling scheme used in \marley\ \version. The integer
labels used in reaction input files are represented in the source code by the
\texttt{marley::Reaction::ProcessType} enumerated type.}
\label{tab:process_types}
\end{table}

For reaction input files describing neutrino-electron elastic scattering
($\text{\texttt{ProcessType}} = 3$), only the header shown above is required to
be present. If the user wishes to enable simulation of this process for
multiple atomic targets, additional values of \texttt{NucPDG} may be included
on subsequent lines. \Cref{lst:reactES} shows an example reaction input file
which instructs \marley\ to simulate neutrino-electron elastic scattering for
the three stable isotopes of argon: \isotope[36]{Ar}, \isotope[38]{Ar}, and
\isotope[40]{Ar}.
\begin{lstlisting}[linewidth=\textwidth,caption={Example \marley\ reaction
input file for neutrino-electron elastic scattering},captionpos=b,
label={lst:reactES}]
3 1000180360
1000180380
1000180400
\end{lstlisting}

For neutrino-nucleus reaction modes, each line of the input file following the
header is used to specify the value of a reduced matrix element describing a
transition to a particular final nuclear level. These lines are
whitespace-delimited and have the format
\begin{center}
\begin{tabular}{c}
\begin{lstlisting}[backgroundcolor=\color{white}, numbers=none]
Ex B type
\end{lstlisting}
\end{tabular}
\end{center}
where \texttt{Ex} is the excitation energy of the final level (\si{\MeV},
measured with respect to the ground state of the residual nucleus), \texttt{B}
is the corresponding value of either $B(\mathrm{F})$ or $B(\mathrm{GT})$, and
\texttt{type} is an integer code that represents whether \texttt{B} should be
interpreted as a value of $B(\mathrm{F})$ ($\text{\texttt{type}} = 0$) or
of $B(\mathrm{GT})$ ($\text{\texttt{type}} = 1$). The matrix elements must
be listed in order of increasing \texttt{Ex} in the reaction input file.
If this is not the case, then \marley\ will halt with the error message
\begin{lstlisting}[backgroundcolor=\color{white}, numbers=none,
backgroundcolor=\color{lightgray},
basicstyle=\small\ttfamily\bfseries, commentstyle=\small\ttfamily\bfseries]
[ERROR]: Invalid reaction dataset. Level energies must be unique and must be given in ascending order.
\end{lstlisting}
when the file is used.

In most cases, the matrix element values given in the reaction input file
should be computed using the full expressions shown in \cref{eq:BF} and
\cref{eq:BGT}. This includes, e.g., any assumed value of the axial-vector
coupling constant $g_A$. The sole exception is the Fermi matrix element for
neutral-current scattering, which should be listed with a factor of $Q_W^2 / 4$
removed. For an NC transition to the ground state of a spin-zero nucleus, $B(F)
= g_V^2 \, Q_W^2 / 4$, but $\text{\texttt{B}} = g_V^2 = 1$.

\Cref{lst:react1} shows an excerpt from the reaction input file
\texttt{data/\allowbreak react/\allowbreak ve40ArCC\allowbreak\_\allowbreak
Bhattacharya\allowbreak 1998.react}, which tabulates nuclear matrix
elements for CC neutrino-argon scattering. Two $B(\mathrm{GT})$ values
(\num{0.90} and \num{1.50}) are given, both for final states between
\SIrange[range-phrase=~and~, range-units=single]{2}{3}{\MeV} above the
\isotope[40]{K} ground state.
\begin{lstlisting}[linewidth=\textwidth,caption={Example \marley\ reaction
matrix element data file},captionpos=b,label={lst:react1}]
0 1000180400
2.289868 0.90 1
2.730357 1.50 1
\end{lstlisting}

\bibliographystyle{elsarticle-num-names}
\bibliography{marleycpc.bib}


\end{document}